\documentclass[superscriptaddress,groupedaddress,nofootnoteinbib,11pt,notoc]{article}
\pdfoutput=1
\usepackage{jcappub}

\usepackage{graphicx,color}
\usepackage{appendix}
\usepackage{latexsym,amsmath,amssymb,graphicx,booktabs}
\usepackage{epsfig,latexsym}
\usepackage{hyperref}
\numberwithin{equation}{section}

\definecolor{MyBlue}{rgb}{0.15,0.15,0.70}

\hypersetup{
colorlinks=true,
citecolor=MyBlue,
linkcolor=MyBlue,
urlcolor=MyBlue
}

\definecolor{lightgray}{gray}{0.9}

\newcommand{\DP}{\Delta_4}

\newcommand{\nn}{\nonumber}

\newcommand{\Lrr}{\Lambda_{\rm\scriptscriptstyle RR}}

\newcommand{\iBox}{\Box^{-1}}

\newcommand{\Fmn}{F_{\mu\nu}}
\newcommand{\FMN}{F^{\mu\nu}}
\newcommand{\Am}{A_{\mu}}

\renewcommand\({\left(}
\renewcommand\){\right)}
\renewcommand\[{\left[}
\renewcommand\]{\right]}

\newcommand\n{{\mbox {\boldmath $\nabla$}}}
\newcommand{\ra}{\rightarrow}

\def\lsim{\raise 0.4ex\hbox{$<$}\kern -0.8em\lower 0.62
ex\hbox{$\sim$}}

\def\gsim{\raise 0.4ex\hbox{$>$}\kern -0.7em\lower 0.62
ex\hbox{$\sim$}}

\def\lbar{{\hbox{$\lambda$}\kern -0.7em\raise 0.6ex
\hbox{$-$}}}

\newcommand\eq[1]{eq.~(\ref{#1})}
\newcommand\eqs[2]{eqs.~(\ref{#1}) and (\ref{#2})}
\newcommand\Eq[1]{Equation~(\ref{#1})}

\newcommand\eqst[2]{eqs.~(\ref{#1})--(\ref{#2})}

\newcommand\pa{\partial}
\newcommand\p{\partial}

\newcommand\ee{\end{equation}}
\newcommand\be{\begin{equation}}
\def\bea{\begin{array}}
\def\eea{\end{array}}\def\ea{\end{array}}
\newcommand\ees{\end{eqnarray}}
\newcommand\bees{\begin{eqnarray}}
\def\nn{\nonumber}

\def\s{\sigma}
\def\g{\gamma}

\def\d{\delta}

\def\dslash{\hspace{-1mm}\not{\hbox{\kern-2pt $\partial$}}}
\def\Dslash{\not{\hbox{\kern-2pt $D$}}}
\def\pslash{\not{\hbox{\kern-2.1pt $p$}}}
\def\kslash{\not{\hbox{\kern-2.3pt $k$}}}
\def\qslash{\not{\hbox{\kern-2.3pt $q$}}}

\newcommand{\vac}{|0\rangle}
\newcommand{\cav}{\langle 0|}

\newcommand{\vk}{{\bf k}}
\newcommand{\vx}{{\bf x}}

\def\p1{{\bf p}_1}
\def\p2{{\bf p}_2}
\def\k1{{\bf k}_1}
\def\k2{{\bf k}_2}

\newcommand{\emn}{\eta_{\mu\nu}}

\newcommand{\gmn}{g_{\mu\nu}}

\newcommand{\gMN}{g^{\mu\nu}}

\newcommand{\gbmn}{\bar{g}_{\mu\nu}}

\newcommand{\hmn}{h_{\mu\nu}}

\newcommand{\pam}{\pa_{\mu}}

\newcommand{\paM}{\pa^{\mu}}

\newcommand{\Rmn}{R_{\mu\nu}}
\newcommand{\Gmn}{G_{\mu\nu}}
\newcommand{\RMN}{R^{\mu\nu}}

\newcommand{\Tmn}{T_{\mu\nu}}
\newcommand{\Smn}{S_{\mu\nu}}

\newcommand{\TMN}{T^{\mu\nu}}

\newcommand{\dddM}{\kern 0.2em \raise 1.9ex\hbox{$...$}\kern -1.0em \hbox{$M$}}
\newcommand{\dddQ}{\kern 0.2em \raise 1.9ex\hbox{$...$}\kern -1.0em \hbox{$Q$}}
\newcommand{\dddI}{\kern 0.2em \raise 1.9ex\hbox{$...$}\kern -1.0em\hbox{$I$}}
\newcommand{\dddJ}{\kern 0.2em \raise 1.9ex\hbox{$...$}\kern-1.0em
\hbox{$J$}}
\newcommand{\dddcalJ}{\kern 0.2em \raise 1.9ex\hbox{$...$}\kern-1.0em
\hbox{${\cal J}$}}

\newcommand{\dddO}{\kern 0.2em \raise 1.9ex\hbox{$...$}\kern -1.0em
\hbox{${\cal O}$}}
\def\dddz{\raise 1.5ex\hbox{$...$}\kern -0.8em \hbox{$z$}}
\def\dddd{\raise 1.8ex\hbox{$...$}\kern -0.8em \hbox{$d$}}
\def\dddbd{\raise 1.8ex\hbox{$...$}\kern -0.8em \hbox{${\bf d}$}}
\def\ddbd{\raise 1.8ex\hbox{$..$}\kern -0.8em \hbox{${\bf d}$}}
\def\dddx{\raise 1.6ex\hbox{$...$}\kern -0.8em \hbox{$x$}}

\newcommand{\mplr}{m_{\rm Pl}}

\newcommand{\oma}{\Omega_{M}}
\newcommand{\ora}{\Omega_{R}}

\newcommand{\ola}{\Omega_{\Lambda}}

\newcommand{\rde}{\rho_{\rm DE}}
\newcommand{\wde}{w_{\rm DE}}

\graphicspath{ {./Graphs/} }

\title{Nonlocal gravity. Conceptual aspects and cosmological predictions}
\author{Enis Belgacem, Yves Dirian, Stefano Foffa and Michele Maggiore}
\affiliation{D\'epartement de Physique Th\'eorique and Center for Astroparticle Physics,  
Universit\'e de Gen\`eve, 24 quai Ansermet, CH--1211 Gen\`eve 4, Switzerland}

\abstract{Even if the fundamental action of gravity is local, the corresponding quantum effective action, 
that includes the effect of quantum fluctuations, is a nonlocal object. 
These nonlocalities are well understood in the ultraviolet regime but much less in the infrared, where they could in principle give rise to important cosmological effects.  Here we systematize and extend previous work of our group, in which it is assumed that a mass scale $\Lambda$ is dynamically generated in the infrared, giving  rise to nonlocal terms in the quantum effective action of gravity. 
We give a detailed discussion of conceptual aspects related to nonlocal gravity (including causality, degrees of freedom, ambiguities related to the boundary conditions of the nonlocal operator, scenarios for the emergence of a dynamical scale in the infrared)
and of the cosmological consequences of these models. The requirement of providing a viable cosmological evolution severely restricts the  form of the  nonlocal terms, and selects a model (the so-called RR model) that corresponds to a dynamical mass generation for the conformal mode. For such a model:
(1) there is a FRW background  evolution, where the nonlocal term acts as an effective dark energy with a phantom equation of state, providing accelerated expansion without a cosmological constant. (2) Cosmological perturbations are well behaved. (3) Implementing the model in a Boltzmann code and  comparing with observations we  find that 
the  RR model  fits the CMB, BAO, SNe, structure formation data and local $H_0$ measurements at a level statistically equivalent to $\Lambda$CDM. (4) Bayesian parameter estimation shows that
the value of $H_0$ obtained  in the RR model   is higher than in $\Lambda$CDM, reducing to $2.0\sigma$ the  tension with the value  from local measurements. (5) The RR model provides a prediction for the sum of neutrino masses that falls within the limits set by oscillation and terrestrial experiments (in contrast to $\Lambda$CDM, where letting  the sum of neutrino masses vary as a free  parameter within these limits, one hits the lower bound).   (6) Gravitational waves propagate at the speed of light, complying with the limit from GW170817/GRB 170817A.}

\begin{document}
\maketitle
\flushbottom

\section{Introduction}\label{sec:intro}

Cosmological models are usually developed starting from a classical action, whether the Einstein--Hilbert action coupled to matter or some modified gravity theory, and studying the background evolution and the cosmological perturbations derived from  it. As a matter of principle, however, once one includes quantum fluctuations the relevant quantity is the quantum effective action. A crucial difference between the classical action and the quantum effective action is that, while the former is local, the latter   has also nonlocal terms whenever the theory contains massless or light particles. In gravity, because of the quantum fluctuations associated to the massless graviton, nonlocal terms are  unavoidably present, and could in principle  significantly affect the infrared (IR) behavior of the theory.

The techniques for computing these nonlocal terms in the ultraviolet (UV) regime are  well understood~\cite{Birrell:1982ix,Barvinsky:1985an,Barvinsky:1987uw,Buchbinder:1992rb,Mukhanov:2007zz,Shapiro:2008sf}.
In the IR, in contrast, the situation is much less clear.  Because of its large symmetry,
de~Sitter space has often been  used has a playground for studying IR effects in gravity. Infrared divergences appear either when we consider  massless matter fields in a fixed de~Sitter background or, in pure gravity, because of graviton fluctuations.
In the case of a massless, minimally-coupled scalar field in de~Sitter space with a $\lambda\phi^4$ interaction, it has been conclusively shown, using several different approaches, that the scalar field develops a dynamical mass $
m_{\rm dyn}^2\propto H^2\sqrt{\lambda}$ (see \cite{Starobinsky:1994bd,Riotto:2008mv,Burgess:2009bs,Rajaraman:2010xd,Beneke:2012kn,Gautier:2013aoa,Nacir:2016fzi} and references therein). In the case of pure gravity the  existence of large  IR effects in gauge-invariant quantities has been the subject of many investigations.  In particular, strong  IR effects appear for the conformal mode  \cite{Antoniadis:1986sb, Antoniadis:1991fa}, but the whole issue of IR effects in de~Sitter space is  quite controversial; 
see e.g. \cite{Tsamis:1994ca,Miao:2011ng,Rajaraman:2016nvv}.

Given the difficulty of a pure top-down approach, an alternative and more phenomenological strategy is to investigate the effect that   nonlocal terms  might have on the cosmological evolution (we will discuss below why IR effects can  manifest themselves as nonlocal terms in the quantum effective action).
This strategy turns out to be fruitful because we will find that it is highly nontrivial to construct a nonlocal model such that: (1) at the background level, the nonlocal term behaves as a dynamical dark energy, giving accelerated expansion even in the absence of a cosmological constant. (2) cosmological perturbation theory is well behaved, and (3) the results fit well the cosmological observations, at a level comparable to $\Lambda$CDM. As we will see, these requirements  allow us to select  a specific  nonlocal term. In turn, with this information in hand, one might then try to go back to the more difficult problem of deriving this nonlocal term   in the quantum effective action from a fundamental action, such as the Einstein-Hilbert action.

To our knowledge, the first nonlocal gravity model was proposed by Wetterich
\cite{Wetterich:1997bz}, who added to the Ricci scalar in the Einstein--Hilbert action a term proportional to $R\iBox R$.\footnote{More precisely, in \cite{Wetterich:1997bz} was considered a term of the form
$R[\Box +\xi R+{\cal O}(R^2)]^{-1}R$ in Euclidean space, where the ${\cal O}(R^2)$ were chosen to ensure positivity of the operator. Observe however that, for cosmological applications, time derivatives are always much larger than spatial derivatives, so even the Minkowskian $\Box$ operator is never singular.}
Since $\iBox R$ is dimensionless, the associated coupling constant is also dimensionless. The model however did not produce a viable cosmological evolution. Deser and Woodard \cite{Deser:2007jk} considered a more general class of models with a nonlocal term of the form $Rf(\iBox R)$, with $f(X)$ a dimensionless function, which is tuned so to obtained the desired background evolution (see \cite{Woodard:2014iga} for review). The model is therefore not predictive as far as the background evolution is concerned but, once fixed $f(X)$ in this way, one can perform cosmological perturbation theory and compare with the data. This has been done for structure formation, once fixed $f(X)$ so to mimic the background evolution of $\Lambda$CDM~\cite{Park:2012cp,Nersisyan:2017mgj,Park:2017zls}, while no  comparison  to CMB data has been performed  yet.  Similar nonlocal models  (again without any explicit mass scale, but rather involving operators such as $\RMN\iBox\Gmn$) have been advocated by Barvinsky~\cite{Barvinsky:2003kg,Barvinsky:2011hd,Barvinsky:2011rk}.

A different approach has been pursued by our group in the last few years, in which we assume that, in the IR,  is  dynamically generated a new mass scale $\Lambda$ associated to nonlocal terms. In Section \ref{sect:dyn} we will discuss some scenarios  suggesting that the emergence of such a dynamical scale is a priori possible. In this paper we will focus in particular on a model that, as we shall see,  is particularly interesting,  the so-called ``RR'' model,  which is defined by the quantum effective action
\bees\label{RR}
\Gamma_{\rm RR}&=&\int d^{4}x \sqrt{-g}\, 
\[\frac{\mplr^2}{2} R-R\frac{\Lambda_{\rm\scriptscriptstyle RR}^4}{\Box^2} R\]\nn\\
&=&
\frac{\mplr^2}{2}\int d^{4}x \sqrt{-g}\, 
\[R-\frac{1}{6} m^2R\frac{1}{\Box^2} R\]\, ,
\ees
where $\mplr$ is the reduced Planck mass, $\mplr^2=1/(8\pi G)$, $\Lambda_{\rm\scriptscriptstyle RR}$ is the fundamental mass scale generated in the IR and, for later convenience, we have also introduced a mass parameter $m$ from $\Lambda_{\rm\scriptscriptstyle RR}^4=(1/12)m^2\mplr^2$. This model was  proposed in \cite{Maggiore:2014sia}, following earlier work in \cite{Maggiore:2013mea}, and  a  detailed comparison with cosmological data has been worked out  in 
\cite{Dirian:2014ara,Dirian:2014bma,Dirian:2016puz,Dirian:2017pwp}. 
In this paper we present a comprehensive and detailed discussion of conceptual aspects and cosmological predictions of this class of nonlocal gravity models, focusing (for reasons that will become clear below) on the RR model, updating and expanding the results that have been reviewed in \cite{Maggiore:2016gpx}. The conceptual aspects are discussed in Section~\ref{sect:conc}, where we will also see how the RR model is singled out within  a potentially large landscape of nonlocal models,
while the cosmological consequences  are presented in
Section~\ref{sect:cosmo}. These two sections are largely independent. In particular, the reader interested uniquely in the cosmological applications could  move directly to Section~\ref{sect:cosmo}.

We use  units $\hbar=c=1$, and MTW conventions~\cite{MTW} for the curvature and signature, so in particular  $\emn=(-,+,+,+)$. 

\section{Conceptual aspects}\label{sect:conc}

\subsection{A reminder of QFT: the quantum effective action}\label{sect:rem}

We  begin by recalling the elementary notion of quantum effective action in QFT (the expert reader might simply wish to skip this section).
Following the standard textbook definitions let us recall that, for a scalar field  $\varphi(x)$ with action $S[\varphi]$ in flat space, the quantum effective action is obtained by introducing first an auxiliary source $J(x)$ and defining the generating functional of the connected Green's function $W[J]$,
\be\label{defWJ}
e^{iW[J]}\equiv\int D\varphi\,\,  e^{iS[\varphi]+i\int J\varphi}\, ,
\ee
where $\int J\varphi\equiv \int d^Dx \, J(x)\varphi(x)$, in $D$ space-time dimensions.
Then 
\be\label{defbarphi}
\frac{\d W[J]}{\d J(x)}=\cav \varphi(x)\vac_J\equiv \phi[J]
\ee
gives the vacuum expectation value of the field $\varphi(x)$ in the presence of the source $J(x)$. The quantum effective action $\Gamma[  \phi]$ is defined as a functional of the expectation value $\phi[J]$ (rather than of the original field $\varphi$), obtained by performing the Legendre transform,
\be
\Gamma[  \phi]\equiv W[J] -\int  \phi J\, ,
\ee
where $J=J[ \phi]$ is obtained inverting $ \phi[J]$ from \eq{defbarphi}. As a consequence,
\bees\label{dGJ}
\frac{\d\Gamma[  \phi]}{\d  \phi(x)}&=&
\int_y \frac{\d W[J]}{\d J(y)}\frac{\d J(y)}{\d \phi(x)}-J(x)
-\int_y  \phi(y)\frac{\d J(y)}{\d \phi(x)}\nn\\
&=&-J(x)\, ,
\ees
where in the second line we used \eq{defbarphi}.
In other words, the variation of the quantum effective action gives the equation of motion for the vacuum expectation value of the field, in the presence of a source. Using \eq{defWJ} we can write
\bees
e^{i\Gamma[ \phi]}&=& e^{iW-i\int  \phi J}\nn\\
&=&\int D\varphi\,\,  e^{iS[\varphi]+i\int (\varphi - \phi) J}\nn\\
&=&\int D\varphi\,\,  e^{iS[ \phi+\varphi]-i\int \frac{\d\Gamma[  \phi]}{\d  \phi}\varphi }
\, ,\label{Gammapathint}
\ees
where in the last line we shifted the integration variable $\varphi\ra \varphi+\phi$ and we used \eq{dGJ}. This exact path integral representation for the quantum effective action can be used to evaluate it perturbatively, replacing to lowest order $\Gamma[ \phi]$ with $S[ \phi]$ on the right-hand side, 
and proceeding iteratively. \Eq{Gammapathint} shows that the quantum effective action is a functional of the vacuum expectation value of the field, obtained by integrating out the quantum fluctuations. The  usefulness of the constructions stems  from  \eq{dGJ}, that  shows that $\Gamma$ is the quantity whose variation gives the exact equations of motion of the expectation values of the field, which by construction include the effect of the quantum fluctuations.

It is useful to stress the difference between the quantum effective action and the Wilsonian effective action. The latter is a functional of the fields (and not of their expectation values) which is obtained by integrating out  massive fields from the theory, and is therefore an effective description valid only at energies low compared to the masses of the fields which have been integrated out. Thus, it can only be used to compute the correlators of the light fields that  have been retained in the effective description. In contrast, in the quantum effective action, we integrate out only the quantum fluctuations of the fields, but $\Gamma$ is still a functional of the vacuum expectation value  of all fields, and can be used to compute, in principle exactly, the equation of motion satisfied by these vacuum expectation values. The quantum effective action is not a low-energy effective action. Rather, it is  valid at all energies for which the original classical  action was valid, and it also includes the effect of quantum fluctuations.
This consideration is important in particular when we deal with massless particles (such as the graviton, or the photon). In a Wilsonian approach, there is no sense in which we can integrate out massless particles. The resulting ``low-energy" theory would have no domain of validity. In contrast, the quantum effective action obtained by integrating over the quantum fluctuations of the massless fields is a well defined object, and indeed it is the integration over the fluctuations of the massless or light fields that induces nonlocalities, as we will review below in some simple cases.

In curved space, treating the metric as a classical field (i.e. not integrating over it in the path integral) the same procedure of Legendre transform gives the quantum effective  action $\Gamma[\gmn;\phi]$,  
\be
e^{i\Gamma[\gmn;\phi]}=e^{iS_{\rm EH}[\gmn]}\, 
\int D\varphi\,\,  e^{iS_m[\gmn; \phi+\varphi]-i\int \frac{\d\Gamma[\gmn;  \phi]}{\d  \phi}\varphi }
\, ,\label{Gammagmnphi}
\ee
where we have included  the Einstein--Hilbert action $S_{\rm EH}$ 
in the definition of $\Gamma[\gmn;\phi]$ (note that, in any case, $S_{\rm EH}$ does not contribute to
$\d\Gamma/\d  \phi$),
and we now denote by $S_m$ the action of the matter fields.
If we are only interested in the situation in which the vacuum expectation values of the matter fields vanish, and there is no external current that excites them, we can set $\phi \equiv \cav \varphi(x)\vac=0$ and $\d\Gamma/\d\phi =J=0$ in \eq{Gammagmnphi}, and we get the quantum effective action of the vacuum,
\bees
e^{i\Gamma[\gmn]}&=&e^{iS_{\rm EH}[\gmn]}\, \int D\varphi\,\,  e^{iS_m[\gmn; \varphi]}\nn\\
&\equiv&e^{iS_{\rm EH}[\gmn]}\,e^{i\Gamma_m[\gmn]}
\, .\label{Gammagmn}
\ees
In particular, from the usual definition of the matter energy-momentum tensor 
\be
\TMN=\frac{2}{\sqrt{-g}}\, \frac{\d S_m}{\d\gmn}\, ,
\ee \
it follows that  
\be\label{Tmnvac}
\cav\TMN\vac=\frac{2}{\sqrt{-g}}\, \frac{\d\Gamma_m}{\d\gmn}\, .
\ee
Therefore, the equations of motion derived from the total quantum effective action $\Gamma=S_{\rm EH}+\Gamma_m$ give the Einstein equations 
$\Gmn=8\pi G \cav\TMN\vac$ where, on the right-hand side,  all quantum fluctuations due to matter fields are automatically included. As a final step, we might wish to quantize also the metric. In the path integral formulation this means that we also integrate over the metrics, with the appropriate gauge fixing and Faddeev-Popov determinant. 
For later purposes, two observations are useful at this point:

(1) In a quantum field theory in the presence of an external classical source $J$, as in \eq{defWJ}, or in the presence of an external classical metric $\gmn$, as in 
\eq{Gammagmnphi}, it is important to distinguish between the ``in" vacuum and the ``out" vacuum. Which expectation value is computed in \eq{Tmnvac} depends on the boundary conditions in the path-integral. Using the standard Feynman path integral we get the in-out vacuum expectation value, e.g.
$\langle 0_{\rm out}|\TMN|0_{\rm in}\rangle$ in the case of the energy-momentum tensor. In-out matrix elements routinely appear as intermediate steps in QFT computations, but by themselves they are not physical quantities. In particular, even if $\hat{\phi}$ is an hermitian operator, 
$\langle 0_{\rm out}|\hat{\phi}|0_{\rm in}\rangle$ is not  real, and furthermore it obeys  equations of motions in which the Feynman propagator appears (and which are therefore  acausal). Similarly, in a theory where gravity is quantized and $\hat{g}_{\mu\nu}$ is an operator, $\langle 0_{\rm out}|\hat{g}_{\mu\nu}|0_{\rm in}\rangle$ is not even real (and does not obey causal equations of motion) and cannot be interpreted as a semiclassical metric.
In contrast, 
using the Schwinger-Keldish path integral gives the in-in expectation values, e.g. $\langle 0_{\rm in}|\TMN|0_{\rm in}\rangle$.
The in-in expectation values are physical quantities, representing the vacuum expectation value of an operator at a given time. In particular,  if $\hat{\phi}$ is  hermitian $\langle 0_{\rm in}|\hat{\phi}|0_{\rm in}\rangle$ is   real, and it obeys  causal equations of motions in which the retarded propagator appears~\cite{Jordan:1986ug,Calzetta:1986ey,Mukhanov:2007zz}, and $\langle 0_{\rm in}|\hat{g}_{\mu\nu}|0_{\rm in}\rangle$ plays the role of a metric.

(2) Once \label{point2} again it is important to understand that, in contrast to a Wilsonian approach, in the quantum effective action (\ref{Gammagmnphi}) we have not ``integrated out" some massive fields, and we are not constructing a low-energy effective theory. Rather, we are integrating out the quantum fluctuations, but all fields are still in principle present, and $\Gamma[\gmn;\phi]$ can be used to study the dynamics of the vacuum expectation values  of all matter fields (or more generally, of all correlation functions), as well as of the metric. The domain of validity of the quantum effective action is the same as that of the original  action. In \eq{Gammagmn} the matter fields no longer appear simply because we have chosen to set them in their vacuum state, i.e. we are studying the effect of the vacuum fluctuations of the matter fields on the dynamics of the metric.

\subsection{Examples of nonlocal terms in quantum effective actions}\label{sect:examples}

As a first simple example, let us consider quantum electrodynamics. Integrating out the quantum fluctuations due to the electron and limiting ourselves for simplicity to the terms involving  the photon field only,  
one finds (see e.g.~\cite{Dalvit:1994gf,Manohar:1996cq,Dobado:1998mr})
\be\label{qed}
\Gamma_{\rm QED}[\Am]=-\frac{1}{4}\int d^4x\, \[F_{\mu\nu}\frac{1}{e^2(\Box)}F^{\mu\nu} 
+ {\cal O}(F^4)\]\, .
\ee
In the limit $|\Box/m_e^2|\gg1$,   i.e. when the electron is light with respect to the relevant energy scales, the form factor $1/e^2(\Box)$ becomes 
\be\label{runninge}
\frac{1}{e^2(\Box)}\simeq\frac{1}{e^2(\mu)}-\beta_0\log\(\frac{-\Box}{\mu^2}\)\, ,
\ee
where  $\mu$ is the renormalization scale, $e(\mu)$ is the renormalized charge at the scale $\mu$, and $\beta_0=1/(12\pi^2)$. The logarithm of the d'Alembertian is a nonlocal operator defined by its integral representation,
\be
\log\(\frac{-\Box}{\mu^2}\)\equiv \int_0^{\infty}dm^2\, \[\frac{1}{m^2+\mu^2}-
\frac{1}{m^2-\Box}\]\, .
\ee
It is clear from \eq{runninge} that 
in this case the nonlocality of the effective action  just reflects the running of the coupling constant, expressed in coordinate space. In the opposite limit,  $|\Box/m_e^2|\ll 1$, so when the electron is heavy compared to the relevant energy scales, the quantum fluctuations due to the electron produce local terms suppressed by powers of $|\Box/m_e^2|$, 
\be
\frac{1}{e^2(\Box)}\simeq\frac{1}{e^2(\mu)}
+\frac{4}{15\, (4\pi)^2}\, \frac{\Box}{m_e^2}\, ,
\ee
reflecting the decoupling of heavy particles.\footnote{More precisely, this is better seen  using a mass-dependent   subtraction scheme, where
the decoupling of heavy particles is explicit; see~\cite{Manohar:1996cq} for review.}
Adding to this also the terms 
of order $\Fmn^4$ gives the well-known local  Euler-Heisenberg effective action  (see e.g. \cite{Dobado:1998mr} for the explicit computation),
\bees\label{Euler}
\Gamma_{\rm QED}[\Am]&\simeq&\int d^4x\, \bigg[-\frac{1}{4e^2(\mu)}F_{\mu\nu}F^{\mu\nu} 
-\frac{1}{15\, (4\pi)^2}\, \frac{1}{m_e^2}\, \Fmn\Box\FMN\nn\\
&&\hspace*{11mm}+\frac{e^2(\mu)}{90 (4\pi)^2}\, \frac{1}{m_e^4}\, \(  (\FMN\Fmn)^2+\frac{7}{4}(\FMN\tilde{F}_{\mu\nu})^2\)
\bigg]\, .
\ees
The higher-order operators in \eq{Euler} are of course  the same that would be obtained in a Wilsonian approach, i.e. constructing a low-energy effective action by integrating out the heavy electron, while the nonlocal corrections given by \eqs{qed}{runninge} are specific to the quantum effective action.

The same kind of computation can be performed in gravity, coupled to matter fields. An explicitly covariant computational method  based on the heat-kernel technique, combined with an expansion in powers of the curvature, gives~\cite{Barvinsky:1985an,Barvinsky:1987uw,Buchbinder:1992rb,Gusev:1998rp,Gorbar:2002pw,Gorbar:2003yt,Shapiro:2008sf,Mukhanov:2007zz}
\be\label{formfact}
\Gamma=\frac{\mplr^2}{2}\int d^4x \sqrt{-g}\,R
+\frac{1}{2(4\pi)^2}\,\int d^4x  \sqrt{-g}\, 
\[  R \,k_R(\Box) R +\frac{1}{2}C_{\mu\nu\rho\sigma}k_W(\Box)C^{\mu\nu\rho\sigma}\]\, ,
\ee
where $C_{\mu\nu\rho\sigma}$ is the Weyl tensor (and we have not written explicitly a similar term for the Gauss-Bonnet). The exact expression for the form factors  $k_{R,W}(\Box)$ induced by particles with generic mass can be found in \cite{Gorbar:2002pw,Gorbar:2003yt}.
Just as in \eq{runninge},
loops of massless particles contribute to the  form factors through logarithmic terms plus finite parts, i.e.  $k_{R,W}(\Box)=c_{R,W}\log (-\Box/\mu^2)$,
where now $\Box$ is the  generally-covariant d'Alembertian, $\mu$ is the renormalization point, and $c_R ,c_W$  are known coefficients that depend on the number of matter species and on their spin. Observe that, even if we start from the Einstein--Hilbert action with no higher-order operators, higher-derivative terms such as $R^2$ are unavoidably generated, with a  coupling constant which, in the UV, is  running logarithmically. 

In the heat-kernel computation the quantum effective action is written as an integral over a parameter $s$ running from zero to infinity (``Schwinger proper time"), such that the limit $s\ra 0^+$ corresponds to the UV regime and $s\ra\infty$ to the IR regime (see e.g. \cite{Birrell:1982ix,Mukhanov:2007zz}). The form (\ref{formfact}) of the quantum effective action, together with the expressions $k_{R,W}(\Box)=c_{R,W}\log (\Box/\mu^2)$ for the form factors, is obtained by using the Schwinger--DeWitt expansion of the heat-kernel, which is an expansion in powers of $s$ around $s=0$, and therefore is only valid in the UV regime.
Note that, even with this logarithmic enhancement,  a term  $R\log (-\Box/\mu^2)R$ can compete with the leading Einstein-Hilbert term $\mplr^2R$ only for values of $R$ not much below $\mplr^2$. Thus, these terms can only be relevant close to the Planck scale, and are totally irrelevant for present-day cosmology.
For cosmological applications, we rather need the large-distance form of the correction, i.e. their IR limit.
However, for massless fields (which are just the fields that can give interesting long-distance effects) the Schwinger--DeWitt  expansion fails in the IR, giving a sequence of IR-divergent integrals over proper time, and  the IR limit of the quantum effective action is poorly understood. 

Non-perturbative information on the quantum effective action of gravity induced by conformal matter fields can be obtained by integrating the conformal anomaly (see \cite{Shapiro:2008sf,Maggiore:2016gpx} for pedagogical discussions). The most famous example is the Polyakov quantum effective action in two dimensions. One starts from  the action of 
2D gravity, including also a cosmological constant $\lambda$, 
\be\label{Skappalambda}
S=\int d^2x\, \sqrt{-g } (\kappa R-\lambda) +S_m\, ,
\ee
where $S_m$  is the  action describing   $N_S$ conformally-coupled massless scalar and $N_F$ massless Dirac fermion fields.
In 2D the Einstein--Hilbert term is a topological invariant and does not contribute to the dynamics. However, at the quantum level the gravitational dynamics is non-trivial, and
the quantum effective action of gravity generated by the quantum fluctuations of the matter fields can be computed exactly, by integrating the conformal anomaly. After dropping the topologically-invariant Einstein--Hilbert term, it is given by
\be\label{SPolyakov26}
\Gamma=-\frac{N-25}{96\pi}\int d^2x\sqrt{-g}\, R\frac{1}{\Box}R -\lambda\int d^2x\, \sqrt{-g } \, ,
\ee
where $N=N_S+N_F$.
This example, in which we know the exact quantum effective action, will be used in Section~\ref{sect:FAQ} to illustrate some conceptual issues on the meaning and use of quantum effective actions.\footnote{The factor $(N-25)$ in \eq{SPolyakov26} comes out because invariance  under diffeomorphisms allows us to fix (locally) $g_{ab}=e^{2\sigma}\bar{g}_{ab}$, where $\bar{g}_{ab}$ is a reference metric. In  a theory with dynamical gravity, where in the path integral we also integrate over $g_{ab}$, this is now a gauge fixing condition, and the corresponding 
reparametrization ghosts give a contribution $-26$ to be added to $N$, while  the conformal factor  $\sigma$ gives a contribution $+1$~\cite{Knizhnik:1988ak,David:1988hj,Distler:1988jt}. In string theory, where 
$\lambda=0$,  beside diff invariance one also has  Weyl invariance on the world-sheet. This  allows one to eliminate also $\sigma$, so  one only has the contribution $-26$ from the reparametrization ghosts, together with the contribution from the   $N=D$ matter fields  $X^{\mu}(\sigma_1,\sigma_2)$ living in the world-sheet,   leading to the critical dimension $D=26$ from the requirement of anomaly cancellation.
\label{foot:anom}}

\subsection{Scenarios for the  emergence of a dynamical mass scale in gravity}\label{sect:dyn}

In nonlocal models such as the RR model (\ref{RR})  enters a mass scale
$\Lrr$ or, equivalently, $m$. The question that we must address is therefore whether in gravity a new mass scale could be generated dynamically in the infrared. This is a difficult question, since  dynamical mass generation is a  non-perturbative phenomenon and, for the moment, the best we can do is to identify some scenarios that indicate that such a dynamical mass generation is a priori possible. 

As we mentioned above, for a minimally coupled massless scalar field in de~Sitter space with a $\lambda\phi^4$ interaction, it has been conclusively shown that a  mass is dynamically generated as a response to the apparent infrared divergences of the massless theory; see in particular the nice Euclidean computation in refs.~\cite{Rajaraman:2010xd,Beneke:2012kn}. Despite the fact that, in de~Sitter space, graviton perturbations  have the same infrared divergences as massless scalars, it is usually believed that the lesson from scalar fields ``[...] is no help at all for the graviton case. The gauge invariance of gravitational perturbations around de~Sitter space precludes the development of a mass, dynamical or otherwise, for the graviton" \cite{Rajaraman:2016nvv}. However, there is a loophole in this argument. When we discuss quantum effects and dynamical mass generation, the appropriate object is of course the quantum effective action, not the fundamental action, and we have seen that the former admits nonlocal terms. With nonlocal terms we can construct diffeomorphism-invariant quantities (or, for gauge fields, gauge-invariant quantities) that have the meaning of a mass term. This was first observed by Dvali~\cite{Dvali:2006su} in the context of massive electrodynamics. Consider indeed the quantum effective action\footnote{Actually, in ~\cite{Dvali:2006su} this expression was considered as a classical action. In that case, the equations of motion only become causal if, after taking the variation of the action to get the equations of motion, we  impose by hand that the $\iBox$ operator that appears in the equations of motion is defined with respect to the retarded Green's function. In contrast, considering this expression as a quantum effective action, causality for the equations of motion of the in-in matrix elements is automatic; see the discussion in Section~\ref{sect:causality}, as well as item   (1) below \eq{Tmnvac}.\label{foot:caus}}
\be\label{Lnonloc}
\Gamma=-\frac{1}{4}\int d^4x\,
 \(\Fmn \FMN -m_{\g}^2 \Fmn \frac{1}{\Box}\FMN\)
\, .
\ee
This effective action is  gauge-invariant, so we can choose the gauge $\pam A^{\mu}=0$. In that gauge, upon integration by parts,
\be
\frac{1}{4}m_{\g}^2 \Fmn \frac{1}{\Box}\FMN=\frac{1}{2}m_{\g}^2 A_{\mu}A^{\mu}\, ,
\ee
so the nonlocal term in \eq{Lnonloc} is just a mass term for the photon, written in a way that preserves gauge-invariance at the price of nonlocality. Indeed, in QCD it  has been advocated the introduction in the quantum effective action of the non-abelian generalization of the above nonlocal term, i.e. of
\be\label{Fmn2}
\frac{m_g^2}{2} {\rm Tr}\, \int d^4x\,   F_{\mu\nu} \frac{1}{\Box}F^{\mu\nu}\, ,
\ee
where  $F_{\mu\nu}=F_{\mu\nu}^aT^a$, $\Box^{ab}=D_{\mu}^{ac}D^{\mu,cb}$ and $D_{\mu}^{ab}=\d^{ab}\pam-gf^{abc}A_{\mu}^c$ is the covariant derivative. This nonlocal term corresponds to giving a mass $m_g$ to the gluons (plus extra  nonlocal interaction terms that, altogether, reconstruct a gauge-invariant quantity), and its introduction correctly reproduces the results on the non-perturbative gluon propagator in the IR, obtained from operator product expansions and from lattice QCD~\cite{Boucaud:2001st,Capri:2005dy,Dudal:2008sp}. 

It is quite interesting to observe that also the nonlocal terms in \eq{RR} has a similar interpretation, as a diffeomorphism-invariant mass term for the conformal mode of the metric~\cite{Maggiore:2015rma,Maggiore:2016fbn}.
Indeed, let us write $\gmn(x)=e^{2\sigma(x)}\gbmn(x)$, where $\sigma(x)$ is 
the conformal mode and $\gbmn$ a  fiducial metric  with fixed determinant. Let us restrict the dynamics to the conformal mode,  choosing
for simplicity $\gbmn=\emn$. Then the Ricci scalar computed from the metric $\gmn=e^{2\sigma(x)}\emn$ is 
\bees
R&=&-6 e^{-2\sigma}\( \Box\sigma +\pam\sigma\paM\sigma\)\nn\\
&=&-6\Box\sigma +{\cal O}(\sigma^2)
\, .
\ees
and, upon integration by parts,
\be\label{m2s2}
 R\frac{1}{\Box^2} R=36 \sigma^2 +{\cal O}(\sigma^3)\, .
\ee
Then, truncating the theory to the conformal mode, writing $\gmn(x)=e^{2\sigma(x)}\emn$ and expanding  to quadratic order in $\sigma$, \eq{RR} gives
\bees
S_2[\sigma]&=& \int d^4x\, \(3\mplr^2\pam\sigma\paM\sigma
-36\Lambda_{\rm\scriptscriptstyle RR}^4\sigma^2\) \nn\\
&=&\frac{1}{2}\int d^4x\, \(\pam\varphi\paM\varphi-m^2\varphi^2\)\, ,\label{massconfmode}
\ees
where $\varphi=\sqrt{6}\,\mplr\sigma$ is a canonically normalized field proportional to the conformal mode (recall that $\sigma$ is dimensionless), and we used 
$\Lambda_{\rm\scriptscriptstyle RR}^4=(1/12)m^2\mplr^2$.
We see that the  $R\Box^{-2}R$ terms is a diffeomorphism-invariant mass term for the conformal mode, plus  higher-order nonlocal interactions  that are required to reconstruct a 
diffeomorphism-invariant quantity.\footnote{Observe that, as usual, the kinetic term of the conformal mode in GR has the ``ghost-like" sign [recall that our metric signature is $\emn=(-,+,+,+)$]. However, in GR only tensor perturbations are true dynamical degrees of freedoms.  In the Lagrangian formulation, when linearizing over a background (such as Minkowski or FRW)  the fact that $\sigma$ is  non dynamical follows once one uses the equation of motion with respect to a second independent scalar perturbation, that here is not apparent since we have truncated the theory to the conformal mode \cite{Jaccard:2012ut}. This conclusion is not altered by the addition of a mass term for $\sigma$; see also \cite{Foffa:2013sma}.} Thus, just as IR fluctuations in space-times such as de~Sitter induce a dynamical mass generation for a scalar field, it is in principle conceivable that the same mechanism generates dynamically a mass for the conformal mode, in the form of a nonlocal term proportional to $R\Box^{-2}R$ in the quantum effective action.

Indications in favor of the dynamical generation of a mass scale also come from non-perturbative studies of Euclidean gravity on the lattice, that
suggest the existence of a nontrivial ultraviolet fixed point, where a continuum limit can be taken. In the vicinity of the fixed point, i.e. in the UV regime,  it is found that Newton's constant runs as
\be\label{Gdik2}
G(k^2)=G_N\[ 1+  \(\frac{m^2}{k^2}\)^{\frac{1}{2\nu}}+
{\cal O} \(\frac{m^2}{k^2}\)^{\frac{1}{\nu}} \, \]\, ,
\ee
where  $\nu$ is a critical index which, within the numerical accuracy, is consistent with $\nu=1/3$, and $m$ is a renormalization-group invariant  mass scale which is dynamically generated, analogous to $\Lambda_{\rm\scriptscriptstyle QCD}$ in QCD  (see \cite{Hamber:1999nu,Hamber:2004ew,Hamber:2005dw,Hamber:2013rb,Hamber:2015jja},
and  \cite{Hamber:2009zz,Hamber:2017pli} for reviews). This result can at least be considered as a proof of principle of the fact that a mass scale can be dynamically generated in gravity. 
More specifically, identifying $-k^2$ with the covariant $\Box$ operator in coordinate space,
at first sight the lattice result suggests to study a model of the form~\cite{Hamber:2005dw,Hamber:2013rb}
\be
\Gamma=\int d^{4}x \sqrt{-g}\, \frac{1}{16\pi G(\Box)} R\, ,
\ee
where, at least in the UV limit $-\Box\gg m^2$,
\be\label{GdiBox}
\frac{1}{16\pi G(\Box)}\simeq \frac{\mplr^2}{2}\[ 1-  \(\frac{m^2}{-\Box }\)^{\frac{3}{2}}\]\, .
\ee
The non-integer powers of the $\Box^{-1}$ operator can be defined using an integral representation, see eq.~(71) of \cite{Hamber:2017pli}.
Actually, it is easy to see by dimensional analysis that, in order for this nonlocal term to be relevant at the present epoch and reproduce the observed dark energy, we need $m^2\sim H_0^2$ (we will confirm this by explicit computations in the nonlocal models studied below). However, on a cosmological background configuration at the present epoch,  also $-\Box$ is of order $H_0^2$.
Thus, for cosmological applications it is not enough to know $G(\Box)$ in the UV limit $-\Box/m^2\gg 1$, but we rather need to know it  for $-\Box/m^2={\cal O}( 1)$. In this IR regime
the expression for $G(\Box)$ will be different, and further form factors, in particular associated to higher-derivative terms, might also come into play (see in particular the discussion in \cite{Maggiore:2015rma} for the possibility of generating this scale through the running of the coupling constant associated to the $R^2$ interaction, and \cite{Einhorn:2014gfa,Tong:2014era} for related ideas involving the coupling of the Gauss-Bonnet term). Once again, the analogy with QCD is instructive: in the UV one has the running of the coupling constant as determined by asymptotic freedom, which already shows the existence of a 
dynamically-generated mass scale $\Lambda_{\rm\scriptscriptstyle QCD}$, 
while in the IR terms such as (\ref{Fmn2}) can emerge.
Thus, in principle the RR model could emerge as the IR limit of the form factors obtained by  Euclidean lattice gravity. It is clear that eventually the exact form of the nonlocal term will have to be fixed from first principles, and the relatively simple models that we will study here are meant first of all to illustrate the general features of nonlocal gravity.

A scenario in which the mass scale $\Lambda_{\rm\scriptscriptstyle RR}$ in \eq{RR} is generated dynamically, say through dimensional transmutation in the coupling associated to the $R^2$ term, would produce an elegant solution to the naturalness problem associated with the cosmological constant~\cite{Maggiore:2015rma}.  In this scenario the scale $\Lambda_{\rm\scriptscriptstyle RR}$ emerges in the same way as $\Lambda_{\rm\scriptscriptstyle QCD}$ in QCD. There is no issue of naturalness for such a quantity, which is a renormalization-group invariant quantity determined by the logarithmic running of a  coupling constant, and does not receive loop corrections proportional to powers of the cutoff (which is the origin of the naturalness problem for the cosmological constant). 
Of course, the value of the scale $\Lambda_{\rm\scriptscriptstyle RR}$ generated dynamically in this way cannot be predicted, just as we cannot predict the value of $\Lambda_{\rm\scriptscriptstyle QCD}$, and can only be obtained by comparison with the observation. 
In our case, as we already mentioned, it is easy to see  that, in order for the nonlocal term in \eq{RR} to generate a dark energy density which is significant today, we need $m=O(H_0)$.  Thus, from 
$\Lambda_{\rm\scriptscriptstyle RR}^4=(1/12)m^2\mplr^2$, it follows that $\Lambda_{\rm\scriptscriptstyle RR}=O(H_0\mplr)^{1/2}=O({\rm meV})$, so the scale which is dynamically generated must be of the order of the milli-eV. More precisely, as we will review in Section~\ref{sect:cosmo}, performing parameter estimation for the RR model we 
get $m\simeq 0.28 H_0$~\cite{Maggiore:2014sia}. This gives  
\be
\Lambda_{\rm\scriptscriptstyle RR}\simeq 0.3 (H_0\mplr)^{1/2}\simeq 0.5\, {\rm meV}\, .
\ee
It should be stressed that in this scenario the fundamental scale  is $\Lambda_{\rm\scriptscriptstyle RR}$, which appears in the first line of \eq{RR}, and determines the IR behavior of the form factor associated to the $R^2$ term. The quantity $m$ which appear in the second line is just a derived quantity, introduced for convenience.
Thus, in this scenario dark energy can be explained   by the dynamical generation of an energy scale which, even if cannot be predicted, still turns out to have a value which is not especially surprising from the point of view of quantum field theory.\footnote{The meV scale is of course the same scale as that of neutrino masses. While this could just be an accident, an intriguing connection is suggested in ref.~\cite{Dvali:2016uhn}, where a neutrino condensate and the neutrino masses emerge from the gravitational anomaly.}
 This should be compared with attempts at explaining dark energy through the introduction of some particle of mass $m$, as for instance in massive gravity, in which case $m$ is the fundamental parameter, and 
should be fixed to an  astonishingly small value $m\sim H_0\sim 10^{-33}$~eV.

\subsection{Nonlocal gravity. Frequently Asked Questions}\label{sect:FAQ}

The introduction of nonlocal  terms, and their use in  cosmology,  raises a number of conceptual issues that  can generate some confusion. In this section, extending the discussion in \cite{Maggiore:2016gpx}, we examine a number of ``frequently asked questions".

\subsubsection{Causality}\label{sect:causality}

Does nonlocality imply loss of causality?
This question arises because indeed, in a fundamental action, nonlocalities do imply a loss of causality. As a simple illustration, consider for instance a nonlocal term proportional to $\int dx \,\phi\iBox\phi$ in the action of a scalar field $\phi$, where  $\iBox$ is defined with respect to some Green's function $G(x;x')$~\cite{Foffa:2013sma}. Then
\bees
&&\frac{\d}{\d\phi(x)}\int dx' \phi(x') (\iBox\phi )(x')=
\frac{\d}{\d\phi(x)} \int dx' dx'' \phi(x') G(x';x'') \phi(x'')\nn\\
&&=\int dx' [G(x;x')+G(x';x)] \phi(x')\, . \label{symGreen}
\ees
Thus, the variation  of the action produces in the equations of motion a  Green's
function symmetric in $(x,x')$. However, the retarded Green's function is not symmetric; rather, 
$G_{\rm ret}(x';x)=G_{\rm adv}(x;x')$, and therefore  cannot be obtained from such a variation. The equations of motion  obtained from a nonlocal classical action are in general acausal. This is one of the reasons why a fundamental action must be local.

The situation is however completely different for the quantum effective action. First of all it is clear a priori  that, since nonlocal quantum effective actions are obtained from fundamental actions which are local and causal, the nonlocality in the quantum effective action cannot be a sign of acausality. Technically, this comes out as follows. As we discussed in point (1) below \eq{Tmnvac}, the variation of the quantum effective action does not give the equations of motion of the fields, but rather the equations of motion of the {\em vacuum expectation values} of the fields. Here  however we  must carefully  distinguish between the in-out and the in-in expectation values.
The standard Feynman path integral gives in-out vacuum expectation value. These vacuum expectation values indeed obey acausal equations involving the Feynman propagator. There is nothing wrong with it, since in-out matrix elements are not physically observable, and are only useful as intermediate steps in the computation, e.g., of scattering cross section. In contrast, in-in matrix elements are observables; e.g.  $\langle 0_{\rm in}|\phi (t,\vx)|0_{\rm in}\rangle$ represents the vacuum expectation value of the quantum field at a given time $t$.
The in-in matrix elements are computed using  
the Schwinger-Keldish path integral,  and  obey  causal equations of motions in which the retarded propagator appears~\cite{Jordan:1986ug,Calzetta:1986ey,Mukhanov:2007zz} (see also \cite{Tsamis:1997rk,Deser:2007jk,Barvinsky:2011rk,Deser:2013uya,Ferreira:2013tqn,Foffa:2013sma,Woodard:2014iga,Cusin:2016nzi} for related discussions).

\subsubsection{Domain of validity}

Another natural question is ``what is the domain of validity of the nonlocal theory?" 
Once again, the important point is that   the nonlocalities  that we consider appear at the level of the  quantum effective action. As such, they are due to the quantum fluctuations of the light or massless fields of the theory, so, in our case, of the graviton (with possibly a contribution from the photon once we include the fields of the Standard Model).
As it is clear from \eq{Gammapathint}, and as discussed in item (2) on page~\pageref{point2}, 
the quantum effective action is different from the Wilsonian low-energy effective action. We are not
integrating out some  fields. Rather, we are taking into account the quantum fluctuations of the massless fields. The quantum effective action therefore has  the same domain of validity of the fundamental theory (with the added virtue that it includes the quantum corrections), and is not a low-energy theory. For a theory such as the RR model we therefore expect that, just as Einstein gravity, it will be valid at all energies below the Planck scale. Furthermore, in the far IR it contains important non-perturbative terms that are generated by the quantum fluctuations of the graviton.

Note in particular that, of course, we are not integrating out massless fields. There is no sense in which massless fields can be integrated out in a Wilsonian approach. We are also not assuming the existence of hypothetical particles beyond the Standard Model. Indeed, already the presence of the massless graviton can in principle generate these long-range effects, as in the scenarios discussed in Section~\ref{sect:dyn}.

\subsubsection{Degrees of freedoms}\label{sect:dof}

Another  standard question is: ``what are the degrees of freedom of the nonlocal theory?", and, in particular, ``does it contain ghosts?" To answer this question it is important first of all to understand that the degrees of freedom of the theory must be read from the fundamental action, and not from the quantum effective action of the vacuum. As an example, consider the Polyakov quantum effective action (\ref{SPolyakov26}). In $D=2$,  invariance under diffeomorphisms allows us to fix locally $g_{ab}=e^{2\sigma}\eta_{ab}$. Then $R= -2e^{-2\sigma}\Box\sigma$ and
\be\label{SPolyakov26sigmatot}
\Gamma=\int d^2x\, \( \frac{N-25}{24\pi}\eta^{ab}\pa_a\sigma\pa_b\sigma 
 -\lambda e^{2\sigma}\)\, ,
\ee
where, in $D=2$, we use $a,b$ to label Lorentz indices.
Thus,  in terms of the conformal mode $\sigma$,
\eq{SPolyakov26} becomes local.
If we were to  read  naively the degrees of freedom from this quantum effective action we would conclude that, in the gravitational sector, for $N\neq 25$ the theory contains one degree of freedom, the conformal mode $\sigma$. Furthermore, this degree of freedom would be a  ghost for $N>25$, and `healthy' for $N<25$. However, in this case we know the fundamental theory, and we know that such conclusions are wrong. The fundamental theory  is just 2D gravity coupled to $N$ matter fields. The $N$ matter fields are not present in \eq{SPolyakov26sigmatot} simply because we are considering the vacuum quantum effective action, i.e. we have implicitly set to zero the vacuum expectation values of the matter fields. These can in principle be reinserted by performing the Legendre transform with respect to the matter fields as in Section~\ref{sect:rem}. As for the gravitational field, in 2D at the classical level the gravitational field has no degree of freedom, since the Einstein--Hilbert action is a topological invariant.  To understand the number of degrees of freedom in the gravitational sector at the quantum level one  
can use a path integral quantization, integrating also over
the metric. From this point of view, the choice $g_{ab}=e^{2\sigma}\bar{g}_{ab}$ is a gauge fixing, and must be supplemented with the appropriate Faddeev-Popov ghosts. The resulting theory has a BRST symmetry, and the physical states are defined as usual as those annihilated by the BRST charge. The study of the BRST condition proves that there are no physical states associated to the conformal mode, as shown explicitly by Polchinski~\cite{Polchinski:1989fn}. Thus, both at the classical and at the quantum level, 2D gravity has no propagating degree of freedom.
Of course, this does not mean that the field $\sigma$  has no physical effects. The situation is  the same  as in electrodynamics, where again the physical-state condition eliminates 
the quanta associated to $A_0$. Still the interaction mediated by $A_0$ generates the Coulomb potential between static charges. In other words, the quanta associated to $\sigma$ (or to $A_0$ in QED) cannot appear in the external lines of Feynman diagram, since there are no physical states associated to them, but they can appear in the internal lines.

A lesson that we learn for our nonlocal gravity models is that the spectrum of the theory cannot be read from the quantum vacuum effective action, simply treating it as if it were a fundamental action and reading off a propagator. The degrees of freedom should be read from the fundamental action. If a nonlocal term such as that in \eq{RR} emerges from IR fluctuations in Einstein gravity (just as the nonlocal term (\ref{Fmn2}) emerges from IR fluctuations in QCD), the fundamental theory behind will simply be Einstein gravity, and in the gravitational sector we will simply have the two degrees of freedom of the graviton. 

The quantum vacuum effective action does not contain information on the possible presence of ghosts in the quantum spectrum of the fundamental  theory, but only  on potential instabilities of the classical equations of motion for the metric.  The linearization over Minkowski space indicates, for  the RR  model,   the existence of an instability associated to a scalar mode~\cite{Maggiore:2014sia}, which corresponds to one of the auxiliary fields introduced in the next subsection and which, as we will see below, is not an additional  dynamical degree of freedom. This instability is controlled by the mass parameter $m$. Since eventually the comparison with the data fixes $m$ to be of order $H_0$, the instability needs a  cosmological timescale to develop, and  to assess its fate we cannot study it  by linearizing over Minkowski space. Rather, 
we must  study the evolution in a  FRW cosmology, where the potential instability  will also compete with the damping due to the Hubble friction. As found in \cite{Maggiore:2014sia},
and as we will review  in Section~\ref{sect:cosmo}, the cosmological evolution obtained from these models is perfectly satisfying and stable, at least up to the present cosmological epoch.

\subsubsection{Localization and spurious degrees of freedom}\label{sect:loca}

A related question on the degrees of freedom of the theory is whether 
nonlocal gravity hides, in the definition of $\iBox$, some extra degrees of freedom, and in particular whether it is  equivalent to a scalar-tensor theory. The issue arises because  quantum effective actions such as
\eq{RR}  can be put into a local form by introducing auxiliary fields (see also \cite{Nojiri:2007uq,Jhingan:2008ym,Koshelev:2008ie,Koivisto:2008dh,Koivisto:2009jn,Barvinsky:2011rk,Deser:2013uya} for related discussions). For instance, the RR model
can be written  in a local form by introducing two auxiliary fields $U$ and $S$, 
defined by~\cite{Maggiore:2014sia}
\be\label{defU}
U=-\iBox R\, ,\qquad 
S=-\iBox U\, .
\ee
This can be implemented at the Lagrangian level by introducing two Lagrange multipliers $\xi_1,\xi_2$, and rewriting \eq{RR} as
\bees\label{S2}
\Gamma_{\rm RR}=\frac{\mplr^2}{2}\int d^4x \sqrt{-g}\, \[ R\( 1-\frac{m^2}{6} S\)-\xi_1(\Box U+R)-\xi_2 (\Box S+U)\]\nn
\, .
\ees
The equations of motion derived performing the variation of this action with respect to $\hmn$ is
\be\label{Gmn}
\Gmn=\frac{m^2}{6} K_{\mu\nu}+8\pi G\Tmn\, ,
\ee
where  $K_{\mu\nu}$ is a tensor that depends on the metric and the auxiliary fields,
\be\label{defKmn}
K^\mu_\nu \equiv 2 S G^\mu_\nu - 2 \nabla^\mu \partial_\nu S + 2 \delta^\mu_\nu \Box_g S + \delta^\mu_\nu \partial_\rho S \partial^\rho U - \frac{1}{2} \delta^\mu_\nu U^2 - \big( \partial^\mu S \partial_\nu U + \partial_\nu S \partial^\mu U \big).
\ee
At the same time, the variation with respect to the  Lagrange multipliers $\xi_1,\xi_2$  implies that $U$ and $S$ satisfy
\be\label{BoxUS}
\Box U=-R\, ,\qquad
\Box S =-U\, .
\ee
Thus, apparently we have a scalar-tensor theory in which, beside the metric, we have two dynamical scalar fields $U$ and $S$. Upon quantization, we would then naively expect to have the quanta of the fields $U$ and $S$. This would be a disaster,  because it can shown that one of these two fields, $U$, when linearizing over flat space, has a wrong-sign kinetic energy, i.e. it would be a ghost~\cite{Foffa:2013sma,Maggiore:2014sia,Maggiore:2016gpx}.

Once again, a neat understanding of the issue can be obtained by going back 
to the Polyakov quantum effective action in 2D, since in this case we  can follow step by step the emergence of the nonlocal term in its derivation from the fundamental theory. Let us briefly recall how the Polyakov effective action is derived (see e.g. \cite{Mukhanov:2007zz}  for pedagogical discussions).
Consider 2D gravity coupled to $N$ conformal matter fields. For such fields, classically the  
trace $T^{a}_{a}$ of the energy-momentum tensor  vanishes. However, at the quantum level 
\be\label{traceT}
\cav T^{a}_{a}\vac=\frac{N}{24\pi} R\, .
\ee
\Eq{traceT} is the trace anomaly. A crucial point about it is that, even if it can be obtained with a one-loop computation, it is actually an exact result.  
We can now find the effective action using its basic property (\ref{Tmnvac}). In $D=2$  we can write  the metric in the form $g_{ab}=e^{2\sigma}\eta_{ab}$ (at least locally). Then \eq{Tmnvac} gives
\bees\label{dGdsD2}
\frac{\d\Gamma}{\d\sigma}&=&2g_{ab} \frac{\d\Gamma}{\d g_{ab} }\nn\\
&=&\sqrt{-g}\,  \cav T^{a}_{a}\vac\nn\\
&=&e^{2\sigma} \frac{N}{24\pi} (-2\Box\sigma)\, ,
\ees
where, in the last line, we used the fact that, on a metric $g_{ab}=e^{2\sigma}\eta_{ab}$, we have 
$\sqrt{-g}=e^{2\sigma}$ and $R=-2\Box\sigma$, where $\Box$ is the covariant d'Alembertian with respect to the metric $g_{ab}$ (related to the 
flat space d'Alembertian $\Box_{\eta}$ by $\Box=e^{-2\sigma}\Box_{\eta}$).
Thus, thanks to the trace anomaly, we know exactly the functional derivative of the effective action with respect to the conformal mode, and we can now integrate it. The term  $2e^{2\sigma}\Box\sigma=
2\Box_{\eta}\sigma$ integrates to $\sigma\Box_{\eta}\sigma$, which is the same as
$e^{2\sigma} \sigma\Box\sigma$. Therefore
\be\label{Ginterm0}
\Gamma[\sigma]-\Gamma[\sigma=0]=-\frac{N}{24\pi}\, \int d^2x\,e^{2\sigma} \sigma\Box\sigma\, .
\ee
Furthermore, $\Gamma[\sigma=0]=0$ since it corresponds to the quantum effective action for the flat metric (this is the step that does not go through in higher dimensions, when the metric contains more degrees of freedom, and not just the conformal mode). For our purposes, the crucial step emerges when one rewrites this expression, which is local but not covariant, as a nonlocal but covariant expression. Let us do it step by step. From $\sqrt{-g}=e^{2\sigma}$ and $R=-2\Box\sigma$, we can immediately write
\be\label{Ginterm1}
\Gamma=\frac{N}{48\pi} \int d^2x\,\sqrt{-g}\, \sigma R\, .
\ee
The point is how to deal with the remaining $\sigma$ factor. Formally, $R=-2\Box\sigma$ can be inverted to give $\sigma=-(1/2)\iBox R$ which, inserted in \eq{Ginterm1}, gives the Polyakov action\footnote{As discussed in \eq{foot:anom}, if we further consider $\gmn$ as a quantum field, integrating also over it in the path integral, performing the gauge fixing 
$g_{ab}=e^{2\sigma}\eta_{ab}$  the corresponding reparametrization ghosts give a further factor $-26$ to be added to $N$, while, if Weyl invariance is broken by the cosmological constant $\lambda$, the conformal mode itself gives a $+1$, leading to $N-25$ instead of $N$ in the prefactor. Adding to this the classical action for 2D gravity with a cosmological constant, and discarding the topological Einstein--Hilbert term, gives \eq{SPolyakov26}.}
\be\label{Ginterm2}
\Gamma[\gmn]=-\frac{N}{96\pi}\, \int d^2x\, \sqrt{-g}\, R\Box^{-1}R\, .
\ee
Suppose that, as in \eq{defU}, we define a field $U$ from $U=-\iBox R$. Once again, at the level of the action, this can be implemented by introducing a Lagrange multiplier $\xi$, and writing\footnote{Of course, for the Polyakov action there is no need to introduce $U$ in order to write it in local form, since we already know from \eq{Ginterm0} that it becomes local when written in terms of the conformal factor. However, it is instructive to see what happens with the Polyakov action when we apply a localization procedure analogous to the one that we will use in four dimensions for the RR model.}
\be \label{SPolyakovloc1}
\Gamma=\frac{N}{96\pi}\int d^2x\sqrt{-g}\, \[ RU  +\xi (\Box U+R)\]\, .
\ee
Indeed, the variation with respect to $\xi$ gives
\be\label{BoxU2D}
\Box U=-R\, ,
\ee
so it enforces $U=-\iBox R$, and we get back \eq{Ginterm2}. We can further manipulate this expression observing that 
the variation with respect to $U$ gives $\Box\xi=-R$ and therefore
$\xi=-\iBox R=U$. This is an algebraic equation that can be put back in the action so that, after an integration by parts,  $\Gamma$ can be rewritten as~\cite{Antoniadis:2006wq}
\be\label{Ganom2}
\Gamma=-\frac{N}{96\pi}\int d^2x\sqrt{-g}\, \[\pa_a U\pa^a U-2UR\]\,.
\ee
Written in this form, the Polyakov action is both covariant and local, and looks like a scalar-tensor theory, which depends on the metric and on the scalar field $U$. Of course, this scalar field cannot be a genuine independent degree of freedom, because we know that the original expression from which we started, \eq{Ginterm0} or \eq{Ginterm1}, is again local, but only depends on the metric. To understand this point observe that
the most general solution of \eq{BoxU2D} is given by a solution of the inhomogeneous equation, which is fixed in terms of $R$ (or, equivalently, of $\sigma$) plus the most general solution of the homogeneous equation.
The latter would be a new degree of freedom, independent of $\sigma$. For instance, in flat space it is given by a superposition of plane waves, with coefficients $a_{\vk}$ and $a_{\vk}^*$ that would be promoted to creation and annihilation operators in the quantum theory. However, if we go back to the original expression (\ref{Ginterm0}) or (\ref{Ginterm1}), we see that  such a degree of freedom is not at all present. The quantum effective action only depends  on the field $\sigma$.  The degree of freedom associated to the most general solution of the homogeneous equation
$\Box U=0$ is spurious, and has been introduced by mistake, when we blindly replace $\sigma$ by
$-(1/2)\iBox R$. In this case, where we know everything explicitly, we see that 
 the solution that we need  of the equation 
$\Box U=-R$, i.e.  of $\Box U=2\Box\sigma$, is simply 
\be\label{U2s}
U(t,\vx)=2\sigma(t,\vx)\, , 
\ee
rather then  
$U=2\sigma +U_{\rm hom}$, where $U_{\rm hom}$ is the most general solution of the associated homogeneous equation $\Box U_{\rm hom}=0$. Indeed, it is only with $U=2\sigma$ that  \eq{Ginterm2} with $\iBox R=-U$ becomes the same as \eq{Ginterm1}. 
In order not to introduce such a spurious degree of freedom, we must specify that, by 
$-\iBox R$, we do not mean the most general function $U$ that satisfies $\Box U=-R$, but a specific  solution of this equation, selected e.g. by  fixing its boundary conditions.

As another example, consider the Proca Lagrangian for a massive photon,
\be\label{Proca}
{\cal L}=
-\frac{1}{4}\Fmn \FMN -\frac{1}{2}m_{\g}^2 A_{\mu}A^{\mu}\, .
 \ee
This theory gives a consistent  description of the three degrees of freedom of a massive photon, even if the  mass term  breaks the gauge invariance (see e.g. Section~4.1 of \cite{Maggiore:2016gpx}). As we already mentioned, it has been shown in
\cite{Dvali:2006su} that the above theory is equivalent to a gauge-invariant but nonlocal Lagrangian given by\footnote{Apart from the subtlety discussed in footnote~\ref{foot:caus}, which actually requires the whole reasoning to be done at the level of quantum effective action, so to obtain retarded Green's function in the equations of motion of the nonlocal formulation.}
\be\label{Lnonloc2}
{\cal L}=-\frac{1}{4} \Fmn \(1-\frac{m_{\g}^2}{\Box}\)\FMN
\, .
\ee
The formulation (\ref{Proca}) breaks gauge invariance but is local, while the formulation (\ref{Lnonloc2}) is gauge-invariant but nonlocal. One might  further localize \eq{Lnonloc2}, introducing for instance a field $U^{\mu\nu}=-\iBox\FMN$, following the pattern described above with Lagrange multipliers, obtaining a formulation of the theory which is both gauge-invariant and local. However, it is clear that $U^{\mu\nu}$ cannot be taken to be the most general solution of $\Box U^{\mu\nu}=-\FMN$, since otherwise we will introduce a spurious degree of freedom, associated to the general solution of the homogeneous equation $\Box U^{\mu\nu}=0$, which is independent of $A^{\mu}$ and is certainly not present in the original theory (\ref{Proca}). Once again, $U^{\mu\nu}$ must be taken to have fixed boundary conditions, and does not represent an extra dynamical field. In particular, once we go to a quantum description, there are no quanta associated to it.

The nonlocal term in eq.~(\ref{RR}) must be understood in the same way. It  represents a dynamically-generated mass term for the conformal mode. We can write it in a local form by introducing auxiliary fields. This is technically convenient because the equations of motion become local, and are formally equivalent to that of a scalar-tensor theory. However, it must be borne in mind that the boundary conditions on these fields are not arbitrary parameters that can be freely varied,  and these auxiliary fields do not represent new degrees of freedom of the theory. In particular, at the quantum level, there are no quanta associated to them.

This is conceptually important, because it shows that there are no quanta associated to these fields (otherwise, the quanta associated to $U$ would be ghost-like particles, see  the discussion in 
Section~6 of \cite{Maggiore:2016gpx}).
It is very important to distinguish between the existence of a ghost in the spectrum of the quantum theory, and the presence of a field, such as $U$, that can potentially induce instabilities at the classical level. The presence of a ghost in the spectrum of a quantum theory is fatal for the consistency of the theory, since the vacuum would be unstable to decay into   negative-energy ghosts plus normal, positive-energy particles. In contrast, the existence of classical instabilities does not necessarily imply any pathology of the theory. 
As we will review in Section~\ref{sect:cosmo}, in this case the classical instability develops on a cosmological timescale, and has the effect of producing an accelerated expansion of the Universe corresponding to an effective  dark energy with a  phantom  equation of state. This results in a perfectly viable cosmological evolution, both at the background level and at the level of cosmological perturbations.

\subsubsection{Boundary conditions on the auxiliary fields}\label{sect:boundary}

In the previous subsection we have seen that the boundary conditions on the two auxiliary fields $U$ and $S$ of the RR model are fixed, rather than representing free degrees of freedom.  In the case of the Polyakov action in $D=2$ we have  full control over the derivation of the nonlocal term, and we see from \eq{U2s} that the initial conditions on $U$ are fixed, in terms of the initial conditions on the metric, by
\be\label{Usigmainit}
U(t_{\rm in},\vx)=2\sigma(t_{\rm in},\vx)\, ,\qquad \dot{U}(t_{\rm in},\vx)=2\dot{\sigma}(t_{\rm in},\vx)\, .
\ee
However, in the RR model we do not currently have a derivation of the nonlocal term from a fundamental theory, so in practice the problem remains of how to choose the initial conditions for these fields. If we had to specify some generic functions $U(t_i,\vx)$, $\dot{U}(t_i,\vx)$, $S(t_i,\vx)$, $\dot{S}(t_i,\vx)$ at an initial time $t_i$, we would be in practice confronted with a large freedom, and the predictivity of the model could be lost. Fortunately, in a cosmological context, things simplify considerably. 

First of all, at the level of background evolution it is clear that, if we set the initial condition at early times, say deep in radiation dominance (RD), the Universe is highly homogeneous so we can set to zero the spatial dependence in these function. This  leaves us with four parameters  $U(t_i)$, $\dot{U}(t_i)$, $S(t_i)$, $\dot{S}(t_i)$, rather than four functions. Furthermore,  
in the RR model
these quantities parametrize one marginal and three  irrelevant directions in the parameter space. 
Indeed, as  we will review in Section~\ref{sect:backgRR},
the solution obtained setting
$U(t_i)=S(t_i)=0$ and $\dot{U}(t_i)=\dot{S}(t_i)=0$ deep in RD is an attractor with respect to three of the four initial conditions, while the marginal direction in the space of initial conditions  effectively corresponds to the introduction of  just one new free parameter~\cite{Maggiore:2013mea,Maggiore:2014sia}.

The problem in principle emerges again at the level of cosmological perturbations, when we expand also  the auxiliary
fields as a background plus perturbations, 
\be
U(t,\vx)=\bar{U}(t)+\d U(t,\vx)\, ,\qquad  
S(t,\vx)=\bar{S}(t)+\d S(t,\vx)\, .
\ee 
At the perturbation level we must 
depart from the assumption of homogeneity, and assign the initial conditions on 
$\d U(t_{\rm in},\vx)$, $\d S(t_{\rm in},\vx)$ and their first time derivatives at an initial time $t_{\rm in}$. 
The fact that the auxiliary fields do not represent arbitrary degrees of freedom but are fixed in terms of the metric means that the initial conditions for the perturbations of the auxiliary fields will be of order of the metric perturbations. One can therefore ask what happens if we start with initial conditions of this order of magnitude.
The initial conditions are set at a time when the modes of cosmological relevance are well outside the horizon. Their behavior in this regime is therefore the same as that of the  mode $\vk=0$, with three decaying modes and a marginal direction.  We will  then see explicitly in Section~\ref{sect:pertRR} that the dependence of our results on  the initial conditions of the perturbations  of the auxiliary fields, chosen to be  of order of the metric perturbations,
is  negligible.

\subsection{Restricting the choice of the  nonlocal term}\label{sect:free}

Another important question is how much freedom we have in choosing the form of the  nonlocal term. 
We have explored several options, and it turns out that it is  quite non-trivial to construct a model with an acceptable cosmological evolution, both at the level of background evolution and of cosmological perturbations. In turn, this gives important hints for the derivation of the nonlocal terms from a fundamental theory.

It can be useful to review the various models that have been studied in this context, and the difficulties that have been met, and that significantly restricted the choice of viable models. The original inspiration for our work came from the degravitation idea~~\cite{ArkaniHamed:2002fu,Dvali:2006su,Dvali:2007kt}, in which the Einstein equations were modified phenomenologically into
\be\label{degrav}
\(1-\frac{m^2}{\Box}\)\Gmn=8\pi G\Tmn\, .
\ee
This was originally proposed as an acausal modification of Einstein equations at the horizon scale but we have seen in Section~\ref{sect:causality} that, interpreting this as the equation of motion for the in-in vacuum expectation value of the metric, derived from the variation of a quantum effective action, automatically ensures causality, i.e. $\iBox$ in the equation of motion will be automatically the one defined with the retarded Green's function. 
However, \eq{degrav} also has the problem that the energy-momentum tensor is no longer automatically conserved, since in curved space the covariant derivative $\n_{\mu}$ does not commute with the covariant d'Alembertian $\Box$, and therefore does not commute with $\iBox$ either. This problem can be fixed by observing that, in a generic curved space-time, any symmetric tensor $\Smn$ can be decomposed as~\cite{Deser:1967zzb,York:1974}
\be\label{splitSmn}
S_{\mu\nu}=S_{\mu\nu}^{\rm T}+\frac{1}{2}(\n_{\mu}S_{\nu}+\n_{\nu}S_{\mu})\, , 
\ee
where $S_{\mu\nu}^{\rm T}$ is the transverse part of the tensor, that satisfies
$\n^{\mu}S_{\mu\nu}^{\rm T}=0$.  The extraction of the transverse part of a tensor is itself a nonlocal operation. Using the possibility of extracting the transverse part of a tensor, in \cite{Jaccard:2013gla} it was proposed to modify \eq{degrav} into
\be\label{GmnT}
\Gmn -m^2\(\iBox\Gmn\)^{\rm T}=8\pi G\,\Tmn\, ,
\ee
so that energy-momentum conservation $\n^{\mu}\Tmn=0$ is automatically satisfied. In \cite{Maggiore:2013mea,Foffa:2013vma} it was however found that the cosmological evolution that follows from this model is unstable, already at the background level. This is due to the fact that, when performing the localization procedure, one of the auxiliary fields has unstable modes  $e^{\beta_{\pm}x}$, with $\beta_+$ and $\beta_-$ positive both in RD and in matter dominance (MD). Even if one would fine-tune these modes to zero (which is very unnatural) these unstable modes would be unavoidably excited by any small perturbation, leading to an unacceptable cosmological evolution. We therefore concluded that this model is not viable. Thus, the stability of the modes associated to the homogeneous equation of the auxiliary fields, in both RD and MD, is a first  stringent requirement for the viability of these nonlocal models.

An initially more successful nonlocal model (the so-called RT model) was then proposed in 
\cite{Maggiore:2013mea},  based on the nonlocal equation
\be\label{RT}
\Gmn -\frac{m^2}{3}\(\gmn\iBox R\)^{\rm T}=8\pi G\,\Tmn\, .
\ee
The homogeneous equations associated to the auxiliary fields have stable solutions in both RD and MD, but not in a previous inflationary de~Sitter epoch. We have then studied this model starting its evolution in RD with the idea that, at the higher energies corresponding to de~Sitter inflation, it should be embedded in a more complete model. In that case the model works very well. In particular, at the background level it has  a self-accelerating solution, i.e. the nonlocal term behaves as  an effective  dark energy density, so we get accelerated expansion without  introducing a cosmological constant~\cite{Maggiore:2013mea,Foffa:2013vma}. This was then the first model of this class that suggested that the nonlocal term can drive the accelerated expansion of the Universe. During RD and MD, as well as in the present DE-dominated era, 
its cosmological perturbations are well behaved~\cite{Dirian:2014ara} and, implementing them into a Boltzmann code and performing Bayesian parameter estimation, one finds that the model fits the data very well, at a level comparable to $\Lambda$CDM~\cite{Dirian:2014bma,Dirian:2016puz}. However, a significant drawback of this model is that it does not give a consistent evolution if it is started from a previous epoch of primordial inflation. In this case there is a growing mode in the solution of the equation associated to an auxiliary field. As shown in   \cite{Cusin:2016mrr}, despite this growing mode, at the level of background  a viable cosmological evolution emerges (basically because the exponential growth of the auxiliary field during inflation never brings the effective DE term to a level competitive with the energy density that drives inflation, so it does not affect the background evolution during inflation, and is then subsequently compensated by an exponential decrease in RD). However, at the level of cosmological perturbations, in a de~Sitter phase  there are growing modes in the perturbations of the auxiliary fields. These induce growing modes in the metric perturbations, which would quickly bring the metric perturbations $\Phi,\Psi$ to a level ${\cal O}(1)$, spoiling the initial values ${\cal O}(10^{-5})$ generated by inflation in the standard scenario. Thus, this model is not viable  as a complete model, derived from a quantum effective action valid at all energy scales below the Planck scale. This makes the RT model less attractive, and here we will not consider it further.

The above models were defined at the level of equations of motion, and no simple form for a corresponding quantum effective action is known. Working at the level of the action, the RR model (\ref{RR}) was then proposed  in \cite{Maggiore:2014sia}. It  has been studied in detail in 
\cite{Dirian:2014ara,Dirian:2014bma,Dirian:2016puz,Dirian:2017pwp} and will be further considered in Section~\ref{sect:cosmo} below. It is related to the RT model by the fact that, when linearized over flat space, the two models become identical. However, they are otherwise different, and in particular their FRW solutions and perturbations are different.  
As we will review in Section~\ref{sect:backgRR},  in the RR model
the auxiliary fields introduced by the localization  have no unstable mode in any cosmological epoch, and the cosmological perturbations are well behaved. Again we have implemented the evolution of the background and of the perturbations into a  Boltzmann code, we have performed Bayesian parameter estimation and compared its performance to that of $\Lambda$CDM. In 
ref.~\cite{Dirian:2016puz} it was found that the RR model is disfavored with respect to $\Lambda$CDM by the comparison with the cosmological observations. However, in ref.~\cite{Dirian:2017pwp} it was then realized that this was an artifact, due to having applied to the RR model the ``{\em Planck} baseline analysis" used by {\em Planck} in the context of $\Lambda$CDM~\cite{Planck_2015_CP}. In this analysis the sum of neutrino masses, which a priori is a free parameter of the cosmological model, is taken fixed to the lower limit allowed by the oscillation experiments,
$\sum_{\nu}m_{\nu}=0.06$~eV. As discussed in \cite{Dirian:2017pwp}, and as we will review below,
this is an adequate choice for $\Lambda$CDM, since in this case, allowing the sum of neutrino masses to vary freely, they are driven toward zero, so imposing  the prior $\sum_{\nu}m_{\nu}\geq 0.06$~eV one finds that they hit the lower limit. However, this is no longer the case when the CMB data are analyzed with the RR model. In that case, the model gives a  prediction for $\sum_{\nu}m_{\nu}$ that nicely sits within the lower limit set by oscillation experiments and the upper limit from Tritium $\beta$-decay. The worse performance of the RR model with respect to $\Lambda$CDM observed in
\cite{Dirian:2016puz} was therefore an artifact due to the fact that one of the fitting parameters, the sum of neutrino masses, had been kept fixed to the value preferred by $\Lambda$CDM. Once the neutrino masses are left free to vary, within the limits of oscillation experiments, the RR model turns out to fit the data at the same level as $\Lambda$CDM.

Another attractive feature of the RR model, which is shared by all nonlocal models of this class, is that, at distances small compared to the inverse of the mass parameter $m$ (i.e., given that eventually $m$ will be fixed to a value $\sim H_0$ by the comparison with observations, at distances small compared to the present Hubble scale $H_0^{-1}$), the model smoothly reduces to GR~\cite{Maggiore:2013mea,Maggiore:2014sia,Kehagias:2014sda}, i.e. there is no van Dam-Veltman-Zakharov discontinuity, in contrast to
massive gravity. Therefore, one recovers all the successes of GR at the solar system and laboratory scale.\footnote{See Appendix~B of \cite{Dirian:2016puz} for  discussion of an issue about Lunar Laser Ranging raised in \cite{Barreira:2014kra}.}

In ref.~\cite{Cusin:2015rex} we have further explored a more general class of models. At the level of terms quadratic in the curvature
the most general quantum effective action involving the $\Box^{-2}$ operator is
\be\label{actionTotal}
\Gamma=\frac{\mplr^2}{2}\int d^4 x \sqrt{-g}\,
\left[R-\mu_1 R\frac{1}{\Box^2}R-\mu_2 C^{\mu\nu\rho\sigma}\frac{1}{\Box^2}C_{\mu\nu\rho\sigma}-\mu_3\RMN\frac{1}{\Box^2}\Rmn
\right]\,,
\ee
where $\mu_1$, $\mu_2$ and $\mu_3$ are  independent parameters with dimension of squared mass.
We  found that the term $\RMN\Box^{-2}\Rmn$ is again ruled out by the existence of growing modes, in RD and MD, of the homogeneous solutions of the auxiliary fields.\footnote{A similar result was  found in \cite{Ferreira:2013tqn}, where it was shown that a term 
$\RMN\Box^{-1}\Rmn$ also produces instabilities in the cosmological evolution. Observe that this term is rather of the Deser-Woodard type, i.e. of the form $\RMN f(\Box^{-1}\Rmn)$, with a dimensionless function $f$ and no explicit mass scale. The same holds if the $\iBox$ operator in the Deser-Woodard model  or the $\Box^{-2}$ operator in $\RMN\Box^{-2}\Rmn$ are replaced by a generic term $\RMN\small{\bigtriangleup^{-1}}\Rmn$, where $\bigtriangleup=m^4+\alpha_1\Box+\alpha_2\Box^2+\beta_1\RMN\n_{\mu}\n_{\nu}+\beta_2 R\Box+\gamma (\n^{\mu}\Rmn)\n^{\nu}$ \cite{Nersisyan:2016jta}.} For the  Weyl-square term the issue is more subtle. At the background level it does not contribute, since the Weyl tensor vanishes in FRW. We also found that, in the scalar sector, its cosmological perturbations are well behaved. However,  its tensor perturbations are unstable, so again this term is ruled out.

These studies identify the RR models has the most promising of this class of nonlocal models in which the nonlocal term is associated to a mass scale.
In an attempt at identifying as precisely as possible the ``correct" nonlocal model, which of course would be important also for understanding how it could be derived from a fundamental theory, we have further investigated variants of the RR model. One is the $\Delta_4$ model, 
defined by 
\be\label{D4}
\Gamma_{\Delta_4}=
\frac{\mplr^2}{2}\int d^{4}x \sqrt{-g}\, 
\[R-\frac{1}{6} m^2R\frac{1}{\Delta_4} R\]\, ,
\ee
where again $m$ is related to a fundamental scale $\Lambda_{\Delta_4}$ by $\Lambda_{\Delta_4}^4=(1/12)m^2\mplr^2$, and
\be\label{defDP}
\DP\equiv\Box^2+2\RMN\n_{\mu}\n_{\nu}-\frac{2}{3}R\Box+\frac{1}{3}\gMN\n_{\mu} R\n_{\nu}\, 
\ee
is called the Paneitz operator.
This operator  is well-known in the  mathematical literature because of its  conformal property: if two metrics $\gmn$ and $\gbmn$ are related by a conformal factor, $\gmn=e^{2\sigma}\gbmn$, then
\be\label{barDP}
\sqrt{-g}\,\DP=\sqrt{-\bar{g}}\, \bar{\Delta}_4\, .
\ee
In this sense, it is the four-dimensional analog  of the two-dimensional Laplacian, and indeed it  appears in the four-dimensional quantum effective action for the conformal mode, derived by integrating the conformal anomaly  (see e.g. \cite{Shapiro:2008sf,Maggiore:2016gpx} for reviews).\footnote{The Paneitz operator was also considered in the context of the Deser-Woodard class of nonlocal models  in \cite{Deser:2007jk}, where the authors considered the possibility of adding to the Ricci scalar in the action a term $R\Delta_4^{-1}R^2$ which, on dimensional ground, does not require the introduction of a mass scale.} 
The model (\ref{D4}) was introduced in \cite{Cusin:2016nzi}, where it was  studied  only at the level of the background evolution. Here we will study it also at the level of cosmological perturbations. As we will see, in the scalar sector the model is well behaved, and broadly consistent with the data, although it does not fit them with an accuracy comparable to  the RR or the $\Lambda$CDM model. 
However, we will show in Appendix~\ref{sect:D4} that it predicts a speed of propagation of gravitational waves (GWs) which is sensibly different from $c$ in the recent epoch, and is therefore ruled out by the recent observation of the binary neutron star coalescence GW170817 and its associated $\gamma$-ray burst GRB 170817A.\footnote{Still, the study of the scalar sector of the $\Delta_4$ model is interesting from a methodological point of view, since this model has a very phantom DE equation of state, which already puts it at the limit of providing an acceptable fit to the cosmological observations. On the other hand, the more phantom is the DE equation of state, the higher is the value of $H_0$ predicted by the model. So, the study of the $\Delta_4$ model is interesting because it gives an idea of the highest possible values that can be obtained for $H_0$ in a modified gravity model. We will then discuss its scalar perturbations and the corresponding Bayesian parameter estimation  in Appendix~\ref{sect:D4}.}
We will see in Section~\ref{sect:cgw} that, in contrast, in the RR model (as well as in the RT model), GWs propagate at the speed of light. 

These results show that it is highly non-trivial to build  nonlocal models that pass all these tests. It is quite interesting to observe that the one that does, the RR model, has a physical interpretation in terms of a mass term for the conformal mode, as we have seen in \eq{m2s2}.\footnote{The same is true for the RT and the $\Delta_4$ models. Indeed, the RT, $\Delta_4$  and RR models coincide when linearized over Minkowski space, but they are different beyond linear order, or when linearized over non-trivial backgrounds~\cite{Maggiore:2014sia}. In contrast, the model
(\ref{GmnT}) corresponds to giving a mass to the spin-2 modes, and indeed emerged in an attempt at writing massive gravity in nonlocal form~\cite{Jaccard:2013gla}.}

Further extensions and variations of the idea, that we will not further explore here, have been proposed.
For instance a  natural possibility, again suggested by conformal symmetry, is to replace $\Box^{-2}$ with 
$(-\Box + R/6)^{-2}$, since $(-\Box + R/6)$ is the operator that enters the action for a conformally-coupled scalar field. This model was studied  in \cite{Cusin:2016nzi} where it was found that, at the background level,  it gives a  viable cosmological model, with an evolution which is much closer to  $\Lambda$CDM compared to the RR model.  It will then be  more difficult to distinguish it observationally from $\Lambda$CDM with current data, so for the moment we will not study it further. A full one-parameter extension of the RR model using the operator $(-\Box + \xi R)$ had already been studied in \cite{Mitsou:2015yfa}. Of course, introducing one extra free parameter reduces the predictivity of the model, so here we will focus on models without this extra free parameters.
Finally, nonlocal models 
with an effective action of the form
\be\label{Gm2n}
\Gamma=\frac{\mplr^2}{2}\int d^{4}x \sqrt{-g}\,\[ R- \(\frac{m^2}{\Box}\)^n R\]\, ,
\ee
corresponding to a running of Newton's constant,
have been studied, at the level of  background evolution,  in \cite{Vardanyan:2017kal} for $n=1$ and in \cite{Amendola:2017qge} for $n=2$, and also appear to be in principle viable, at least at the background level. In particular the model with $n=1$ has an evolution very close to that of the RR model, up to the present epoch.

\section{Cosmological  consequences of the RR model}\label{sect:cosmo}

In this section we discuss the cosmological predictions of the RR  model,  we  test it against   CMB, BAO, SNe, local $H_0$ measurements and structure formation data, and we compare its performances to that 
of $\Lambda$CDM. We also investigate in detail the dependence of the predictions on the initial conditions of the auxiliary fields, at the background level as well as at the level of cosmological perturbations. 

\subsection{Cosmological background evolution and self-acceleration}\label{sect:backgRR}

To study the background evolution we specialize the equations of motion (\ref{Gmn})--(\ref{BoxUS}) to  a flat FRW metric  
$ds^2=-dt^2+a^2(t)d\vx^2$. We use $x\equiv\ln a$ to parametrize the temporal evolution, and we 
introduce the auxiliary fields $U$ and $S$ according to \eq{defU}. We also  define the dimensionless variables 
\be
W(x)\equiv H^2(x)S(x)\, ,
\ee 
and $h(x)\equiv H(x)/H_0$, where  $H(t)=\dot{a}/a$ and $H_0$ is the present value of the Hubble parameter. 
We use a prime to denote the derivatives with respect to $x$.   One then obtains~\cite{Maggiore:2014sia}
\bees
&&h^2(x)=\Omega_M e^{-3x}+\Omega_R e^{-4x}+\g Y\label{syh}\\
&&U''+(3+\zeta) U'=6(2+\zeta)\, ,\label{syU}\\
&&W''+3(1-\zeta) W'-2(\zeta'+3\zeta-\zeta^2)W= U\, ,\label{syW}
\ees
where $\gamma= m^2/(9H_0^2)$, $\zeta=h'/h$ and
\be\label{defY}
Y\equiv \frac{1}{2}W'(6-U') +W (3-6\zeta+\zeta U')+\frac{1}{4}U^2\, .
\ee
\Eq{syh} is a modified Friedmann equation with an effective dark-energy density 
\be\label{rdeY}
\rde=\rho_0\gamma Y\, ,
\ee
where, as usual, $\rho_0=3H_0^2/(8\pi G)$ is the critical density. 

Let us discuss first the potential ambiguity related to the choice of boundary conditions on the auxiliary fields, at the background level.
The initial conditions on the auxiliary fields $U$ and $W$ are in one-to-one correspondence with the solutions of the homogeneous equations associated to
\eqs{syU}{syW}, i.e. 
\bees
&&U''+(3+\zeta) U'=0\,  ,\\
&&W''+3(1-\zeta) W'-2(\zeta'+3\zeta-\zeta^2)W=0\, .
\ees 
In any given cosmological  epoch $\zeta$ has an approximately constant value $\zeta_0$, with  
$\zeta_0=\{0,-2,-3/2\}$  in de~Sitter (dS),  RD and  MD, respectively. Taking $\zeta$ constant  the homogeneous solutions are obtained analytically,
\bees
U_{\rm hom}&=& u_0+u_1 e^{-(3+\zeta_0)x}\, ,\\
W_{\rm hom}&=&w_0e^{-(3-\zeta_0)x}+w_1 e^{2\zeta_0x}\, .
\ees
In the early Universe we have $-2\leq \zeta_0\leq 0$ and  the terms associated to $u_1,w_0$ and $w_1$ 
are exponentially decreasing in $x$ (i.e., as a power-law in the scale factor $a$), and therefore correspond to irrelevant directions in the parameter space. Any solution that starts, deep in RD, with a non-vanishing value of $u_1,w_0$ or $w_1$ will quickly approach the solution obtained by setting $u_1=w_0=w_1=0$.
We also observe that there is no exponentially growing solution, so no instability at this level of the analysis. The remaining parameter $u_0$ is in principle a free parameter of the model. We will begin by studying the model with initial conditions $u_0=u_1=w_0=w_1=0$ at an initial time deep in RD (the ``minimal model"), and we will later study how the results change varying  $u_0$.

At the background level, the minimal mode  depends on the reduced Hubble parameter $h_0$, on the matter fraction $\oma$ and on the mass parameter $m$, that replaces the cosmological constant that appears in $\Lambda$CDM. In $\Lambda$CDM, assuming flatness, the energy fraction $\Omega_{\Lambda}$ associated to the cosmological constant is a derived quantity, fixed in terms of the matter energy density $\oma$  by $\ola+\oma=1$ (apart from the small contribution from  the radiation density $\ora$). Similarly, in the nonlocal models $m$ is a derived quantity, fixed again by the flatness condition in terms of $\oma$ (and $\ora$). The values of $\omega_M\equiv h_0^2\oma$ and of $h_0$ are determined in each model by developing cosmological perturbation theory and performing Bayesian parameter estimation, as will be discussed in Section~\ref{sect:Bay}. Let us anticipate that, using the CMB+BAO+SNa datasets specified in Section~\ref{sect:Bay},  for the RR model with $u_0=0$ one finds $\oma\simeq 0.299$ and $h_0\simeq 0.695$~\cite{Dirian:2017pwp}, see Table~\ref{tab:res1}. Once fixed the value of these parameters, it is straightforward to integrate numerically \eqst{syh}{syW}~\cite{Maggiore:2014sia}. In particular, we can then study the evolution of the dark energy density defined by \eqs{defY}{rdeY}, and of the corresponding DE  equation of state parameter $w_{\rm DE}(x)$, defined by

\begin{figure}[t]
\centering
\includegraphics[width=0.42\columnwidth]{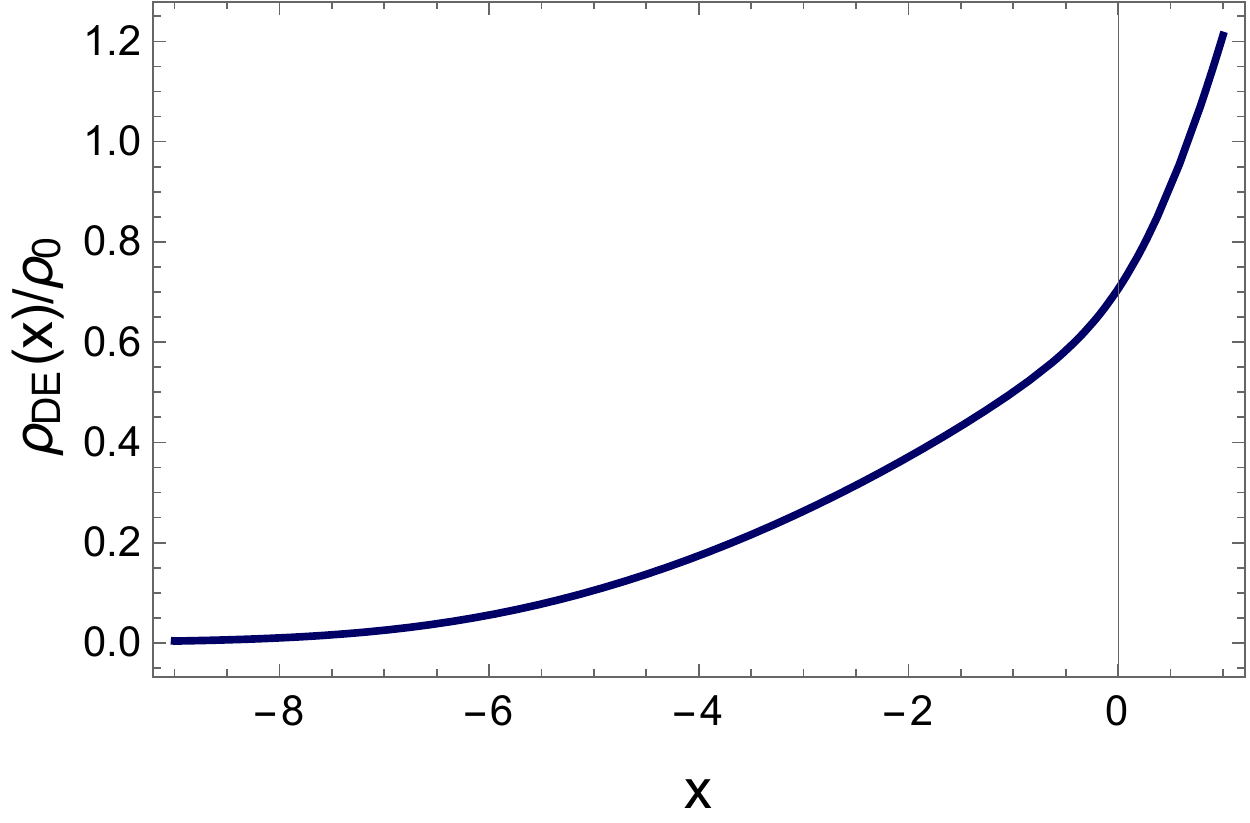}
\includegraphics[width=0.42\columnwidth]{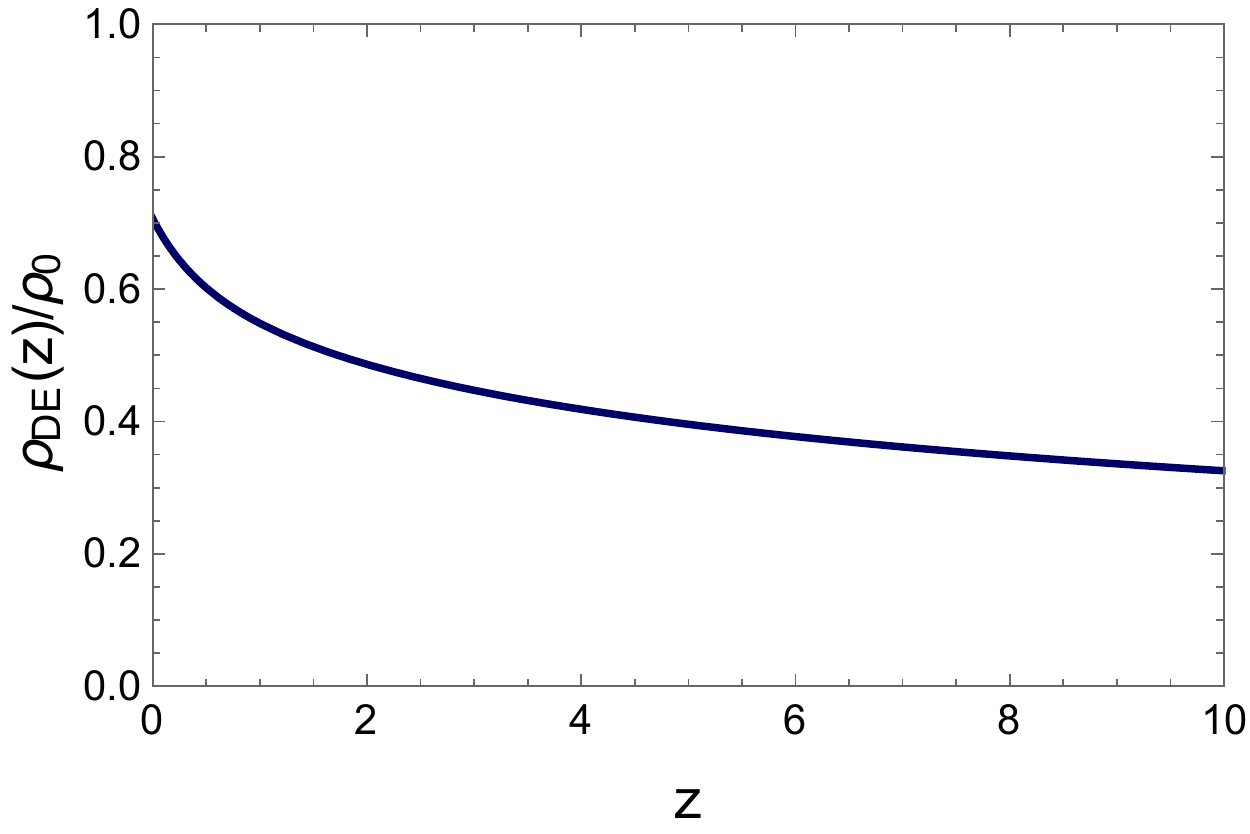}
\includegraphics[width=0.42\columnwidth]{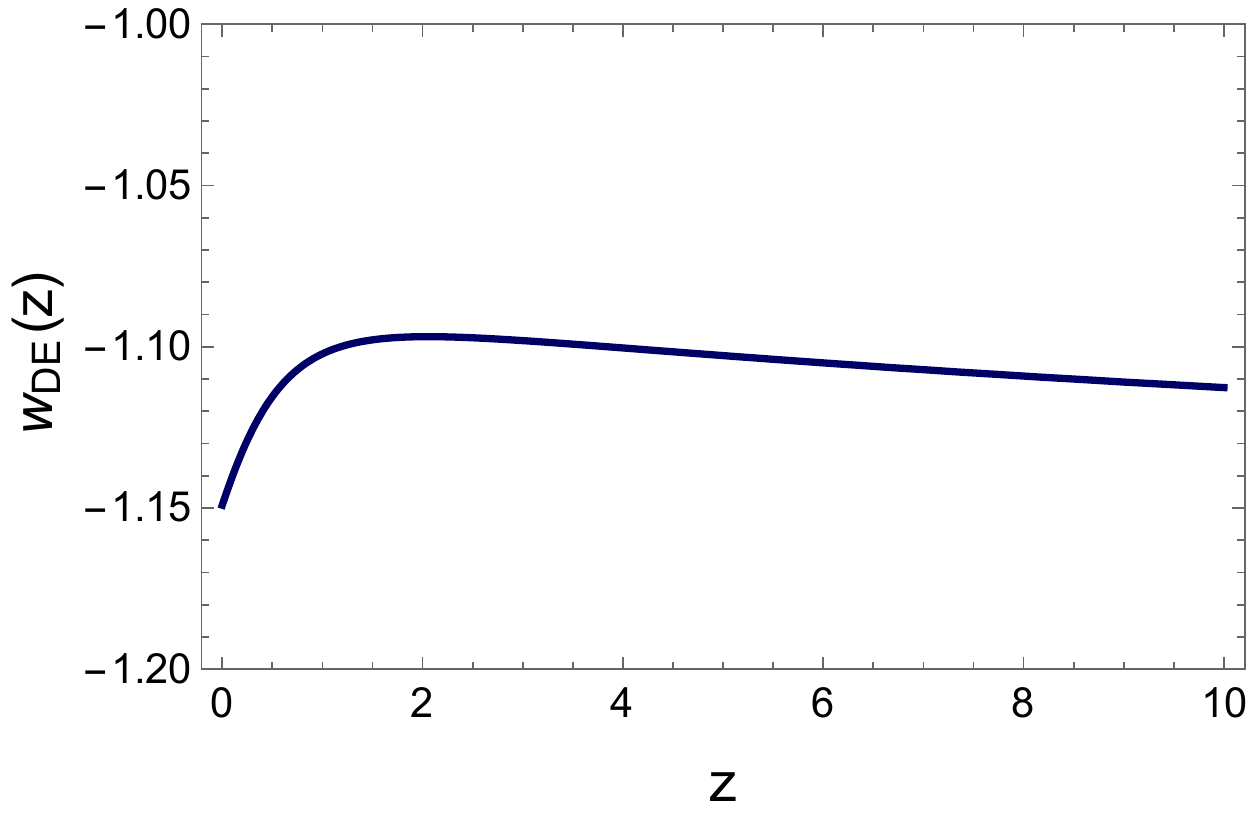}
\includegraphics[width=0.42\columnwidth]{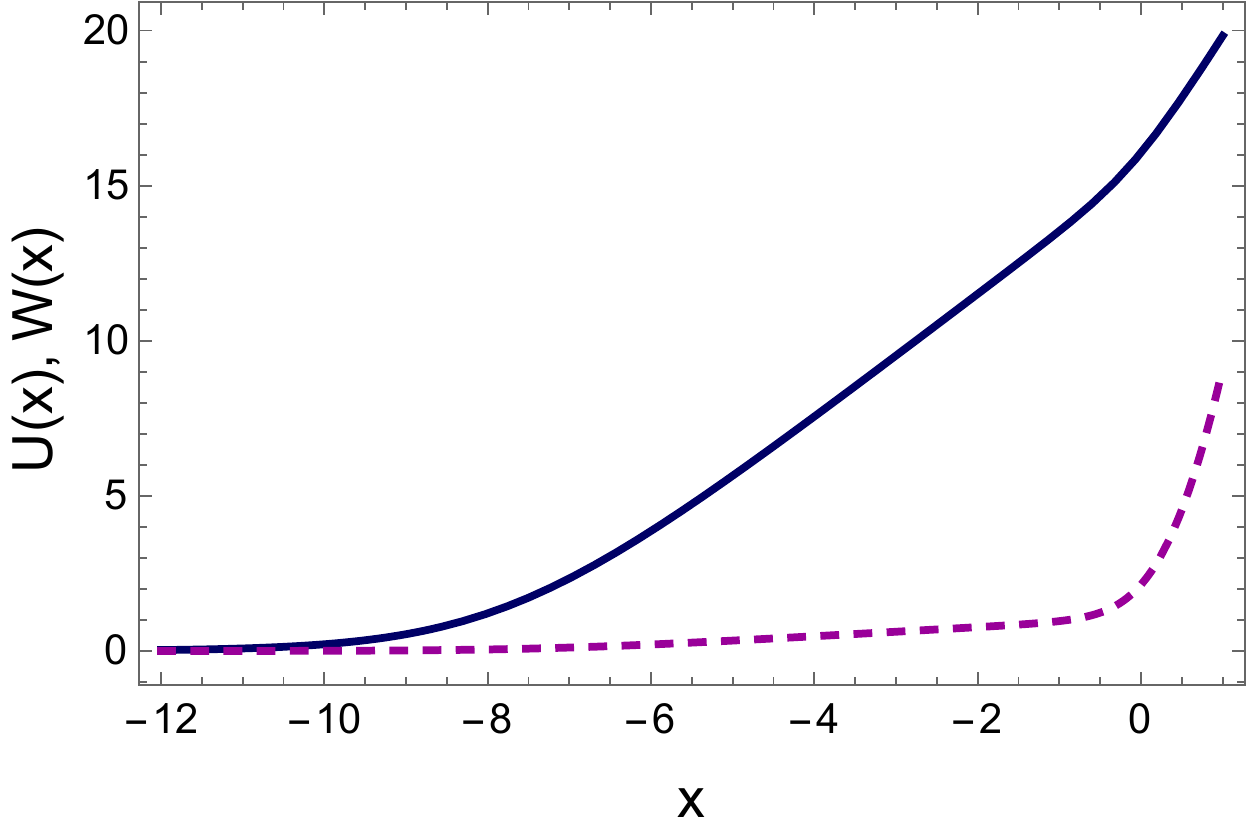}
\caption{Upper left panel: the function $\rde(x)/\rho_0$ for the RR model (setting $\oma\simeq 0.294$ and $h_0\simeq 0.695$) against $x\equiv \ln a$.
Upper right panel:  the same quantity shown against the redshift $z$.
Lower left panel: the DE equation of state $w_{\rm DE}(z)$. Lower right panel: the auxiliary fields $U(x)$ (blue, solid line) and $W(x)$ (red, dashed).
\label{fig:backgRR}
}
\end{figure}

\be\label{consrho}
{\rho}'_{\rm DE}+3(1+w_{\rm DE})\rho_{\rm DE}=0\, .
\ee
The upper-left panel in Fig.~\ref{fig:backgRR} shows $\rde(x)/\rho_0$ as a function of $x=\log a$. Observe that RD--MD equilibrium takes place at $x=x_{\rm eq}\simeq  -8.1$, while the present epoch corresponds to $x=0$, since we set $a(t_0)=1$. Thus $\rde(x)$ vanishes during RD, and then starts to grow as we enter in the MD epoch. At the present time, $x=0$, $\rde(0)/\rho_0\simeq 0.701$, as determined by the value of $\oma$, and in the cosmological future ($x>0$) it continues to grow.
The upper-right panel shows $\rde(z)/\rho_0$ as a function of redshift $z$, on a scale that emphasizes the recent epoch $z<10$. The corresponding equation of state (EoS) function $w_{\rm DE}(z)$ is shown in
the lower-left panel. Observe that  the EoS is phantom, $w(z)<-1$. This is clear from \eq{consrho}: since  $\rde>0$ and also its derivative $\rde'>0$, we must have $(1+w_{\rm DE})<0$. With the chosen values of the parameters, we get $w_{\rm DE}[z=0]\simeq -1.149$. 
Comparing with the commonly used parametrization of the form \cite{Chevallier:2000qy,Linder:2002et}
\be\label{ChevLind}
w_{\rm DE}(a)= w_0+(1-a) w_a\, ,
\ee
(where $a(x)=e^x$) in the region $-1<x<0$,  for  $\oma\simeq 0.299$ and $h_0\simeq 0.695$ we get
\be
w_0\simeq -1.15, \qquad w_a\simeq 0.09\, .
\ee 
In the lower-right panel of Fig.~\ref{fig:backgRR} we plot the evolution of the auxiliary fields. Having set $u_0=0$, the field $U=-\iBox R $ starts from zero and remains zero during RD, as a consequence of the fact that, in RD, the Ricci scalar $R=0$, so the equation $\Box U=-R$ with initial condition $U(x_{\rm in})=0$ gives $U(x)=0$. However, as we enter in the MD phase,  it starts to grow, driving the growth of $\rde$ according to \eq{defY}. In a sense, this behavior is the reflection of a classical instability, related to the fact that $U$ is a ``ghost-like" field. It is however a welcome instability, since it generates an effective dark energy and self-acceleration. A consequence of this behavior is that $\rde>0$ and $\rde'>0$ and therefore, as discussed above, a phantom EoS of dark energy, $\wde(z)<-1$. Indeed, it was realized 
long ago~\cite{Caldwell:1999ew} that a phantom dark energy requires a ghost-like field. Here this behavior is indeed generated by the field $U$ which, classically, indeed has a wrong-sign kinetic term~\cite{Foffa:2013sma}. However, as discussed in Section~\ref{sect:loca}, there are no quanta associated to this field. Thus, in a sense, the $U$ field is a ``benign" ghost, that classically does the job of inducing an instability that gives a phantom EoS for the dark energy, but does not create consistency problems at the quantum level. We see that the inclusion of nonlocal terms in the quantum effective action naturally produces, in a consistent field-theoretical framework, the ``exotic" behavior advocated in \cite{Caldwell:1999ew} as the origin of a phantom dark energy EoS. In this sense, the observation of a phantom dark energy would be a very strong indication in favor of nonlocal models of this class.

\begin{figure}[t]
\centering
\includegraphics[width=0.45\columnwidth]{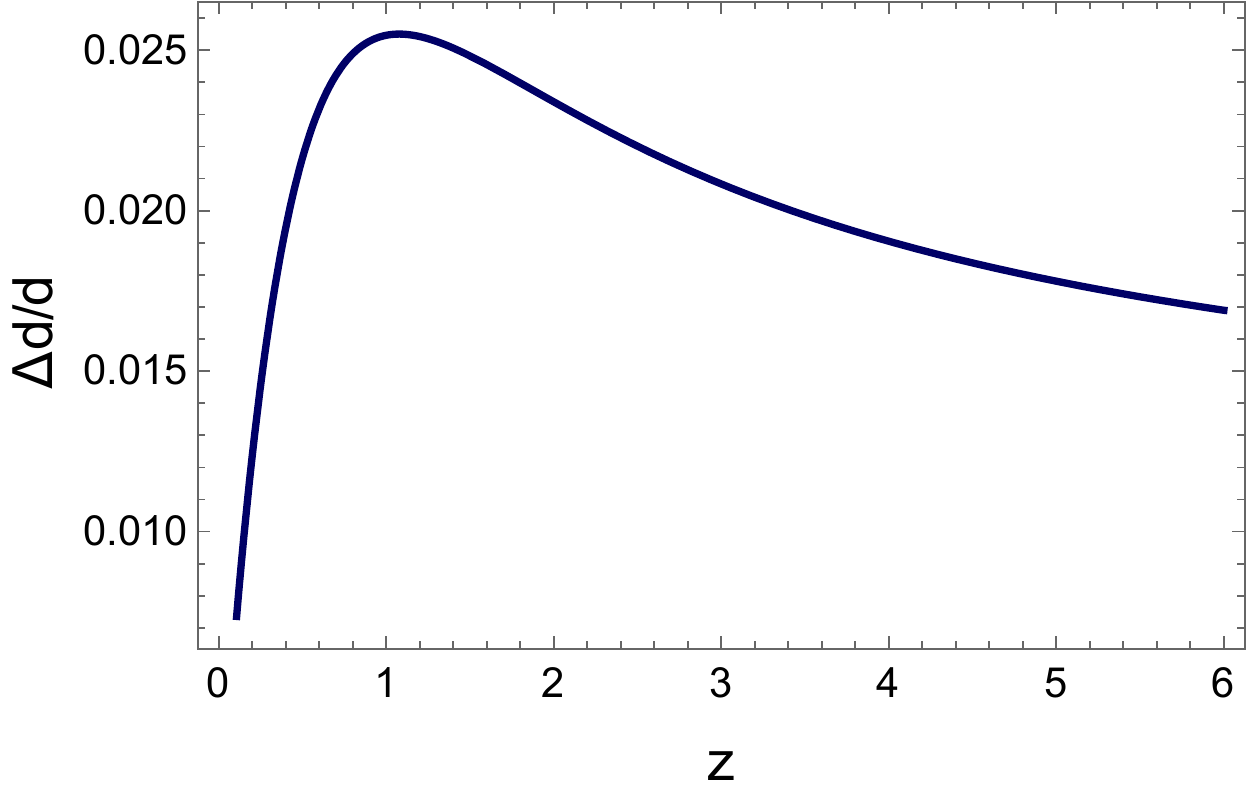}
\includegraphics[width=0.45\columnwidth]{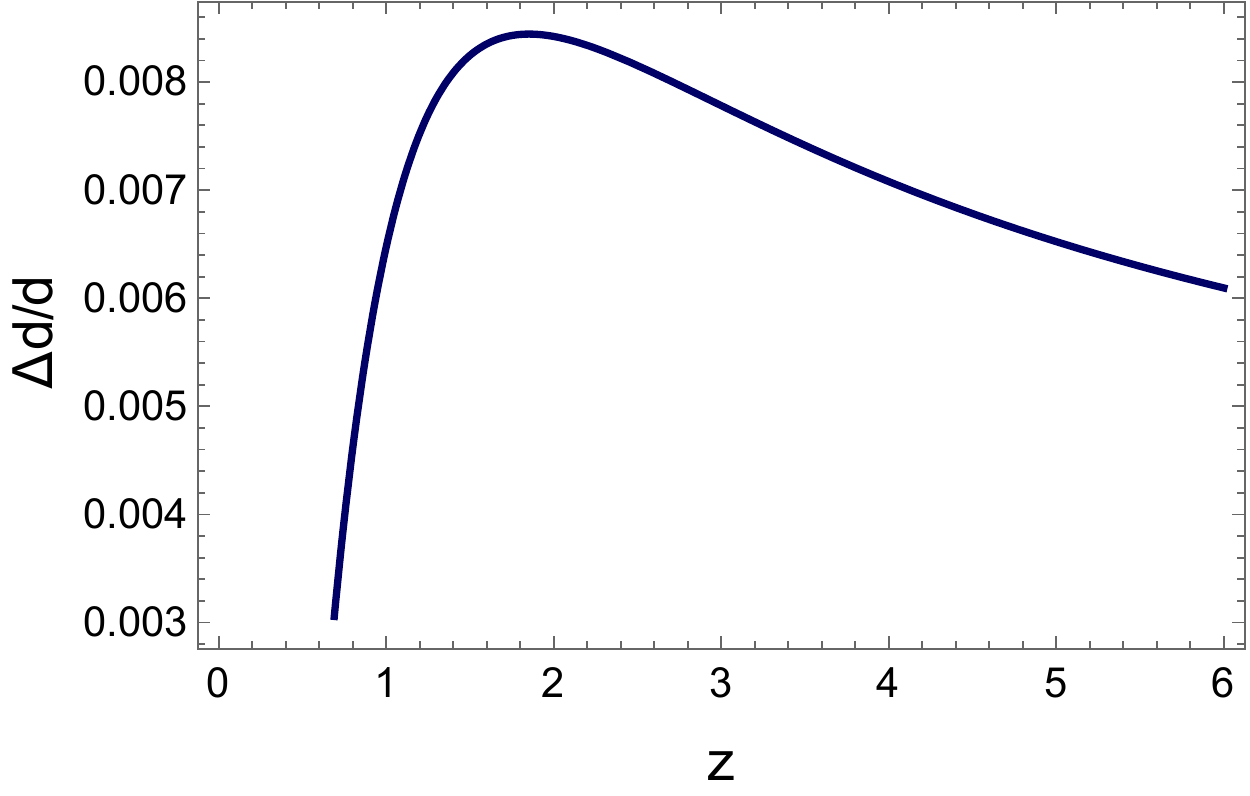}
\caption{Left panel: the relative difference $\Delta d/d=[d^{\rm RR}_{\rm com}-d^{\Lambda{\rm CDM}}_{\rm com}]/d^{\Lambda{\rm CDM}}_{\rm com}$ of comoving distances, using the same values of $h_0$ and $\oma$ for both the RR  and $\Lambda$CDM models.
Right panel: $\Delta d/d$ using the best-fit values $\oma\simeq 0.299$ and $h_0\simeq 0.695$ for RR,
and  $\oma\simeq 0.309$ and $h_0\simeq 0.677$  for $\Lambda$CDM (see Table~\ref{tab:res1}). 
\label{fig:comdistRR}
}
\end{figure}

It is also interesting to compare  the comoving distance 
 $d_{\rm com}(z)$ in the RR model,
\be\label{4comovdistzRR}
d^{\rm RR}_{\rm com}(z)=\frac{1}{H_0}\int_0^z\, 
\frac{d\tilde{z}}{\sqrt{\ora (1+\tilde{z})^4+\oma (1+\tilde{z})^3+\rde(\tilde{z})/\rho_0 }}\, ,
\ee
to the comoving distance in $\Lambda$CDM,
\be\label{4comovdistzLCDM}
d^{\Lambda{\rm CDM}}_{\rm com}(z)=\frac{1}{H_0}\int_0^z\, 
\frac{d\tilde{z}}{\sqrt{\ora (1+\tilde{z})^4+\oma (1+\tilde{z})^3+\ola}}\, ,
\ee
as a function of redshift.  In Fig.~\ref{fig:comdistRR} we show the relative difference 
\be
\frac{\Delta d}{d}\equiv
\frac{d^{\rm RR}_{\rm com}-d^{\Lambda{\rm CDM}}_{\rm com}}{d^{\Lambda{\rm CDM}}_{\rm com}}\, .
\ee
This is of course the same as the relative difference  of the luminosity distances $d_L(z)=(1+z)d_{\rm com}(z)$, or  of the angular diameter distances $d_A(z)=(1+z)^{-1}d_{\rm com}(z)$.\footnote{As we will discuss in Section~\ref{sect:cgw}, this notion of  luminosity distance 
is only appropriate for electromagnetic signals; in the RR model the  luminosity distance associated to GW sources is different. We will expand on this in a companion paper~\cite{Belgacem:2017ihm}.}
In the left panel we use
the same values of $h_0$ and $\oma$ in both \eqs{4comovdistzRR}{4comovdistzLCDM}, to show how the different functional dependence of $\rde(z)$, with respect to the constant dark energy density of $\Lambda$CDM, affects the result. In the right panel we use for each model its own best-fit values obtained by parameter estimation,  namely $\oma\simeq 0.299$ and $h_0\simeq 0.695$ for RR,
and  $\oma\simeq 0.309$ and $h_0\simeq 0.677$  for $\Lambda$CDM 
(see Table~\ref{tab:res1}). 
We see that, for the same values of the parameters, at $z\simeq 1$  the comoving distance in RR is higher than in $\Lambda$CDM
by about 2.5 \%. However, we also see from the right panel that parameter estimation partially compensates, bringing the relative difference well below $1\%$. This happens because  the model parameters are fitted so to reproduce fixed distance rulers provided by CMB, BAO and SNe.

\begin{figure}[t]
\centering
\includegraphics[width=0.42\columnwidth]{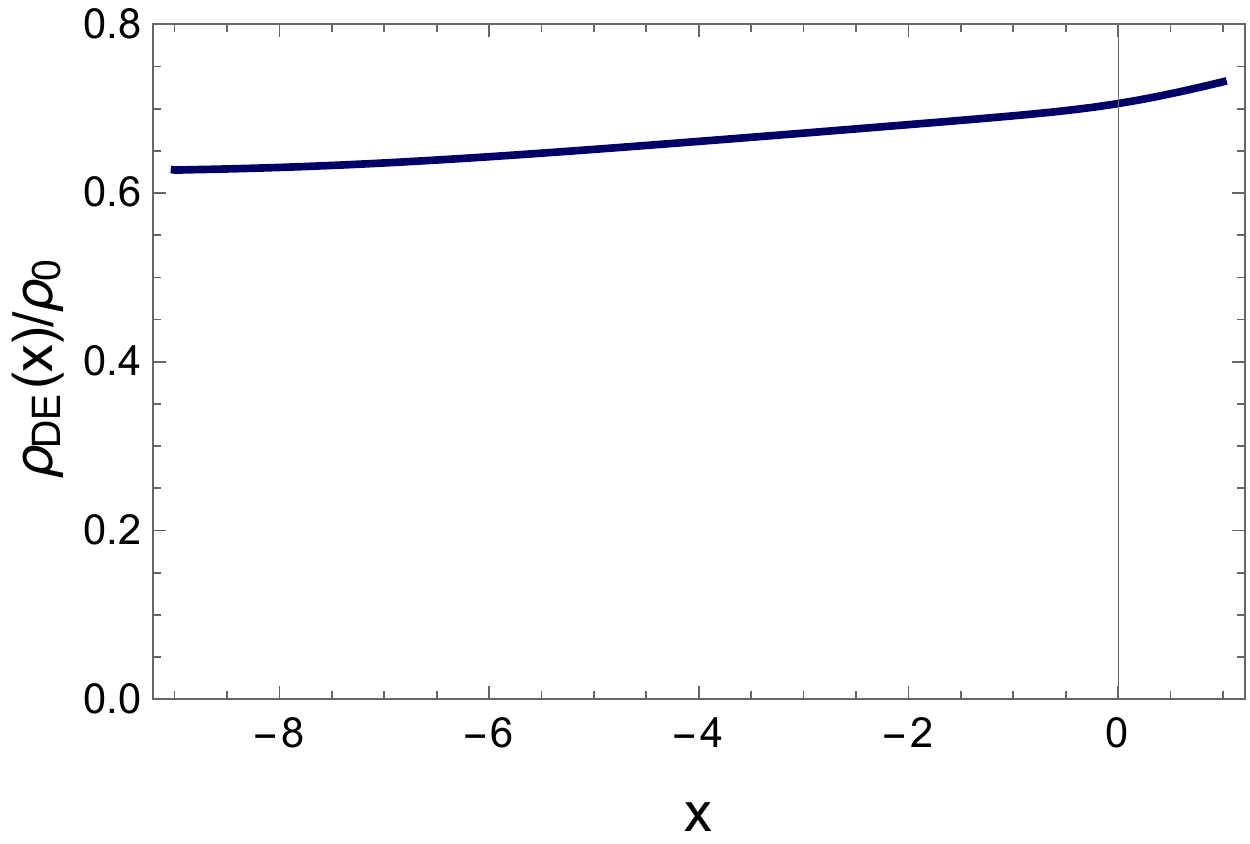}
\includegraphics[width=0.42\columnwidth]{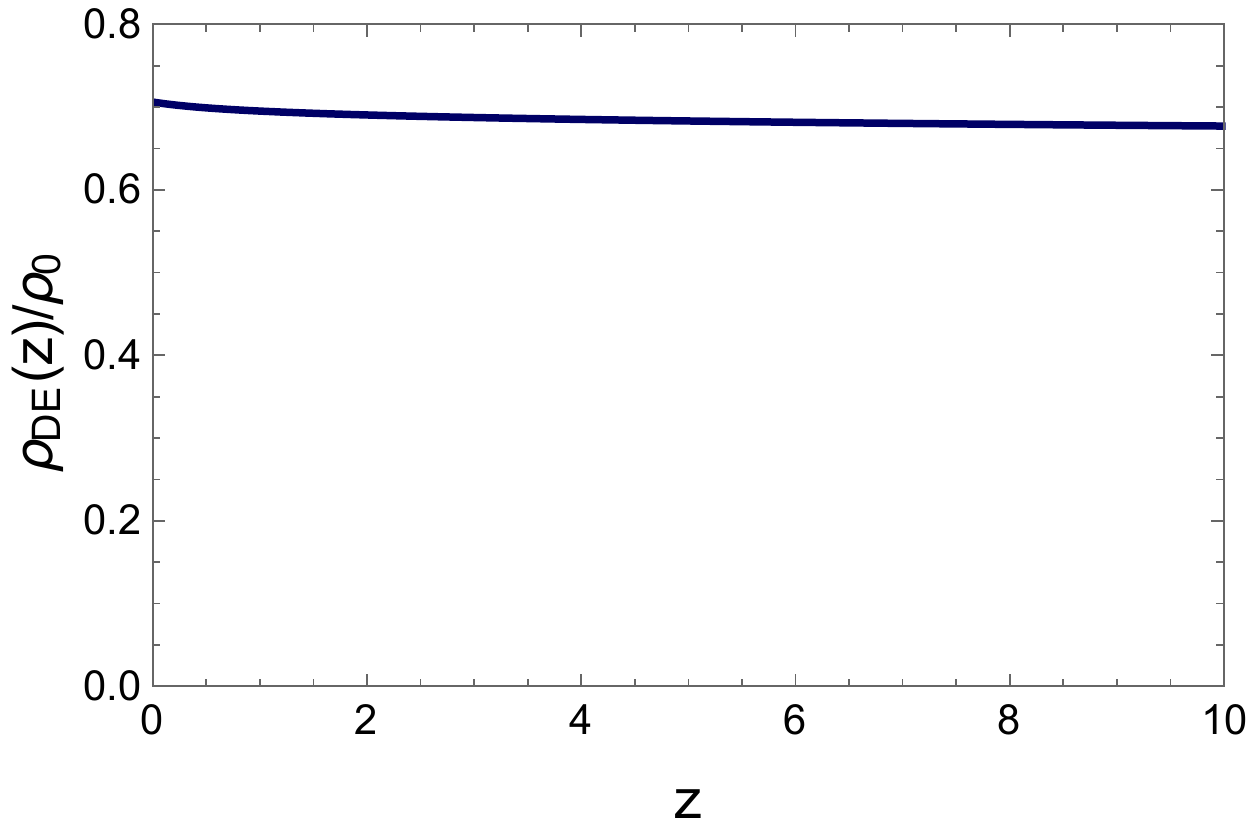}
\includegraphics[width=0.42\columnwidth]{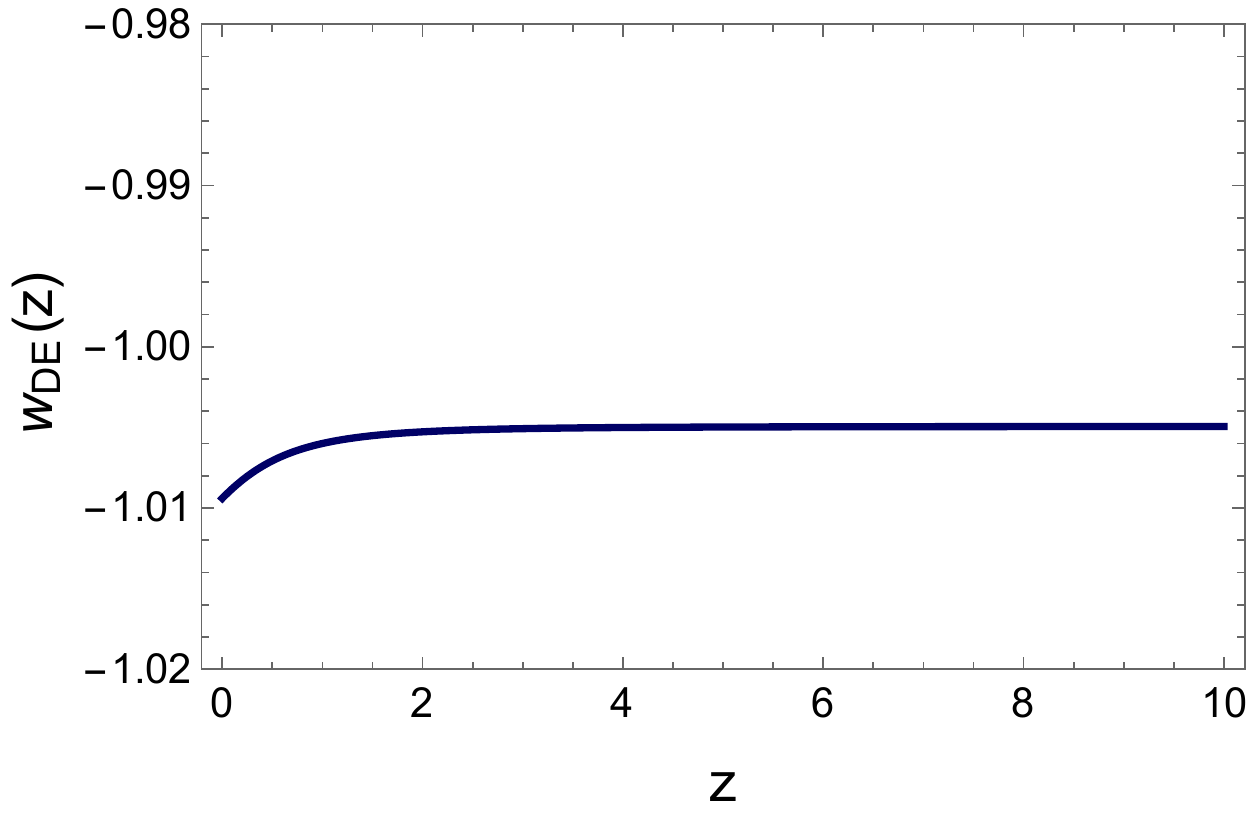}
\includegraphics[width=0.42\columnwidth]{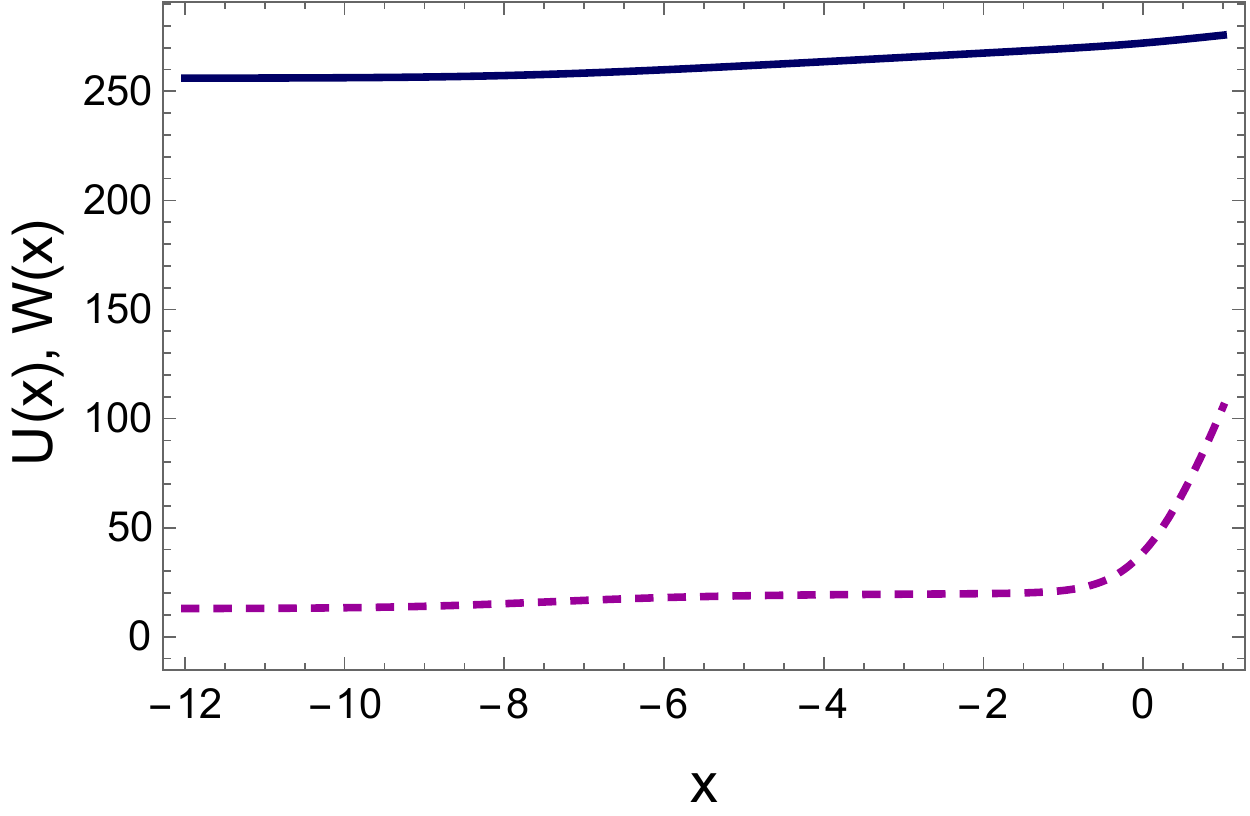}
\caption{As in Fig.~\ref{fig:backgRR}, for the RR model with initial conditions $u_0=250$.\label{fig:RRu0256}
}
\end{figure}

We next discuss how the results change if we set $u_0\neq 0$ as initial condition in RD (see also~\cite{Cusin:2016mrr,Maggiore:2016gpx}). In principle, a large positive value of $U$ during RD could be generated by a previous inflationary era.
In any given cosmological era, the function $\zeta(x)$ has an approximately constant value $\zeta_0$, with  $\zeta_0=0$ in dS, $\zeta_0=-2$ in RD and $\zeta_0=-3/2$ in MD. In the approximation of constant $\zeta$ \eq{syU} can be integrated analytically
\cite{Maggiore:2013mea}, 
\be \label{pertU}
U(x)=\frac{6(2+\zeta_0)}{3+\zeta_0}x+u_0
+u_1 e^{-(3+\zeta_0)x}\, ,
\ee
where the first term on the right-hand side is a solution of the inhomogeneous equations, to which we add  the most general solution of the homogeneous equation.  Even if, at the beginning of the inflationary era, $u_{0,1}$ were small and
$U(x_{\rm in})$ were at most of order one, at the end of inflation $U(x)$ is large because of the linear dependence on $x$ of the inhomogeneous term. 
Setting in \eq{pertU} $\zeta_0=0$, as in a de~Sitter inflationary phase, we see that in a de~Sitter epoch
\be \label{pertUdS}
U^{\rm dS}(x)= 4x+u^{\rm dS}_0
+u^{\rm dS}_1 e^{-3x}\, .
\ee
Thus, at the end of inflation   $U(x_{\rm end}) \simeq 4 (x_{\rm end}-x_{\rm in})\equiv
4\Delta N$, where $\Delta N$ is the number of inflationary e-folds (recall that $x=\ln a$), and
the field $U$ can enter the RD phase with an initial value ${\cal O}(10^2)$. Matching this solution which 
the RD solution obtained setting $\zeta_0=-2$ in \eq{pertU}
\be
U^{\rm R}(x)\simeq u^{\rm R}_0
+u^{\rm R}_1 e^{-x}\, ,
\ee
we see that this  corresponds to setting $u^{\rm R}_0={\cal O}(10^2)$ at an initial time deep in RD. In contrast, the other auxiliary field enters the RD phase with a negligible value~\cite{Maggiore:2016gpx}.
Fig.~\ref{fig:RRu0256} shows the results for $u_0=250$, corresponding to $\Delta N\simeq 63$ in a simple inflationary model that ignores reheating [since $U(x)$ is constant to great precision during RD, it is irrelevant the exact point in RD when we impose this initial condition].  We see that the DE density is almost constant, and hardly distinguishable from  the result in $\Lambda$CDM and, for this value of $u_0$, the DE equation of state $w_{\rm DE}(z)$ differs from $-1$ by less than $1\%$, again on the phantom side.
Observe also that a different branch of solutions, also potentially viable, at least at the background level, exists if $u_0$ is negative and smaller than a critical value \cite{Nersisyan:2016hjh}.

In the following, when discussing the parameter estimation for the RR model, we will first focus on the $u_0=0$ case, since this is the case that can be more easily  distinguished from $\Lambda$CDM with current or near-future observations, but we will also compare with the results for a large value of $u_0$, chosen to be $u_0=250$. The latter is of course more difficult to distinguish from $\Lambda$CDM with near-future  
cosmological observations. However, the RR model with a large value of $u_0$ is also conceptually interesting because it gives an example of a  model that  generates an effective dark energy that, at least up to the present epoch, behaves almost like a cosmological constant, without however relying on a vacuum energy term, and therefore without suffering from the lack of technical naturalness associated to the cosmological constant.

\subsection{Cosmological perturbations}\label{sect:cosmopert}

The next step is the study of the cosmological perturbations of the nonlocal model.
When performing Bayesian parameter estimation and comparison with $\Lambda$CDM we will insert the cosmological perturbation equations in a modified version of the CLASS Boltzmann code.
The details on the implementation of the cosmological perturbation equations for the RR  model in CLASS have been  discussed in detail in appendix~A of \cite{Dirian:2016puz}. In this section we  discuss a simpler treatment of the cosmological equations, in which we consider radiation and non-relativistic matter as  perfect fluids, neglecting anisotropic stress, neutrino effects, etc. The corresponding equations, which can be quickly integrated numerically, are useful for a first understanding of the behavior of the perturbations.
For the  RR model the perturbations were studied in great detail   
in~\cite{Dirian:2014ara} and, for completeness,  we will recall the main equations  in Section~\ref{sect:pertRR}
(see also Section~7.2 of \cite{Maggiore:2016gpx} for review). We will also extend the 
results of ref.~\cite{Dirian:2014ara}, studying the dependence of the results on the initial conditions for the perturbations of the auxiliary fields.

We consider here the scalar sector, deferring the discussion of tensor perturbations to Section~\ref{sect:cgw}. We work in the  Newtonian gauge, so  the metric perturbations in the scalar sector are written as
\be\label{defPhiPsi}
ds^2 =  -(1+2 \Psi) dt^2 + a^2(t) (1 + 2 \Phi) \delta_{ij} dx^i dx^j\, .
\ee
Similarly,  we  expand perturbatively the auxiliary fields. In the RR model we have two auxiliary fields $U$ and $S$, defined in \eq{BoxUS}, or rather $U$ and $W=H^2S$.
We then expand the auxiliary fields, writing 
\be
U(t,\vx)=\bar{U}(t)+\d U(t,\vx)\, ,\qquad W(t,\vx)=\bar{W}(t)+\d W(t,\vx)\, ,
\ee
(in this section we use an overbar to denote background quantities), 
and we work in terms of the Fourier modes $\Psi_{\vk}(t)$, $\Phi_{\vk}(t)$,
$\d U_{\vk}(t)$, $\d W_{\vk}(t)$.\footnote{Our conventions on the volume factors in the Fourier transform are the standard ones in the cosmological context, such that, e.g., $k^{3/2} \Psi_{\vk}$ is dimensionless, see footnote~12 in
\cite{Dirian:2014ara}.} In the simplified treatment of this section
we use the standard definitions for the perturbations of the energy-momentum tensor of matter (i.e. non-relativistic matter and radiation)
\bees
T^0_0 &=& -(\bar{\rho} +\delta \rho)\, ,\\ 
T^0_i & =&  (\bar{\rho} + \bar{p})v_i\, ,\\
T^i_j &=& (\bar{p} + \delta p) \delta^i_j + \Sigma^i_j\, ,
\ees
where $\bar{\rho}$ and $\bar{p}$ are the unperturbed energy density and pressure. 
Thus, in the energy-momentum tensor the perturbation variables are $\delta \rho, \delta p$, $v_i$, and 
the anisotropic stress tensor $\Sigma^i_j$. In the perfect-fluid approximation we take $\Sigma^i_j=0$.
The pressure perturbations can be written as $\d p =c_s^2\d \rho$, where $c_s^2$ is the speed of sound of the fluid, and we define as usual $\delta \equiv \delta \rho / \bar{\rho}$ and $\theta \equiv \delta^{ij} \partial_i v_j$; $\d_R,\theta_R$ will refer to radiation, and
$\d_M,\theta_M$ to matter.

\subsubsection{Initial conditions and perturbations of the auxiliary fields}

The initial conditions on the metric and matter perturbations are the usual adiabatic initial conditions derived from inflation. For modes that were well outside the horizon at the initial time, when the initial conditions are imposed, they are given by\footnote{Actually, for performing the numerical work, we use the more accurate expressions that also include  the  corrections of order $\hat{k}_{\rm in}^2$, where
$\hat{\vk}_{\rm in}=\vk/(a_{\rm in} H_{\rm in})$, see  eqs.~(6.1)--(6.3) of \cite{Dirian:2014ara}.}
\bees
\Phi_{\vk}&=&-\Psi_{\vk}=A(k)\, ,\label{PsiPhiini}\\
(\d_R)_{\vk}&=&\frac{4}{3}(\d_M)_{\vk}=A(k)\, ,\\
(\theta_R)_{\vk}&=&(\theta_M)_{\vk}=-\frac{\hat{k}_{\rm in}^2}{2}A(k)\, ,
\ees
set at a time $t_{\rm in}$ deep in RD. We also defined
\be
\hat{\vk}(t)\equiv \frac{\vk}{a(t) H(t)}\, ,
\ee
so in particular
$\hat{\vk}_{\rm in}= \vk/(a_{\rm in} H_{\rm in})$. The function
$A(k)$ is related to the amplitude $A_s$ and tilt $n_s$ of the scalar perturbations (at a pivot scale $k_*$) by
\be\label{Ak2}
A^2(k)=\frac{8\pi^2}{9k^3} A_s\, \(\frac{k}{k_*}\)^{n_s-1}\, .
\ee
The use, in the nonlocal model,  of the standard adiabatic initial conditions derived from inflation is justified by the fact   that the energy density associated to the nonlocal term is totally irrelevant during an earlier phase of inflation~\cite{Maggiore:2016gpx,Cusin:2016mrr}, so the inflationary dynamics in a nonlocal model supplemented by an inflationary sector is exactly the same as in $\Lambda$CDM supplemented by the same inflationary sector. This holds because in the RR model there is no instability during an inflationary epoch, neither at the level of background evolution, nor at the level of 
perturbations~\cite{Cusin:2016mrr}. As we mentioned in Section~\ref{sect:free}, this is not the case for the RT model, which is the reason why we do not consider it further.  

We must also assign the initial conditions on the perturbations of the auxiliary fields.
In refs.~\cite{Dirian:2014bma,Dirian:2016puz,Dirian:2017pwp}
the analysis  was limited to vanishing initial conditions, i.e.  
\be
\d U_{\vk}=\d W_{\vk}=\d U'_{\vk}=\d W'_{\vk}=0
\ee
for all Fourier modes (set at some initial time  $t_{\rm in}$ deep in RD). 
More generally,
in Section~\ref{sect:backgRR} we have seen that, at the background level, i.e. for the homogeneous mode $\vk=0$, out of the four  quantities $\bar{U}_{\vk=0}(t_{\rm in})$, $\bar{W}_{\vk=0}(t_{\rm in})$, $\bar{U}'_{\vk=0}(t_{\rm in})$ and   $\bar{W}'_{\vk=0}(t_{\rm in})$, three  parametrize irrelevant directions, since the corresponding solutions are quickly attracted toward that obtained with vanishing initial conditions. There is however a marginal direction in parameter space, corresponding to 
a constant shift $U(t_{\rm in},\vx)\ra U(t_{\rm in},\vx)+u_0$, which can also be seen as  a shift of the zero mode of the perturbations,
$\d U_{\vk=0}\ra \d U_{\vk=0}+u_0$. We now wish to understand what is the effect of changing the initial conditions of the modes
$\d U_{\vk}$, $\d W_{\vk}$, $\d U'_{\vk}$ and $\d W'_{\vk}$, with $\vk$ non-vanishing and of the order of the typical comoving momenta relevant for cosmology. 
We have seen in Sections~\ref{sect:loca} and \ref{sect:boundary} that the initial conditions of the auxiliary fields are not free parameters corresponding to new degrees of freedom, but rather are in principle  fixed in terms of the initial conditions of the metric. In the case of the Polyakov action in $D=2$, where we have an explicit derivation of the nonlocal quantum effective action from the fundamental theory, we have seen  in \eq{Usigmainit} how the initial conditions on the auxiliary fields are related to that of the metric. In particular, if one  expands 
$U(t,\vx)=\bar{U}(t)+\d U(t,\vx)$ and $\sigma(t,\vx)=\bar{\sigma}(t)+\d \sigma(t,\vx)$, \eq{Usigmainit} fixes the initial conditions on the Fourier modes of the perturbation $\d U_{\vk}(t)$ of the auxiliary field, in terms of the initial conditions on the metric perturbation $\d\sigma_{\vk}(t)$, 
\be\label{deltaUdsigmainit}
\d U_{\vk}(t_{\rm in})=2\,\d\sigma_{\vk}(t_{\rm in})\, ,\qquad 
\d U'_{\vk}(t_{\rm in})=2\,\d\sigma'_{\vk}(t_{\rm in})\, .
\ee
For the RR  model we do not have a similar  derivation of the nonlocal term from a fundamental theory. At first sight, one might fear that our ignorance of the initial conditions on the perturbations of the auxiliary fields will lead to a significant loss of predictivity. However, first of all
one must not forget that this effect only enters at first order in cosmological perturbation theory. Furthermore, the explicit example with the $D=2$ Polyakov action allows us to understand that $\d U$ and $\d W$ are driven by the metric perturbations $\Phi$ and $\Psi$. In practice, given that anisotropic stresses are negligible and $\Phi\simeq -\Psi$ all along the evolution, we can take just $\Phi$ as the typical scale for the  metric perturbations. From the definition $U=-\iBox R$, together with the fact that the perturbations of the Ricci scalar are, parametrically, of order of  two derivatives of the metric perturbation $\Psi$,  it is natural to 
assign  initial conditions on $\d U$ of the form
\be\label{iniconddU}
\d U(t_{\rm in},\vx)={\cal O}\[ \Phi(t_{\rm in},\vx)\]\, ,\qquad \d U'(t_{\rm in},\vx)=
{\cal O}\[\Phi'(t_{\rm in},\vx)\]\, .
\ee
From the definition $S=-\iBox U$ together with the fact that, in a cosmological setting, $\Box$ is parametrically of order $H^2$, it is also natural to take
$\d S={\cal O}(\Phi/H^2)$, i.e. 
\be\label{iniconddW}
\d W(t_{\rm in},\vx)={\cal O}\[ \Phi(t_{\rm in},\vx)\]\, ,\qquad \d W'(t_{\rm in},\vx)=
{\cal O}\[\Phi'(t_{\rm in},\vx)\]\, .
\ee
We next observe that, for all modes of cosmological relevance, the initial conditions are set at a time when they are well outside the horizon. Thus, the subsequent evolution is the same as that of the $\vk=0$ mode. In particular, $\d U'_{\vk}(t_{\rm in})$, $\d W_{\vk}(t_{\rm in})$ and $\d W'_{\vk}(t_{\rm in})$ parametrize irrelevant directions in parameter space. Any initial value ${\cal O}(1)$ assigned to them is immediately washed out, and the solution quickly reduces  to that obtained setting 
$\d U'_{\vk}(t_{\rm in})=\d W_{\vk}(t_{\rm in})=\d W'_{\vk}(t_{\rm in})=0$, as we have also checked explicitly by numerical integration.\footnote{Note that, furthermore, for modes well outside the horizon $\Phi$ is constant to great accuracy, so
$\Phi'(t_{\rm in},\vx)\simeq 0$, which further renders  irrelevant the initial conditions of 
$\d U'$ and $\d W'$.}
The only direction in parameter space that is marginal, rather than irrelevant, corresponds to $\d U$. We can then consider initial conditions of the form
\be\label{iniconddUk}
\d U_{\vk}(t_{\rm in})=c^U_{k}\Phi_{\vk}(t_{\rm in})\, .
\ee
Actually, the example of the $D=2$ Polyakov action even suggests to take the same value
of the  constant $c^U_k$ for all Fourier modes:  we see from \eq{deltaUdsigmainit}  that in this case $\d U_{\vk}=c^U\d \sigma_{\vk}$ with
$c^U=2$ for all $k$. In any case, we can also easily check what happens  varying independently the constants $c^U_k$, for different modes. For some selected modes of cosmological relevance, we will study below how our results depend on $c^U_k$, by comparing the case $c^U_k=0$ with, e.g., 
the cases $c^U_k=\pm 6$ or $c^U_k=\pm 10$. As we will see below, the predictions of the RR model turn out to be basically  insensitive to variations of $c^U_k$ in this range (and in fact, even in a much wider range). This is an important result, because it implies that our ignorance on the exact initial conditions for the perturbations of the auxiliary fields does not spoil the predictivity of the model.

\subsubsection{Perturbation equations}\label{sect:pertRR}

The perturbation equations for the RR model have been written down in \cite{Dirian:2014ara}, and we recall them here. At the level of perturbations we find convenient to write the equations in terms of 
\be\label{defVH0S}
V(t,\vx)\equiv H_0^2S(t,\vx)\, ,
\ee 
instead of 
$W(t,\vx)=H^2(t)S(t,\vx)$.
Recall also, from  Section~\ref{sect:backgRR}, that the prime denotes the derivative with respect to $x\equiv \log a$, while $\gamma= m^2/(9H_0^2)$. 
Perturbing \eq{BoxUS} one finds  
\bees
&& \hspace{-10mm}\delta U'' + (3 + \zeta)  \delta U' + \hat{k}^2 \delta U - 2 \Psi \bar{U}'' - \big[ 2 (3+\zeta) \Psi  +  \Psi' - 3  \Phi' \big]  \bar{U}'  \nn\\
&&  = 2 \hat{k}^2 (\Psi + 2 \Phi) + 6 \big[\Phi'' +(4+\zeta)  \Phi' \big] 
- 6 \big[ \Psi' + 2(2+ \zeta) \Psi \big], \label{eqlinU}\\
&& \hspace{-10mm}\delta V'' + (3 + \zeta) \delta V' + \hat{k}^2 \delta V - 2 \Psi  \bar{V}'' - 
\[ 2 (3+\zeta) \Psi  +  \Psi' - 3  \Phi' \]  \bar{V}' = h^{-2} \delta U\, .\label{eqlinV}
\ees
(For readability, we omit here and in the following equations the label $\vk$ from $U_{\vk}$, $\Psi_{\vk}$, etc.)
Perturbing \eqs{Gmn}{defKmn}, using radiation and non-relativistic matter in the energy-momentum tensor, and projecting onto the scalar sector gives
\bees
&&\hspace*{-1.5cm}\( 1 - 3 \gamma \bar{V} \) \( \hat{k}^2 \Phi + 3  \Phi' -3 \Psi  \) + \frac{3 \gamma}{2} \bigg[  - \frac{1}{2 h^2} \bar{U} \delta U + \big( 6 \Psi - 3 \Phi' - \Psi \bar{U}' \big)  \bar{V}' 
 \nn \label{ModPoisson}\\
&&\hspace*{-1.5cm} + \frac{1}{2} \big( \bar{U}'  \delta V' +  \bar{V}'  \delta U' \big) - 3 \delta V  - 3 \delta V' - \hat{k}^2 \delta V \bigg] =  \frac{3}{2 h^2} \big(  \Omega_{\textsc{R}} e^{-4x} \delta_{\textsc{R}} + \Omega_{\textsc{M}} e^{-3x} \delta_{\textsc{M}} \big)  , \label{MEE1x}\\
&&\hspace*{-1.5cm}\( 1 - 3 \gamma \bar{V} \)  \hat{k}^2 (  \Phi' - \Psi) - \frac{3 \gamma \hat{k}^2}{2}
\[ \delta V' -  \bar{V}' \Psi - \delta V + \frac{1}{2} \(  \bar{U}' \delta V +\bar{V}' \delta U \)  \] \nn\\
&&\hspace*{-1.5cm}=
 - \frac{3}{2 h^2} \bigg(  \frac{4}{3} \Omega_{\textsc{R}} e^{-4x} \hat{\theta}_{\textsc{R}} +  \Omega_{\textsc{M}} e^{-3x} \hat{\theta}_{\textsc{M}} \bigg), \label{MEE2x}\\
&&\hspace*{-1.5cm}( 1 - 3 \gamma \bar{V} )\[   \Phi'' + (3+\zeta) \Phi' -  \Psi' - (3+2 \zeta) \Psi + \frac{\hat{k}^2}{3}(\Phi + \Psi) \] 
 \nn \\ 
&&\hspace*{-1.5cm} - \frac{3 \gamma}{2} \bigg\{ \frac{1}{2 h^2} \bar{U} \delta U - 2 \Psi  \bar{V}'' + 
\[ 2 \Phi' - 2 (2 + \zeta) \Psi -  \Psi' - \Psi  \bar{U}' \]  \bar{V}' +  \delta V'' + (2+\zeta)  \delta V' \nn \\
&&  + \frac{2 \hat{k}^2}{3} \delta V + ( 3 + 2 \zeta ) \delta V + \frac{1}{2} \big(  \bar{U}'  \delta V'  +  \bar{V}'   \delta U' \big) \bigg\}
= - \frac{1}{2 h^2}  \Omega_{\textsc{R}} e^{-4x} \delta_{\textsc{R}}\, ,\label{MEE3x}\\
&&\hspace*{-1.5cm}
( 1 - 3 \gamma \bar{V})  (\Psi + \Phi)  - 3 \gamma \delta V =0\, .\label{MEE4x}
\ees
which corresponds, respectively, to the perturbations of the $(0,0)$ component of the modified Einstein equations, the divergence of the $(0,i)$ component, the trace of the $(i,j)$ component, and  the result of applying the operator $(\n^{-2}\pa_i\pa_j-\frac{1}{3}\d_{ij})$ to the $(ij)$ component.
Finally, perturbing  the energy-momentum conservation equation we get
\begin{align}
\delta_M' &= - (3  \Phi' + \hat{\theta}_M),\label{dM1} \\
\hat{\theta}'_M &= - ( 2 + \zeta) \hat{\theta}_M + \hat{k}^2 \Psi\, ,\label{dtheta1}\\
\delta_R' &= - \frac{4}{3} (3  \Phi' + \hat{\theta}_R), \\
 \hat{\theta}'_R &= - (1 + \zeta)\hat{\theta}_R + \hat{k}^2 \( \Psi + \frac{\delta_R}{4} \). \label{dtheta2}
\end{align}
Of course, because of diffeomorphism invariance, these equations are not all independent. A convenient choice for the numerical integration consists in eliminating $\Psi$ from
\eq{MEE4x} and then 
using \eqref{eqlinU}, \eqref{eqlinV}, \eqref{MEE3x} and \eqref{dM1}--\eqref{dtheta2} as 7 independent equations for the 7 variables $\Psi,U,V,\d_M,\theta_M,\d_M,\theta_M$. We then use the modified Poisson equation (\ref{ModPoisson}) as a test of the numerical integration, verifying that it is satisfied to high accuracy (one part in $10^6$ to one part in $10^8$, depending on the value of $k$) all along the integration.
As in \cite{Dirian:2014ara},
we introduce 
\be
\kappa \equiv k/k_{\rm eq}\, , 
\ee
where
$k_{\rm eq}=a_{\rm eq} H_{\rm eq}$ is the wavenumber of the mode that enters the horizon at matter-radiation equilibrium, and we illustrate our numerical results displaying the results 
for  $\kappa = 0.1$, $\kappa=1$ and $\kappa =5$. The mode with 
 $\kappa =5$  re-entered  the horizon  during RD,
the mode with $\kappa=1$ re-entered at matter-radiation equality,  and  the mode with $\kappa=0.1$  was  outside the horizon during RD and most of MD, and re-entered at $z\simeq 1.5$. Overall, these three values of $k$ illustrate well  the $k$ dependence of the results, for a large range of scales relevant for cosmology.

\begin{figure}[t]
\centering
\includegraphics[width=0.42\columnwidth]{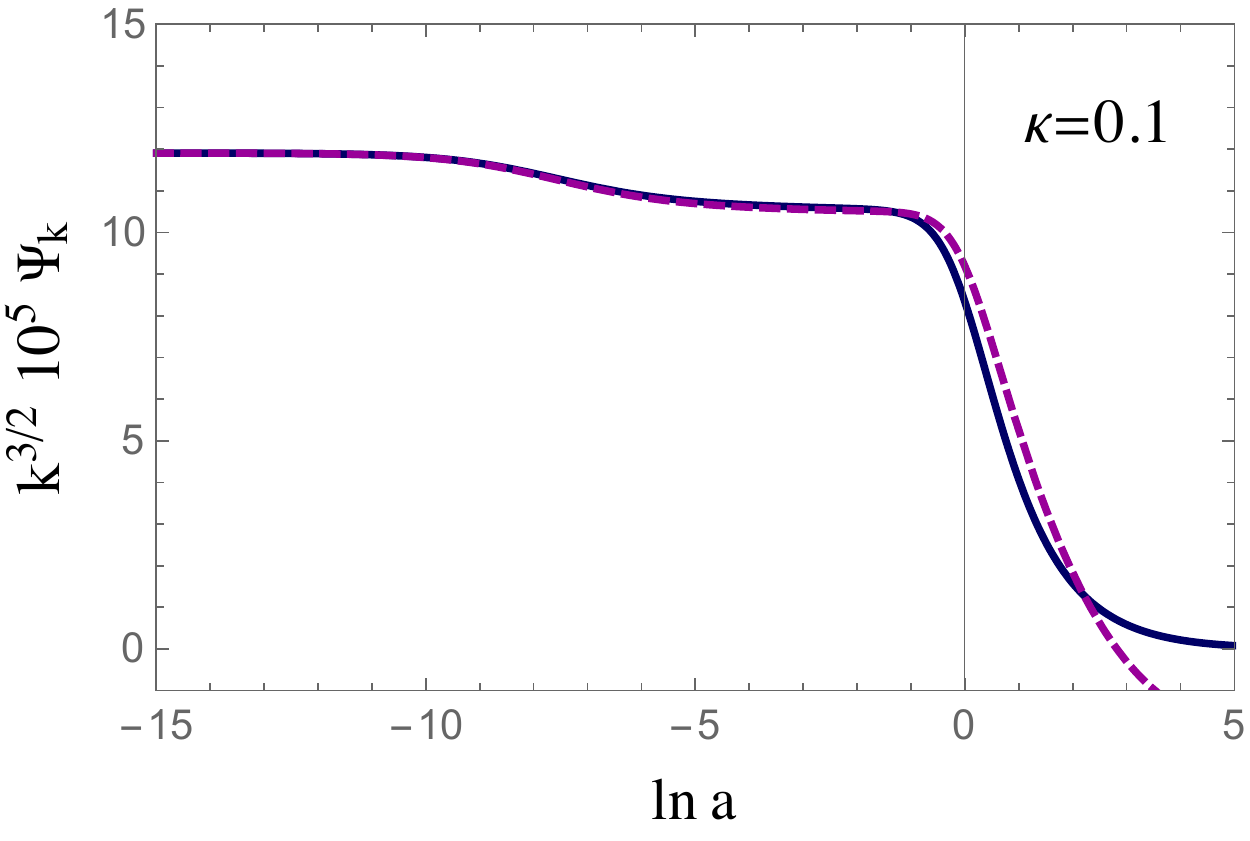}
\includegraphics[width=0.42\columnwidth]{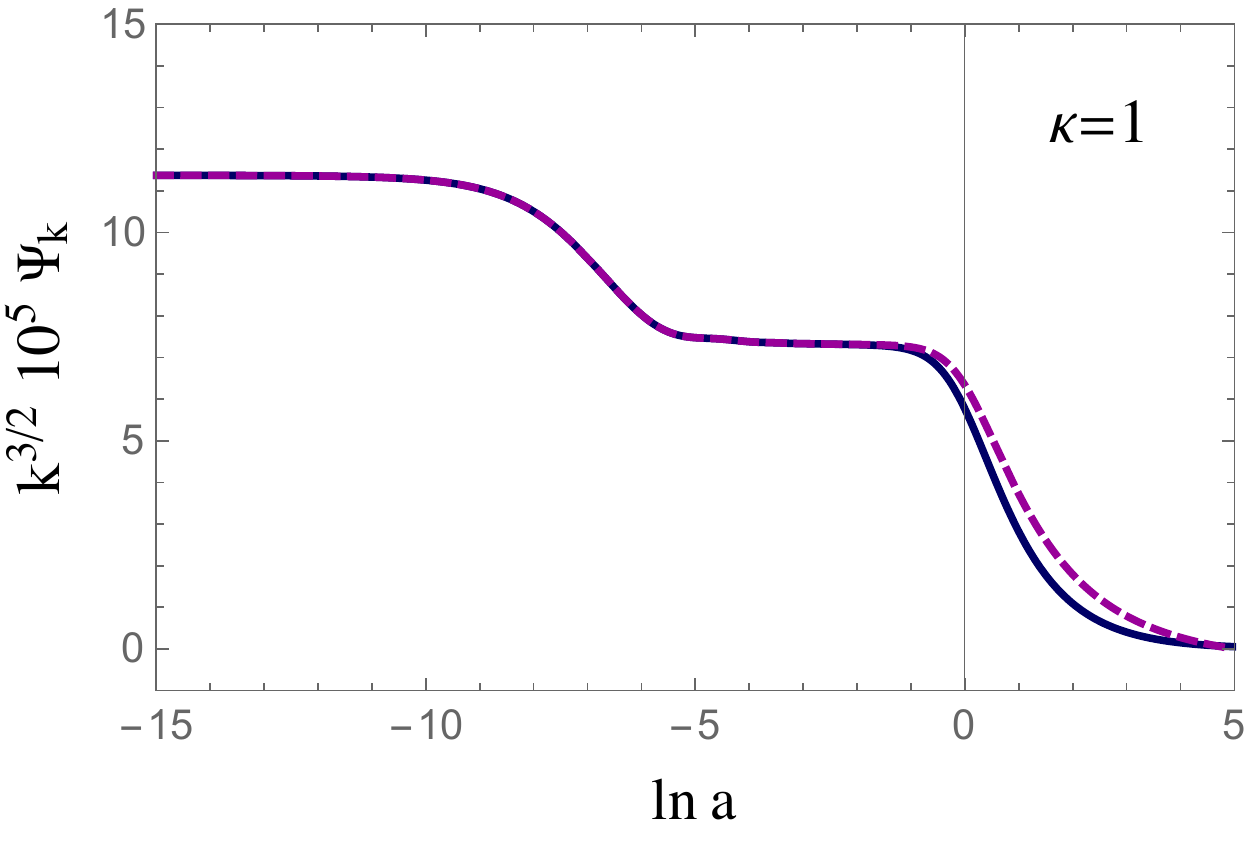}
\includegraphics[width=0.42\columnwidth]{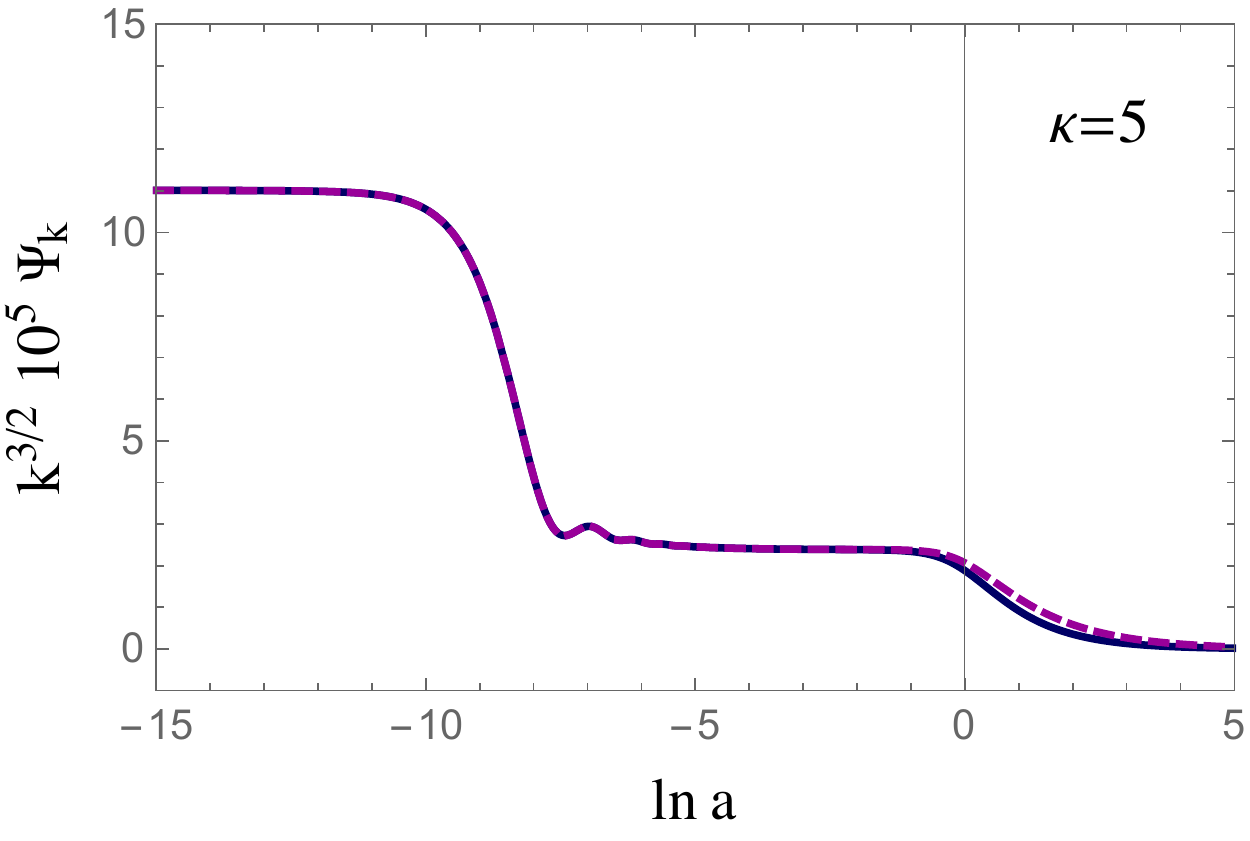}
\caption{The dimensionless quantity $k^{3/2}10^5\Psi_k$ against $\ln a$, for $\kappa=0.1$ (upper left panel), $\kappa=1$ (upper right panel) and $\kappa =5$ (lower panel). In each figure the blue solid line is the result in $\Lambda$CDM and the red dashed line is the result in the minimal RR model ($u_0=0$) with $c^U_k=0$. 
For this comparison, we have used the same fiducial values of the cosmological parameters in $\Lambda$CDM and in the RR model.
On the scale of these plots, the results obtained with  $c^U_k=\pm 10$ would be indistinguishable from the line with $c^U_k=0$.
\label{fig:pertRRPsik}
}
\end{figure}

\begin{figure}[th]
\centering
\includegraphics[width=0.42\columnwidth]{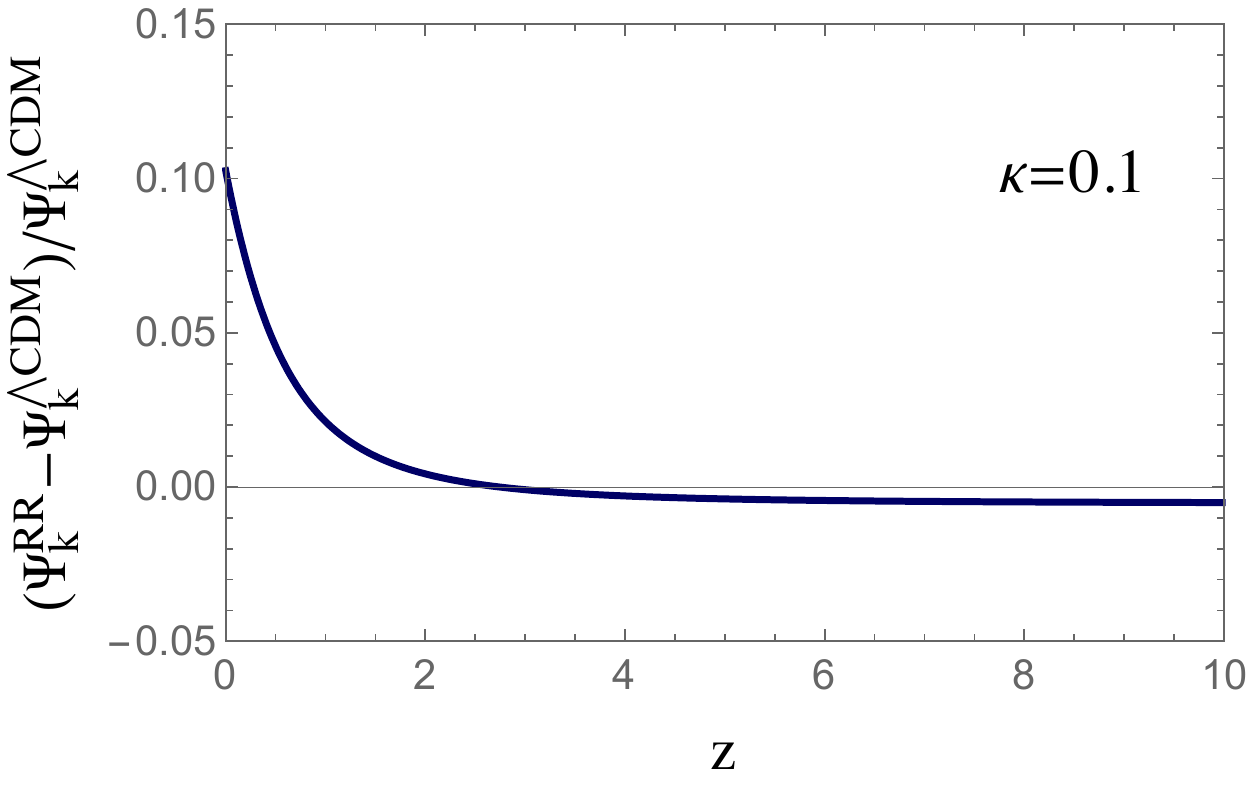}
\includegraphics[width=0.42\columnwidth]{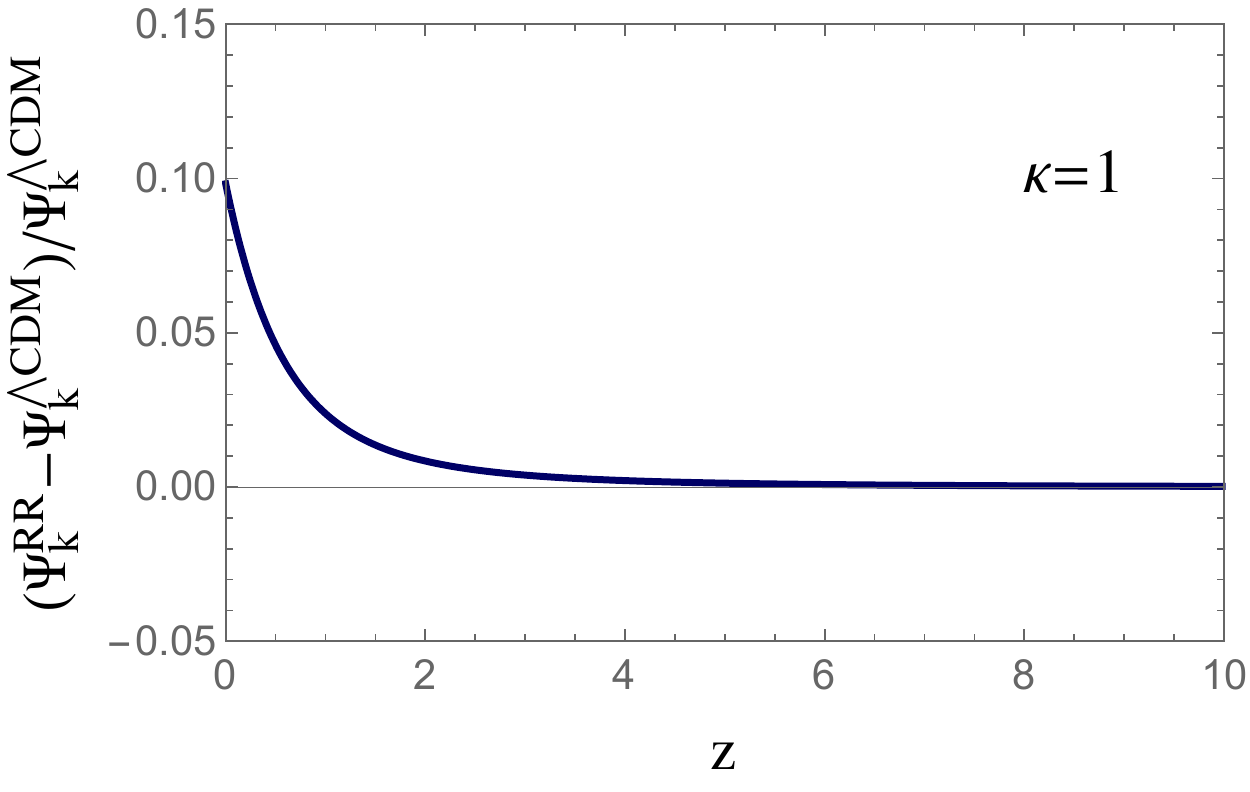}
\includegraphics[width=0.42\columnwidth]{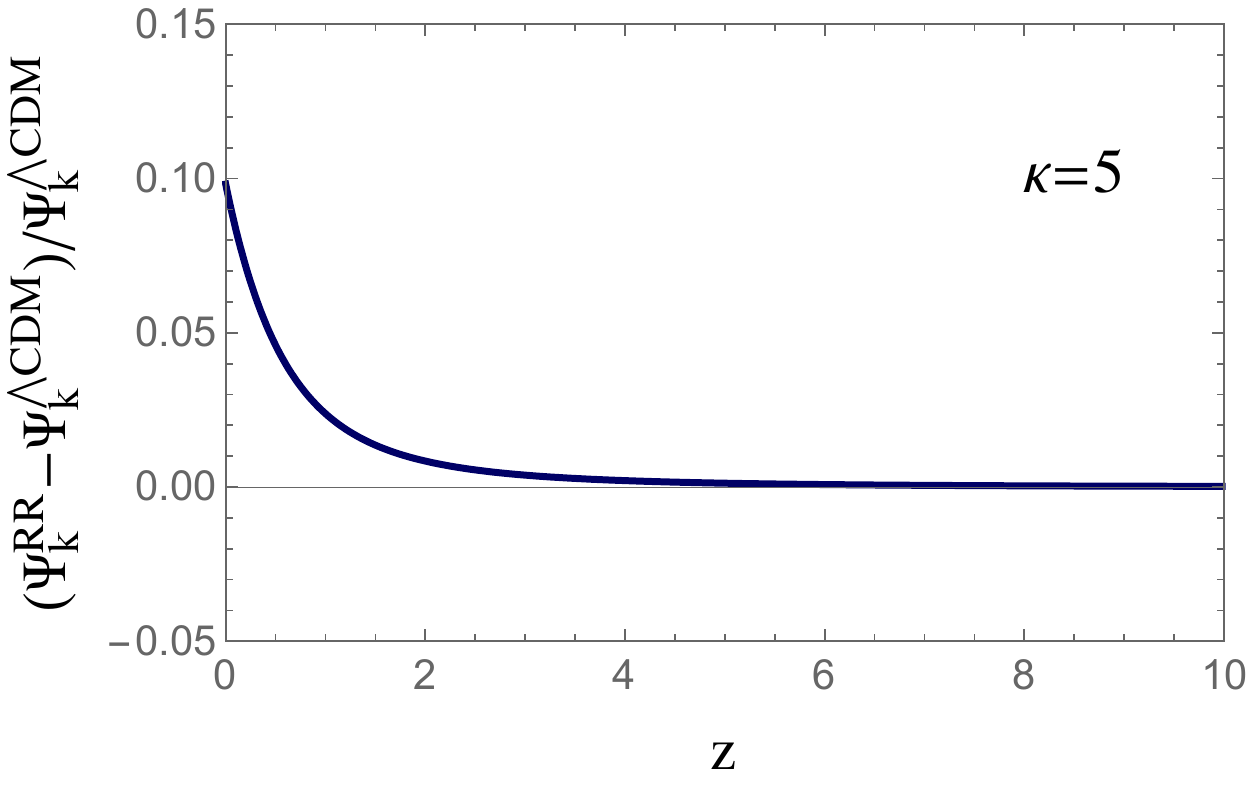}
\caption{The relative differences 
$[\Psi_k^{\rm RR}-\Psi_k^{\Lambda{\rm CDM}}]/\Psi_k^{\Lambda{\rm CDM}}$ for $\kappa=0.1,1,5$, again 
setting $c^U_k=0$, against redshift, for the RR model. 
\label{fig:pertDeltaRRPsik}
}
\end{figure}

\begin{figure}[ht]
\centering
\includegraphics[width=0.42\columnwidth]{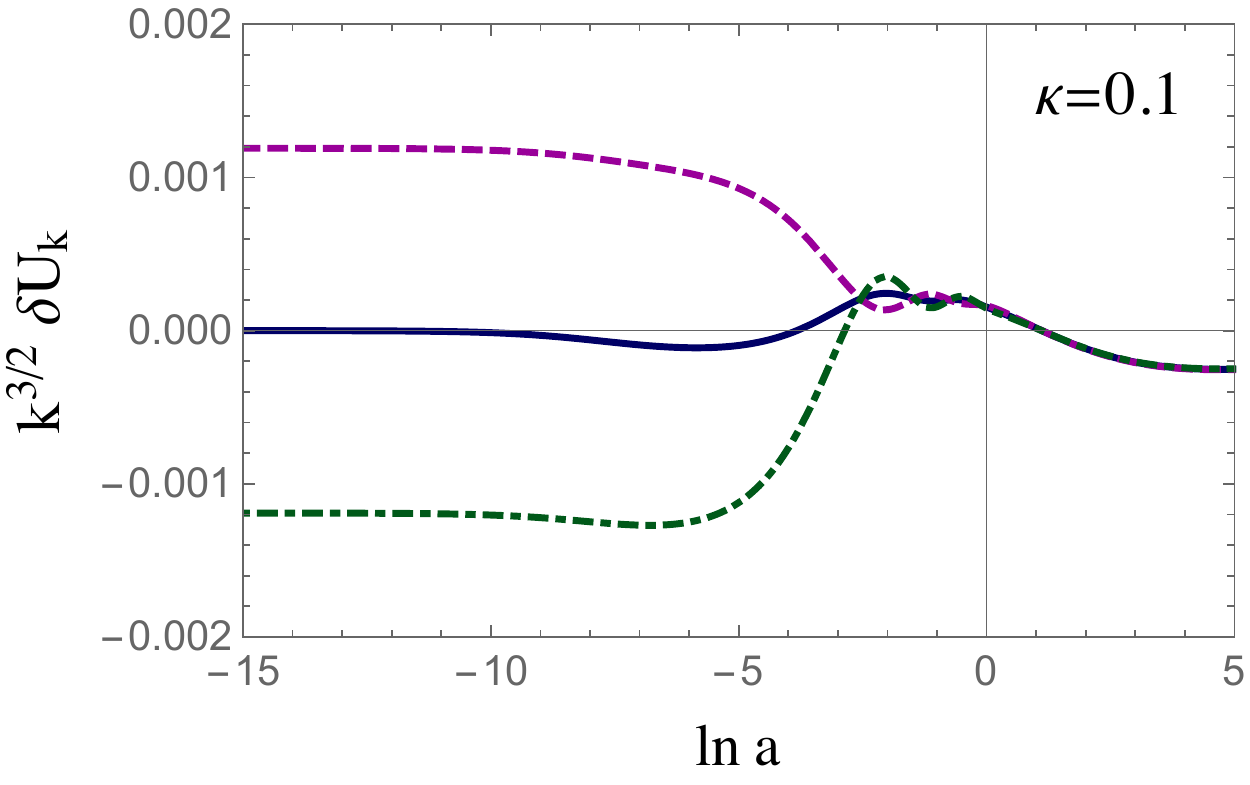}
\includegraphics[width=0.42\columnwidth]{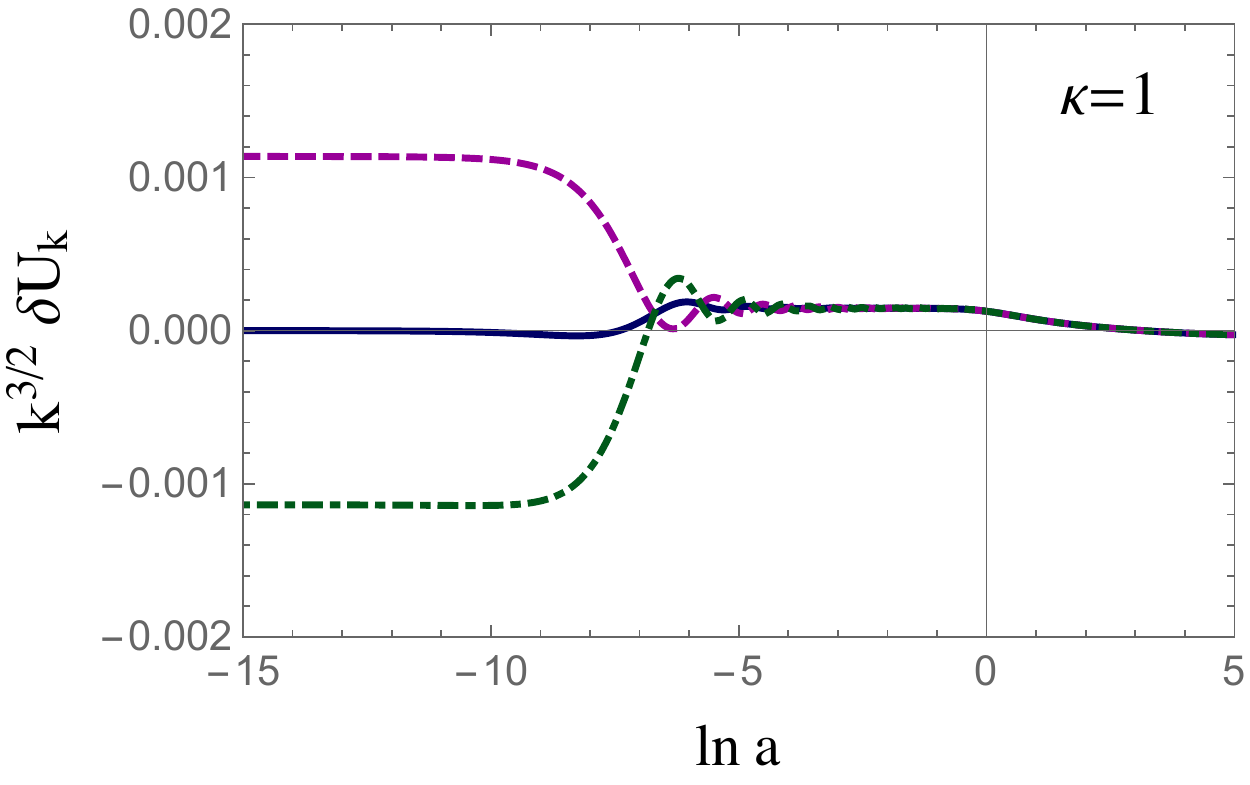}
\includegraphics[width=0.42\columnwidth]{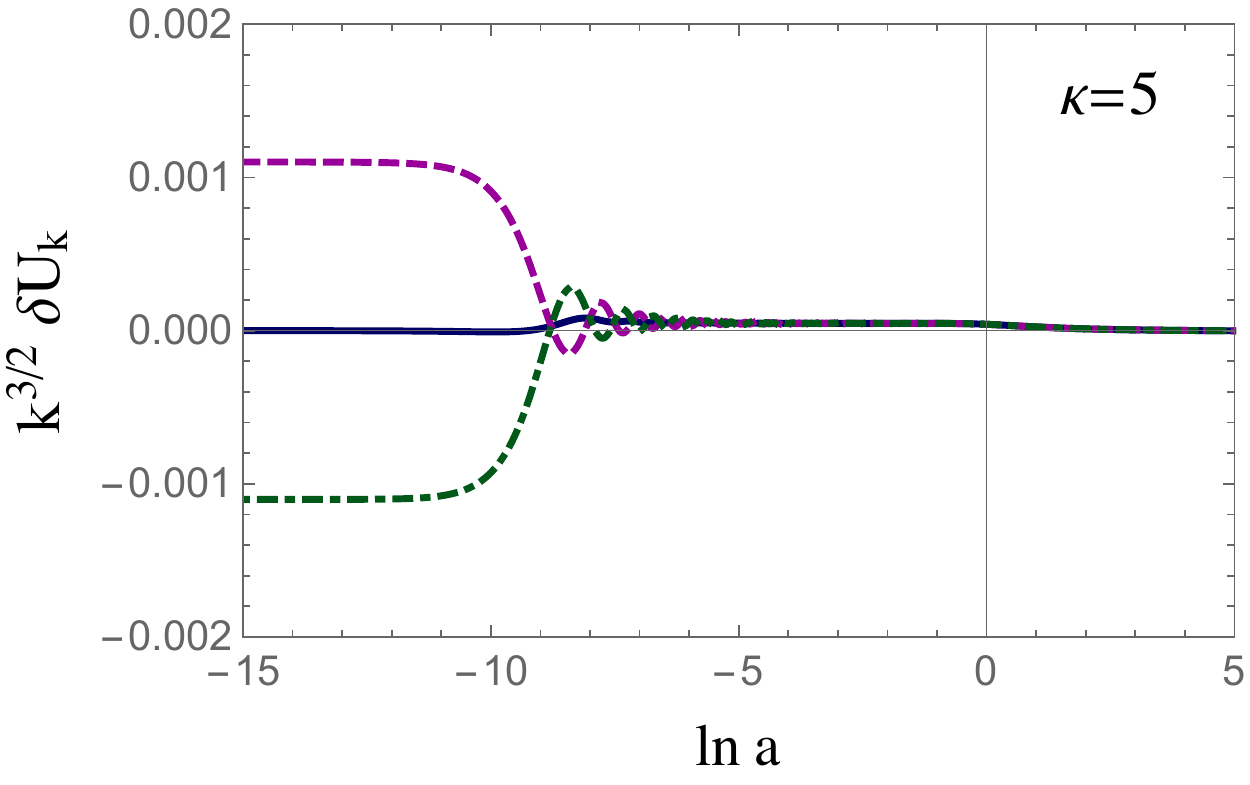}
\caption{The evolution of the perturbations $k^{3/2}\d U_k$ for different initial conditions $c^U_k$ and
$\kappa=0.1,1,5$. In each figure we show the evolution obtained with  $c^U_k=0$ (blue solid line), $c^U_k=+10$ (red dashed line) and  $c^U_k=-10$ (green dot-dashed line).
\label{fig:pertRRdUforThreecUk01}
}
\end{figure}

\begin{figure}[ht]
\centering
\includegraphics[width=0.42\columnwidth]{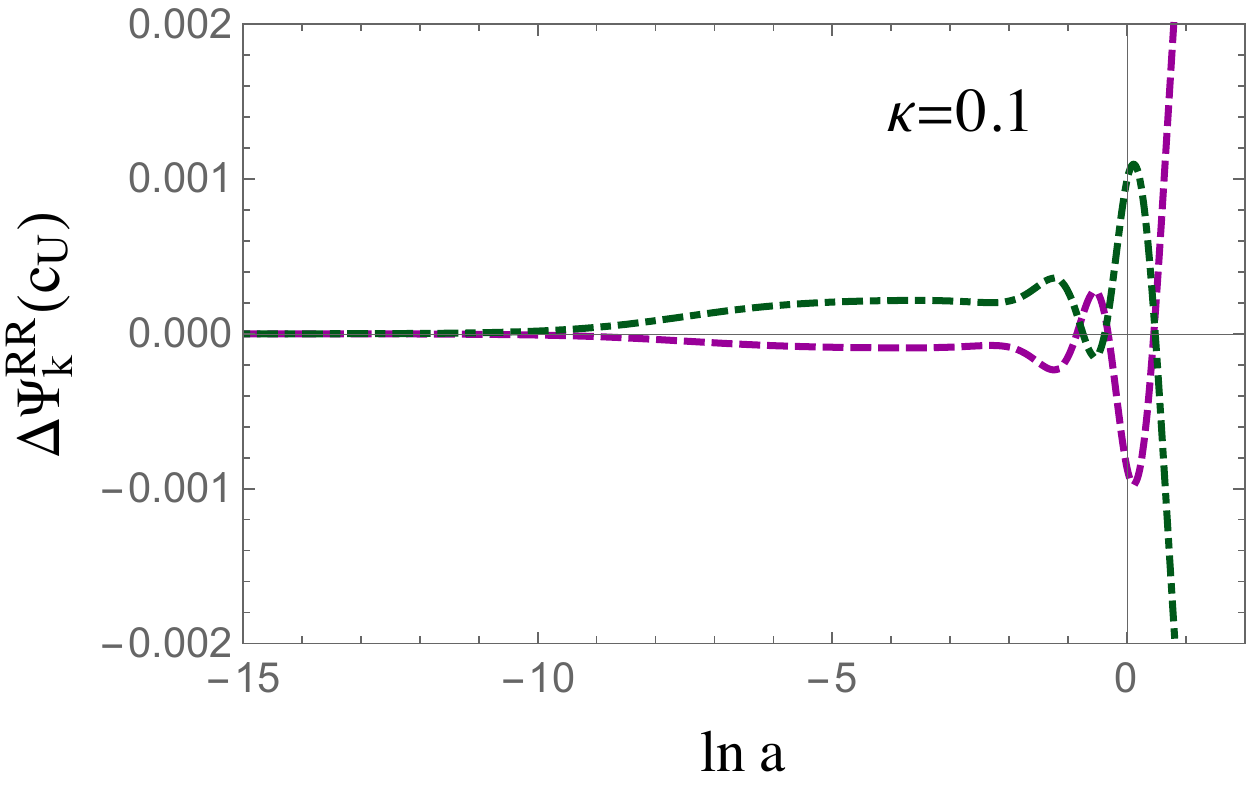}
\includegraphics[width=0.42\columnwidth]{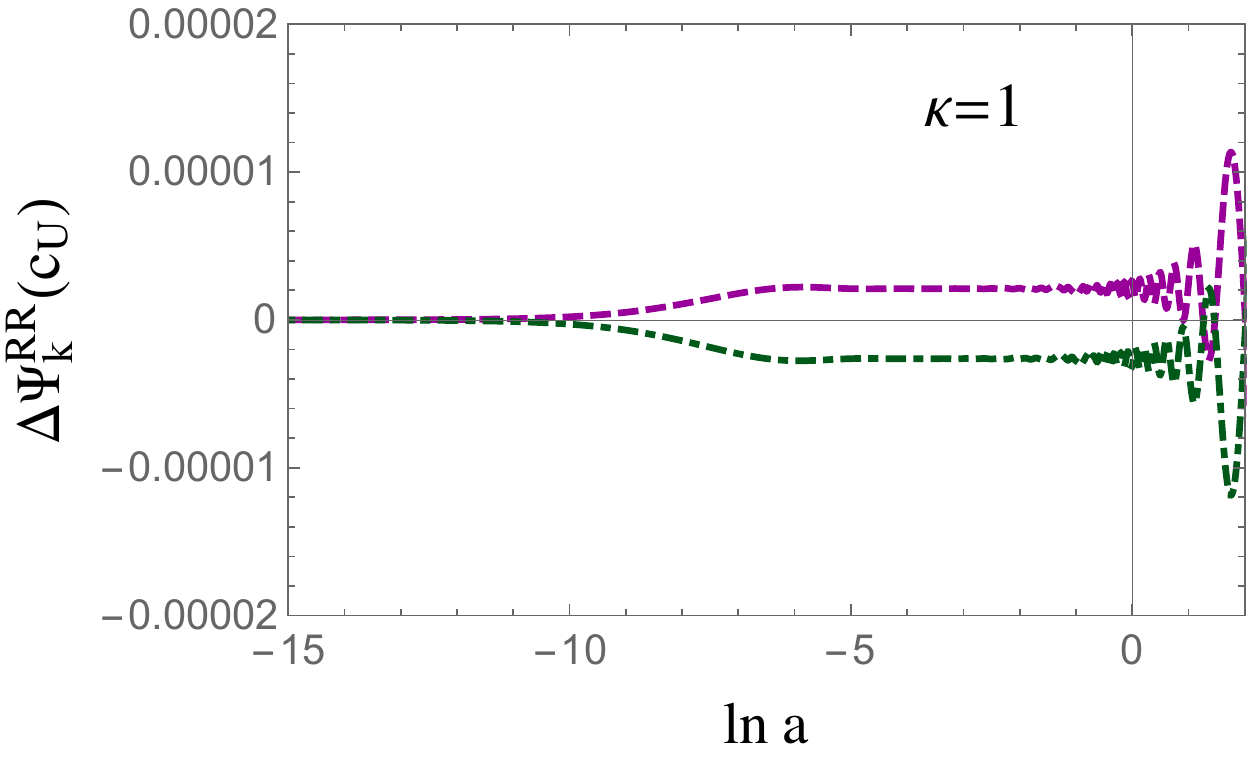}
\includegraphics[width=0.42\columnwidth]{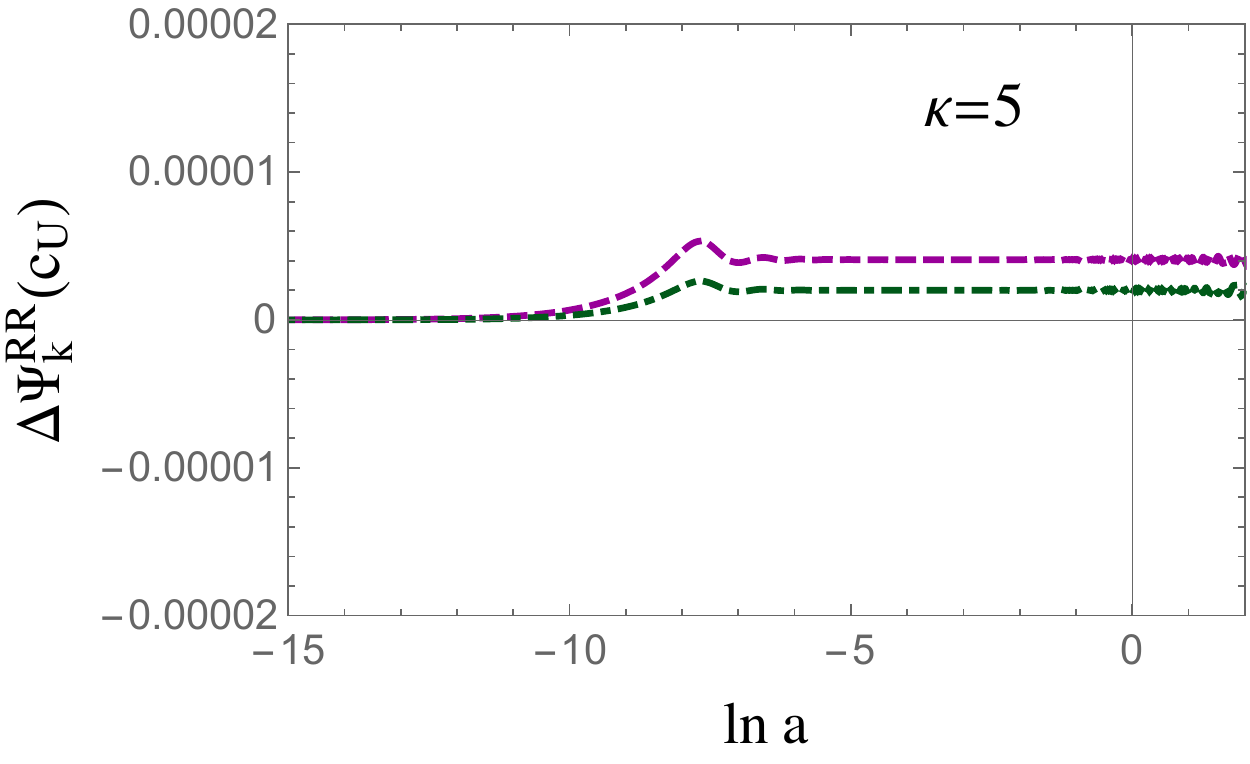}
\caption{The relative differences 
$[\Psi^{\rm RR}_k(c^U_k)-\Psi^{\rm RR}_k(c^U_k=0)]/\Psi^{\rm RR}_k(c^U_k=0)$ for $\kappa=0.1,1,5$. In each figure we show the cases   $c^U_k=+10$ (red dashed line) and  $c^U_k=-10$ (green dot-dashed line).
\label{fig:pertRRPsi_cUmenoc0}
}
\end{figure}

The result of the integration of the perturbation equations  is shown in 
Figs.~\ref{fig:pertRRPsik}--\ref{fig:pertRRPsi_cUmenoc0}.
In particular, in Fig.~\ref{fig:pertRRPsik} we show the dimensionless quantity $k^{3/2}10^5\Psi_k$ against $x=\ln a$, for our reference values of $\kappa$, in $\Lambda$CDM (blue solid line) and in the minimal RR model (i.e. setting $u_0=0$), using $c^U_k=0$ for the initial conditions on the perturbations (red dashed line). We see that, up to the present time $x=0$, the cosmological perturbations of the RR model are very close  to that of $\Lambda$CDM, and will only start to be significantly different in the cosmological future. Fig.~\ref{fig:pertDeltaRRPsik} shows the corresponding relative differences
$[\Psi_k^{\rm RR}-\Psi_k^{\Lambda{\rm CDM}}]/\Psi_k^{\Lambda{\rm CDM}}$ as a function of redshift, on a scale that emphasizes the recent cosmological past. We see  that the deviations from $\Lambda$CDM raise up to about $10\%$ at $z=0$, with very little dependence on momentum (at least in the cosmological past; from Fig.~\ref{fig:pertRRPsik} we see that in the cosmological future the differences, as well as the momentum dependence,  can be more significant). This result, already discussed in \cite{Dirian:2014ara}, indicates that the perturbations in the RR model are sufficiently close to that in $\Lambda$CDM to be in the right ballpark for fitting the data, yet sufficiently different to be potentially detectable.

We next explore how the results depend on the initial conditions on the cosmological perturbations, i.e. on $c^U_k$.
Fig.~\ref{fig:pertRRdUforThreecUk01} shows, for each of our three reference values of $k=\kappa k_{\rm eq}$, the evolution of the perturbation $\d U_{\vk}$ with initial conditions $c^U_k=0$ (blue solid line), $c^U_k=+10$ (red dashed line) and  $c^U_k=-10$, set deep in RD. We see that, by the time that the dark-energy density becomes important, the differences between these solutions have become very small.
Fig.~\ref{fig:pertRRPsi_cUmenoc0} shows how this reflects on the evolution of the metric perturbations. Here we show
the relative difference between the evolution of $\Psi_k$ in the RR model, comparing the result obtained with a non-vanishing value of $c^U_k$ and the result for $c^U_k=0$, i.e. we plot
\be
\Delta\Psi_k^{\rm RR}(c^U_k)\equiv \frac{\Psi_k^{\rm RR}(c^U_k)-\Psi_k^{\rm RR}(c^U_k=0)}{\Psi_k^{\rm RR}(c^U_k=0)}\, .
\ee 
For each of our three reference values of $\kappa$, we show the results for $c^U_k=+10$ (red dashed line) and for $c^U_k=-10$ (green dot-dashed line). We see that, depending on $\kappa$, the {\em relative} differences are of order $10^{-5}$ to $10^{-3}$. Such relative differences on a quantity, such as $\Psi_{\vk}$, which is already of first order in cosmological perturbation theory, are totally negligible at the present level of accuracy of the analysis (they are rather of the order of terms of second order in perturbation theory). This is a very encouraging result, because it means that our current lack of derivation of the nonlocal model from a fundamental theory, which implies a lack of knowledge of these initial conditions, in practice does not affect the comparison with the data and the predictivity of the model. We will confirm these result in the next section,  performing a full Bayesian parameter estimation for the RR model with different initial conditions on the perturbations.

\clearpage

\subsection{Bayesian parameter estimation and model comparison}\label{sect:Bay}

The results of the previous sections show, first of all,  that the cosmological perturbations of the RR model  are  stable~\cite{Dirian:2014ara}.
This is already a non-trivial result. For instance   the DPG model~\cite{Dvali:2000hr}, which opened the way to the study of IR modifications of GR, has a  self-accelerated solution \cite{Deffayet:2000uy,Deffayet:2001pu} but  succumbed to fatal instabilities at the level of perturbations~\cite{Luty:2003vm,Nicolis:2004qq,Gorbunov:2005zk,Charmousis:2006pn,Izumi:2006ca}. 
The construction of a consistent theory of  massive gravity~\cite{deRham:2010ik,deRham:2010kj,Hassan:2011hr} and of bigravity~\cite{Hassan:2011zd}  has provided lasting field-theoretical understanding of the dynamics of massive spin-2 fields. However, once again, their application to cosmology has shown how non-trivial  is to build an IR modification of GR  with a stable evolution at the background level, as well as at the level of cosmological perturbations; indeed, massive gravity
has difficulties already in obtaining a viable background FRW evolution~\cite{DAmico:2011eto}. In bigravity  background FRW solutions exist, but, in a  branch of solutions that has a dynamical dark energy, the cosmological perturbations are  plagued by instabilities in both the scalar and  tensor sectors;   in a second branch, taking the limit in which the Planck mass associated to the second metric is small, the scalar instabilities can be pushed to unobservably early times, but in this limit  the background evolution becomes indistinguishable from that of $\Lambda$CDM   \cite{Konnig:2014xva,Lagos:2014lca,Cusin:2014psa,Akrami:2015qga,Cusin:2015pya,Schmidt-May:2015vnx,Cusin:2015tmf}.

Beside being stable,  we have seen that, up to the present cosmological epoch, the perturbations in the minimal RR  model turn out to be quite close to that of $\Lambda$CDM, with differences that (for the minimal model with $u_0=0$) are at most of order 10\%; see \cite{Dirian:2014ara} for several further plots of metric and matter perturbations.
This already tells us that this model is in the right ballpark for fitting the cosmological data, yet sufficiently different to be potentially distinguishable from $\Lambda$CDM with present and near-future observations. 
To make more quantitative statements it is necessary to implement these perturbations in a Boltzmann code, perform Bayesian parameter estimation, and see quantitatively how the model fits the data in comparison with $\Lambda$CDM. For the RR model this has been done in
\cite{Dirian:2014ara,Dirian:2014bma,Dirian:2016puz,Dirian:2017pwp}, and here we will expand on these results. 
We have implemented the cosmological equations in CLASS \cite{Class}, and
constrained the nonlocal model with observations using the Markov Chain Monte Carlo (MCMC) code Montepython \cite{MP}. The details on the implementation of the RR model in CLASS have been discussed in appendix~A of \cite{Dirian:2016puz}.

It should also be appreciated that we are considering a model in which  we simply have a new parameter, $\Lrr$, that replaces the cosmological constant in $\Lambda$CDM, and which has a well-defined physical meaning, as a dynamically-generated mass term for the conformal mode; see \eq{massconfmode}  [plus a hidden  parameter, $u_0$ that, as we have seen, enters in the initial conditions of the auxiliary field $U$, and whose value could be  related to the duration of a primordial inflationary phase; see the discussion below \eq{pertUdS}]. In contrast, modified gravity models that have been intensely studied in the literature in recent years involve either a free function of $R$, as in $f(R)$ theories \cite{DeFelice:2010aj}, or a free function of $\iBox R$, as 
in the Deser-Woodard model~\cite{Deser:2007jk}, or even a set of  arbitrary functions of a (hypothetical) scalar or  massive vector field, as for instance in Horndesky  \cite{Horndeski:1974wa,Deffayet:2009mn} and beyond-Horndeski  theories~\cite{Gleyzes:2014dya},  or  in
generalized Proca theories \cite{Heisenberg:2014rta}. There is no theoretical clue on the form of these functions,  
leading to highly redundant (and, sometimes, quite baroque) alternatives to $\Lambda$CDM.
Furthermore, it should be observed that these modified gravity models have  usually been compared to $\Lambda$CDM using, rather than the full available CMB information, only the ``shift parameter" (see e.g. \cite{deFelice:2017paw}), which is just an approximate indicator of the position of the first peak. However, it is not difficult to adjust a model to this single data point, especially if one can play with extra free parameters (and even  free functions). The test to which we are submitting our nonlocal model, involving the full information on the CMB multipoles of temperature (and also of polarization) is much more stringent. Furthermore, the minimal RR model with $u_0=0$ has the same number of parameter as $\Lambda$CDM. Here we will just consider two cases, $u_0=0$ and $u_0$ equal to a large value, that we will choose as $u_0=250$, and we will not vary $u_0$ as a free parameter. Even if one would optimize the results with respect to $u_0$, this would just give a model with one parameter more than $\Lambda$CDM, which is still much more economical than models in which one can play with free functions.

\subsubsection{Datasets}\label{sect:datasets}
For CMB, SNe and BAO we  use the same cosmological datasets that were used in \cite{Dirian:2016puz,Dirian:2017pwp}, namely:

\begin{itemize}

\item{\em CMB.} We use the 2015 \textit{Planck}  \cite{Planck_2015_1} measurements of the angular (cross-)power spectra of the CMB. In particular, we take the full-mission lowTEB data for low multipoles ($\ell \leq 29$) and the high-$\ell$ Plik  TT,TE,EE (cross-half-mission) ones for the high multipoles ($\ell > 29$) of the temperature and polarization auto- and cross- power spectra \cite{Ade:2015rim}. 
We also include the temperature$+$polarization (T$+$P) lensing data, using  only the conservative multipole range $\ell =40-400$  \cite{Planck_2015_Lkl,Planck_2015_lens}.

\item {\em Type Ia supernovae.} We use the JLA data for  SN~Ia provided by the SDSS-II/SNLS3 Joint Light-curve Analysis~\cite{Betoule:2014frx}.

\item {\em Baryon Acoustic Oscillations} (BAO). We use the isotropic constraints provided by 6dFGS at $z_{\rm eff}=0.106$ \cite{Beutler:2011hx}, SDSS-MGS DR7 at $z_{\rm eff}=0.15$ \cite{Ross_SDSS_2014} and BOSS LOWZ at $z_{\rm eff}=0.32$  \cite{Anderson:2013zyy}, as well as  the anisotropic constraints from CMASS at $z_{\rm eff}=0.57$ \cite{Anderson:2013zyy}.

\end{itemize}

Initially we will compare the models  to  CMB+BAO+SNa data, without including 
a prior on $H_0$, since it is interesting to see how the prediction of the RR model obtained just from  CMB+BAO+SNa compares with the local $H_0$ measurement. We will see that the tension that exists in $\Lambda$CDM is significantly reduced in the RR model. We will then add to our datasets also  the value of  $H_0$ obtained from local measurements, using \cite{Riess:2016jrr}
\be\label{H0Riess}
H_0 = 73.24 \pm 1.74\, .
\ee
in the usual units of 
${\rm km}\, {\rm s}^{-1}\, {\rm Mpc}^{-1}$.

\subsubsection{Free parameters}\label{sect:Free}
The {\em Planck} baseline analysis for $\Lambda$CDM uses six independent
cosmological parameters:  the Hubble parameter today
$H_0 = 100 h \, \rm{km} \, \rm{s}^{-1} \rm{Mpc}^{-1}$, the physical baryon and cold dark matter density fractions today $\omega_b = \Omega_b h^2$ and $\omega_c = \Omega_c h^2$, respectively, the amplitude  $A_s$ and tilt $n_s$ of the primordial scalar perturbations [defined in \eq{Ak2}],    and  the reionization optical depth  $\tau_{\rm re}$. Note that, assuming flatness, $\ola$ is a derived parameter, fixed by the flatness condition. In the  nonlocal models we have a  mass scale $m$ which replaces the cosmological constant, and again can be taken as  
a derived parameter, fixed by the flatness condition. Thus, for the  RR model, we can take the same six independent cosmological parameters, as in $\Lambda$CDM.

An important point is, however, the treatment of the sum of neutrino masses, $\sum_{\nu}m_{\nu}$. Oscillation experiments provide a lower bound $\sum_{\nu}m_{\nu}\, \gsim \, 0.06$~eV~\cite{GonzalezGarcia:2012sz},
assuming a normal mass hierarchy dominated by the heaviest neutrino mass eigenstate. The {\em Planck} baseline analysis sets  $\sum_{\nu}m_{\nu}$ to a fixed value corresponding to the lower limit,  $\sum_{\nu}m_{\nu}=0.06$~eV. 
As discussed in the {\em Planck} paper~\cite{Planck_2015_CP}, there is actually no compelling theoretical reason for this choice, and there are other possibilities, including a degenerate hierarchy with 
$\sum_{\nu}m_{\nu}\,\gsim\, 0.1$~eV.

If one  rather leaves $\sum_{\nu}m_{\nu}$ as a free parameter when analyzing the cosmological data, in $\Lambda$CDM one finds that the one-dimensional marginalized likelihood for $\sum_{\nu}m_{\nu}$ has its  maximum at $\sum_{\nu}m_{\nu}=0$ (see Fig.~30 of \cite{Planck_2015_CP}), and this marginalized likelihood  implies an upper bound $\sum_{\nu}m_{\nu}\,\lsim\, 0.23$~eV (at 95\% c.l.)  (combining {\em Planck} TT+lowP+lensing+$H_0$) \cite{Planck_2015_CP}.
In other words, if one performs the analysis in $\Lambda$CDM letting free the neutrino masses with  a prior $\sum_{\nu}m_{\nu}\geq 0.06$~eV, as required by oscillation experiments, the {\em Planck} data drive the value of $\sum_{\nu}m_{\nu}$ back to the lower limit $0.06$~eV.

As observed in \cite{Dirian:2017pwp}, the situation is different in the RR nonlocal model. In that case, letting the neutrino masses as free parameters, one finds a one-dimensional marginalized likelihood for 
$\sum_{\nu}m_{\nu}$  that is peaked at a nonzero value, which is between the lower limit set by oscillation experiments and the upper limit from $\beta$-decay experiments.\footnote{Tritium  $\beta$-decay experiments give an upper limit $m_{\nu_e}<2.2$~eV, which can be translated into
 $\sum_{\nu}m_{\nu}<6.6$~eV assuming three species of degenerate neutrinos (in any case, the precise value of the upper limit will be of no relevance in the following, since  the value of $\sum_{\nu}m_{\nu}$ predicted by the nonlocal models is anyhow well below these upper limits).}
This fact has two important implications: 

\begin{enumerate}

\item The (minimal) RR  model provides a {\em prediction} for the value of the sum of the  neutrino masses. In contrast, in $\Lambda$CDM for the best-fit value one simply gets back the value of the prior that one has put in, and one can only obtain an upper limit on $\sum_{\nu}m_{\nu}$. 

\item When comparing the performances of the RR model to that of $\Lambda$CDM, it is essential to let the neutrino masses as free parameters. Indeed, in \cite{Dirian:2016puz}, performing Bayesian parameter estimation and fitting   to CMB+BAO+SNa data, it was found that 
the RR model appeared to be  disfavored with respect to $\Lambda$CDM, with moderate-to-strong evidence. We now understand this as an artifact due to the fact that, in \cite{Dirian:2016puz}, the {\em Planck} baseline analysis was applied also to the nonlocal models, 
fixing $\sum_{\nu}m_{\nu}=0.06$~eV.  However, this amounts to arbitrarily fixing a parameter, which a priori is free, or at least constrained by oscillation and $\beta$-decay experiments within some range, to the value preferred by  $\Lambda$CDM. Obviously, this has the effect of  favoring 
$\Lambda$CDM over RR. Once the sum of neutrino masses is included among the free parameters, both in the RR  model and in $\Lambda$CDM, the situation changes. In $\Lambda$CDM it will go toward zero or, if we impose a prior   $\sum_{\nu}m_{\nu}\geq 0.06$~eV, it will hit the prior. In contrast,  in the RR model it goes to its own best-fit value,  which is different. Then, from the corresponding chi-squares or Bayes' factors, one finds that   the RR and $\Lambda$CDM model are statistically equivalent from the point of view of fitting the data~\cite{Dirian:2017pwp}. 

\end{enumerate}

Thus, in the following, we will perform Bayesian parameter estimation for $\Lambda$CDM and the RR model, and we will compare their performances using, as free parameters, the set
\be \label{base_param}
\theta = \left\{H_0, \omega_b, \omega_c, A_s, n_s, \tau_{\rm re}, \mbox{$\sum_{\nu}$}m_{\nu}\right\} \, .
\ee
For $\Lambda$CDM we will actually examine two cases, corresponding to fixing $\sum_{\nu}m_{\nu}=0.06$~eV, which allows us to make contact with the {\em Planck} baseline analysis 
(column labeled $\Lambda$CDM in the following tables) 
and letting it as a free parameter 
(column labeled $\nu\Lambda$CDM), which will allow us to make  a more homogeneous comparison with the RR model, in which neutrino masses are always taken as free parameters. For the RR  model, in contrast, we only give the results for the case  in which the sum of neutrino masses is allowed to vary freely, since fixing it to $\sum_{\nu}m_{\nu}=0.06$~eV for this model as little  meaning. For the RR model, we show the results for the ``minimal model"  with $u_0=0$, as well as for a model with a large value, $u_0=250$, as suggested by the evolution during a previous inflationary phase.

\subsubsection{Results using  {\em Planck}+BAO+JLA}

In Table~\ref{tab:res1} we show the Bayesian parameter estimation and the resulting $\chi^2$ for  
$\Lambda$CDM, $\nu\Lambda$CDM and the RR  model, using the dataset combination 
{\em Planck}+BAO+JLA.  In this table, for the RR model the initial condition of the perturbation $\d U$ is set to zero. We will show later that varying the parameter $c_U$ in \eq{iniconddUk} has basically no effect. Beside the values of the seven fundamental independent parameters (\ref{base_param}), we also give some useful derived parameters, namely  $\oma, \sigma_8$ and the reionization redshift $z_{\rm re}$, and we show the corresponding $\chi^2$ as well as the differences in $\chi^2$, taken here with respect to the value for $\nu\Lambda$CDM. 
Two main conclusions  can be drawn from these results.

\begin{table}[t]
\centering
\begin{tabular}{|l||c|c|c|c|} 
 \hline 
\multicolumn{1}{|l||}{ } & \multicolumn{4}{|c|}{CMB+BAO+SNe}  \\ \hline
Parameter & $\Lambda$CDM  & $\nu\Lambda$CDM & RR $(u_0=0)$ &  RR $(u_0=250)$    \\ \hline 
$H_0$ \phantom{\Big|}& $67.67^{+0.47}_{-0.50}$ & $67.60^{+0.66}_{-0.55}$ & $69.49^{+0.79}_{-0.80}$ &  $67.74_{-0.58}^{+0.66}$ \\  
$\sum_{\nu}m_{\nu}\ [{\rm eV}]$ & $0.06\ {\rm (fixed)}$ & $< 0.10$ (at $1\sigma$) & $0.219^{+0.083}_{-0.084}$ & $< 0.09$  (at $1\sigma$)\\
$\omega_c$ \phantom{\Big|}& $0.1190^{+0.0011}_{-0.0011}$ & $0.1189^{+0.0011}_{-0.0011}$ & $0.1197^{+0.0012}_{-0.0012}$ & $0.1189_{-0.0011}^{+0.0011}$ \\
100$\omega_b$ \phantom{\Big|}& $2.228^{+0.014}_{-0.015}$ & $2.229^{+0.014}_{-0.015}$ & $2.221^{+0.014}_{-0.015}$ &  $2.228_{-0.014}^{+0.014}$ \\
$\ln (10^{10} A_s)$\phantom{\Big|} & $3.066^{+0.019}_{-0.026}$ & $3.071^{+0.026}_{-0.029}$ & $3.071^{+0.032}_{-0.032}$ & $3.072_{-0.028}^{+0.028}$ \\
$n_s$ \phantom{\Big|}& $0.9656^{+0.0041}_{-0.0043}$ & $0.9661^{+0.0043}_{-0.0043}$ & $0.9635^{+0.0043}_{-0.0045}$ & $0.9660_{-0.0042}^{+0.0042}$ \\
$\tau_{\rm re}$ \phantom{\Big|}& $0.06678^{+0.01096}_{-0.01345}$ & $0.06965^{+0.01393}_{-0.01549}$ & $0.06880^{+0.01709}_{-0.01718}$ & $0.06988_{-0.01481}^{+0.01503}$ \\
\hline
$\Omega_M$\phantom{\Big|} & $0.3085_{-0.0065}^{+0.0065}$ & $0.3109_{-0.0084}^{+0.0069}$ & $0.2989_{-0.0088}^{+0.0084}$ & $0.3099_{-0.0083}^{+0.0071}$ \\
$z_{\rm re}$ & $8.893_{-1.186}^{+1.121}$ & $9.150_{-1.355}^{+1.396}$ & $9.097_{-1.499}^{+1.743}$ & $9.174_{-1.321}^{+1.468}$ \\
$\sigma_8$ \phantom{\Big|}& $0.8170^{+0.0076}_{-0.0095}$ & $0.8157^{+0.0135}_{-0.0104}$ & $0.8215^{+0.0169}_{-0.0167}$ & $0.8173_{-0.0113}^{+0.0147}$ \\
\hline
$\chi_{\rm min}^2$ \phantom{\Big|}& 13631.04 & 13630.78 & 13634.40 & 13630.68 \\
$\Delta\chi_{\rm min}^2$\phantom{\big|} & 0.26 & 0 & 3.62 & -0.10 \\
\hline
\end{tabular}
\caption{\label{tab:res1} Mean values of the parameters  and $\chi^2$, for $\Lambda$CDM, $\nu\Lambda$CDM, and the RR model with two different values of $u_0$, using the CMB, BAO and SNe datasets discussed in Section~\ref{sect:datasets}. $H_0$ is in  units of 
${\rm km}\, {\rm s}^{-1}\, {\rm Mpc}^{-1}$.}
\end{table}

\begin{enumerate}

\item Using {\em Planck}+BAO+SNe data, the minimal RR  model predicts
\be
H_0= 69.49\pm 0.80\, ,
\ee
in the usual units ${\rm km}\, {\rm s}^{-1}\, {\rm Mpc}^{-1}$.
Compared to the value $H_0 = 73.24 \pm 1.74$ from local measurements \cite{Riess:2016jrr},  the difference   is  now only $2.0\sigma$, which hardly qualifies as a tension.  In contrast, in $\Lambda$CDM 
and in $\nu\Lambda$CDM we find  a $3.1\sigma$ discrepancy.\footnote{When deriving the  value of a parameter and its error through a MCMC, there is unavoidably a fluctuation in the central value due to the stochastic nature of a MCMC. The central value for $H_0$ that we get from our MCMC for $\nu\Lambda$CDM is however in excellent agreement with the value $H_0=67.7\pm 0.6$ found by {\em Planck} marginalizing over neutrino masses and  using Planck+BAO+JLA+$H_0$, see eq.~(58) of 
\cite{Planck_2015_CP}. Note that here we have not yet used $H_0$ in our dataset.}

\item  The {\em Planck}+BAO+JLA dataset, interpreted within $\Lambda$CDM,  provides no evidence for non-vanishing neutrino masses, and  only gives an upper limit $\sum_{\nu}m_{\nu}<0.10$~eV.\footnote{This bound is slightly stronger than the bound  $\sum_{\nu}m_{\nu}<0.17$~eV given
in eq.~(54d) of \cite{Planck_2015_CP}, where  only  {\em Planck}+BAO data were used.} In contrast the same dataset, interpreted within the minimal RR  model, provides clear evidence for  non-vanishing neutrino masses, and the prediction 
\be
\sum_{\nu}m_{\nu}=0.219^{+0.083}_{-0.084}\hspace{3mm}{\rm eV}\, ,
\ee
that nicely falls within the window $0.06~{\rm eV}\,\lsim \,\sum_{\nu}m_{\nu}\, \lsim\, 6.6~{\rm eV}$ provided by oscillation and beta-decay experiments.

\end{enumerate}

\begin{figure}[t]
\centering
\includegraphics[width=0.8\columnwidth]{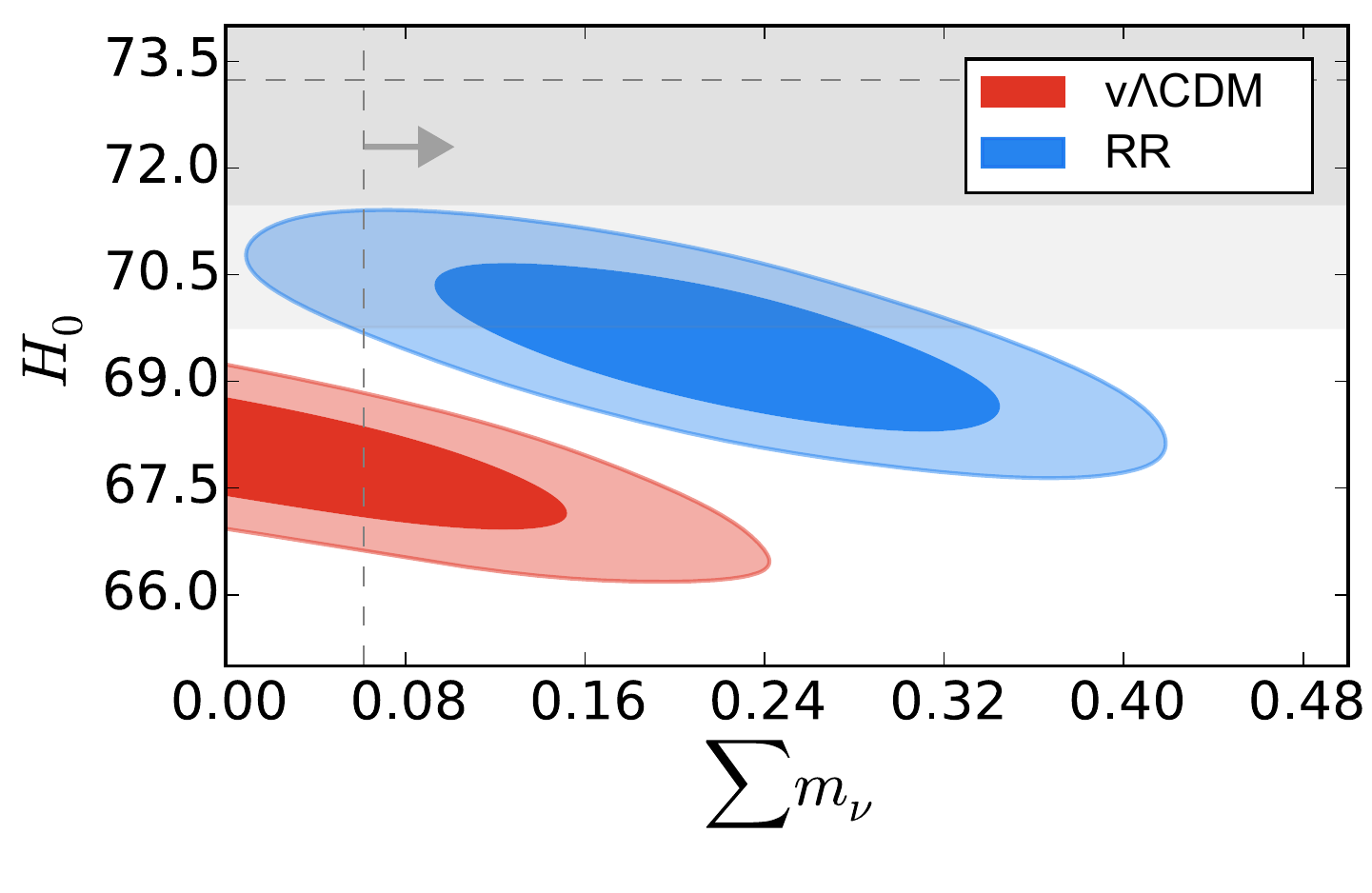}
\caption{The $1\sigma$ and $2\sigma$ contours  in the plane $(H_0,\sum_{\nu} m_{\nu})$, 
obtained from the {\em Planck}+BAO+JLA dataset, for the minimal RR  model (blue) and 
$\nu\Lambda$CDM (red). The horizontal dashed line is the central value of the local $H_0$ measurement \cite{Riess:2016jrr}, and the 
gray areas are the corresponding $1\sigma$ and $2\sigma$ limits (whose upper parts extend above the scale of the figure). The dashed vertical line marked by an arrow is the lower limit on the the sum of neutrino masses from oscillation experiments.
\label{H0_mnu} 
}
\end{figure}

\noindent
The predictions for $H_0$ and $\sum_{\nu}m_{\nu}$ are the two most significant phenomenological results
of the minimal RR model. Fig.~\ref{H0_mnu} shows the two-dimensional marginalized likelihood in the plane $(H_0,\sum_{\nu} m_{\nu})$ for the minimal RR model, compared to the prediction of $\nu\Lambda$CDM, to the lower limit on neutrino masses, and to the value of $H_0$ from local measurements. We see that the predictions of the RR model are more consistent both with the lower limit on the sum of neutrino masses, and with the local $H_0$ measurement.

From Table~\ref{tab:res1} we also see that the RR model with a large value of $u_0$ becomes very close to $\Lambda$CDM, a result that was expected already from the background evolution shown in Fig.~\ref{fig:RRu0256}.

The differences in the $\chi^2$ are reported in the last line of Table~\ref{tab:res1}. We recall that, for models with the same number of free parameters, as $\nu\Lambda$CDM and the minimal RR model, the conventional interpretation is that a  difference $|\Delta \chi^2| \leq  2$ implies  statistical equivalence between the two models, while $2\,\lsim\, |\Delta \chi^2|\,\lsim\, 6$ suggests ``weak evidence'' in favor of the model with lower $\chi^2$, and $|\Delta \chi^2|\gtrsim 6$ indicates ``strong evidence'' in favor of the model with lower $\chi^2$.  The value $\Delta\chi^2=3.62$ in Table~\ref{tab:res1} indicates  weak evidence in favor of $\nu\Lambda$CDM.

Finally, we study the effect of varying the constant $c_k^U$ that determines the initial conditions on $\d U_k$, see \eq{iniconddUk}.  As discussed there, the natural size for the initial conditions on $\d U_k$ is fixed by the metric perturbation $\Phi$, so we expect $c_k^U=O(1)$. For instance, for the Polyakov action we found that $c_k=2$; see \eq{deltaUdsigmainit}. Here, for definiteness,
we have run some further MCMC setting $c_k^U=+6$ or $c_k^U=-6$. The results are shown  in Table~\ref{tab:rescU}, and confirm that variation of the initial conditions of this order have a negligible  effect on the results.

\begin{table}[t]
\centering
\begin{tabular}{|l||c|c|c|c|} 
 \hline 
\multicolumn{1}{|l||}{ } & \multicolumn{3}{|c|}{CMB+BAO+SNe}  \\ \hline
\phantom{\Big|}Parameter & RR $(c^U_k=-6)$  & RR $(c^U_k=0)$ &  RR $(c^U_k=+6)$   \\ \hline 
\phantom{\Big|}$H_0$ & $69.45^{+0.85}_{-0.86}$ & $69.49^{+0.79}_{-0.80}$ & $69.50^{+0.87}_{-0.87}$  \\ 
\phantom{\Big|}$\sum_{\nu}m_{\nu}\ [{\rm eV}]$ & $0.219^{+0.094}_{-0.092}$ & $0.219^{+0.083}_{-0.084}$ & $0.216^{+0.088}_{-0.097}$ \\
\phantom{\Big|}$\omega_c$ & $0.1198^{+0.0013}_{-0.0013}$  & $0.1197^{+0.0012}_{-0.0012}$ & $0.1197^{+0.0013}_{-0.0014}$  \\
\phantom{\Big|}100$\omega_b$ & $2.221^{+0.016}_{-0.016}$  & $2.221^{+0.014}_{-0.015}$ & $2.221^{+0.016}_{-0.017}$ \\
\phantom{\Big|}$\ln (10^{10} A_s)$ & $3.070^{+0.035}_{-0.036}$  & $3.071^{+0.032}_{-0.032}$ & $3.070^{+0.035}_{-0.036}$ \\
\phantom{\Big|}$n_s$ & $0.9637^{+0.0047}_{-0.0047}$  & $0.9635^{+0.0043}_{-0.0045}$ & $0.9637^{+0.0049}_{-0.0048}$  \\
\phantom{\Big|}$\tau_{\rm re}$ & $0.06826^{+0.01864}_{-0.01885}$ & $0.06880^{+0.01709}_{-0.01718}$  & $0.06847^{+0.01897}_{-0.01882}$ \\
\hline
\phantom{\Big|}$\Omega_M$ & $0.2994_{-0.0097}^{+0.0088}$  & $0.2989_{-0.0088}^{+0.0084}$ & $0.2988_{-0.0097}^{+0.0092}$ \\
\phantom{\Big|}$z_{\rm re}$ & $9.035_{-1.646}^{+1.903}$ & $9.097_{-1.499}^{+1.743}$ & $9.052_{-1.633}^{+1.955}$  \\
\phantom{\Big|}$\sigma_8$ & $0.8212^{+0.0189}_{-0.0176}$  & $0.8215^{+0.0169}_{-0.0167}$ & $0.8220^{+0.0188}_{-0.0175}$ \\
\hline
\phantom{\Big|}$\chi_{\rm min}^2$ & 13634.34  & 13634.40 & 13634.28 \\
\phantom{\big|}$\Delta\chi_{\rm min}^2$ & 0.06  & 0.12 & 0 \\
\hline
\end{tabular}
\caption{\label{tab:rescU} Mean values of the parameters  and $\chi^2$,  for the RR model with $u_0=0$ and three different values of the parameter $c^U_k$ for the initial conditions of perturbations, using the CMB, BAO and SNe datasets discussed in Section~\ref{sect:datasets}.}
\end{table}

\subsubsection{Results using  {\em Planck}+BAO+JLA+$H_0$}

Until now, when performing Bayesian parameter estimation, we have not included the local $H_0$ measurements, since an important question that we wanted to answer was whether, in the nonlocal model, the prediction for $H_0$ obtained from the 
{\em Planck}+BAO+JLA dataset gives a result consistent  with that of the local measurements. 
When one looks at the data through the ``glasses" of $\Lambda$CDM there is a natural tendency not to mix 
the local $H_0$ measurement with the {\em Planck}+BAO+JLA dataset,
since they are  in tension, and  one rather appeals to   the possible existence of  some unaccounted systematic effect. However,
when comparing  the performances of two models, it is correct to insert also this data point in the analysis. Discarding it a priori, assuming that it might be due to some hitherto unexplained systematics, would be a form of bias in favor of  $\Lambda$CDM. If we want to test $\Lambda$CDM against other models, we cannot perform a ``cherry picking" of the observations, excluding those that produce tensions in $\Lambda$CDM, unless  clear evidence for  systematic errors is unveiled. We have therefore performed further runs of the MCMC, adding also the value
(\ref{H0Riess}) to the {\em Planck}+BAO+JLA dataset.
The results are shown in Table~\ref{tab:res2}.  The value of $H_0$ in the minimal RR model now further raises to  $H_0=70.13_{-0.72}^{+0.76}$.
We also see that now 
the best chi-square is provided by the minimal RR model, although the difference is not statistically significant. 

\begin{table}[t]
\centering
\begin{tabular}{|l||c|c|c|c|} 
 \hline 
\multicolumn{1}{|l||}{ } & \multicolumn{2}{|c|}{CMB+BAO+SNe+($H_0 = 73.24 \pm 1.74$)}  \\ \hline
Parameter  & $\nu\Lambda$CDM &RR $(u_0=0)$    \\ \hline 
$H_0$ \phantom{\Big|} & $68.11_{-0.51}^{+0.57}$ & $70.13_{-0.72}^{+0.76}$  \\ 
$\sum_{\nu}m_{\nu}\ [{\rm eV}]$  \phantom{\Big|}& $<0.07$ (at $1\sigma$)& $0.168_{-0.084}^{+0.078}$ \\
$\omega_c$ \phantom{\Big|}  & $0.1183_{-0.0011}^{+0.0011}$ & $0.1196_{-0.0012}^{+0.0012}$   \\
100$\omega_b$ \phantom{\Big|}  & $2.235_{-0.015}^{+0.014}$ & $2.224_{-0.015}^{+0.015}$  \\
$\ln (10^{10} A_s)$ \phantom{\Big|}  & $3.074_{-0.027}^{+0.026}$ & $3.063_{-0.034}^{+0.032}$ \\
$n_s$   \phantom{\Big|}& $0.9676_{-0.0043}^{+0.0043}$ & $0.9639_{-0.0045}^{+0.0044}$ \\
$\tau_{\rm re}$ \phantom{\Big|} & $0.07159_{-0.01428}^{+0.01360}$ & $0.06473_{-0.01817}^{+0.01677}$ \\
\hline
$\Omega_M$  \phantom{\Big|} & $0.3045_{-0.0071}^{+0.0064}$ & $0.2922_{-0.0081}^{+0.0075}$  \\
$z_{\rm re}$  \phantom{\Big|}& $9.305_{-1.236}^{+1.334}$ & $8.682_{-1.599}^{+1.777}$ \\
$\sigma_8$  \phantom{\Big|} & $0.8205_{-0.0099}^{+0.0119}$ & $0.8309_{-0.0150}^{+0.0164}$  \\
\hline
$\chi_{\rm min}^2$  \phantom{\Big|} & 13639.26 & 13638.26 \\
$\Delta\chi_{\rm min}^2$  \phantom{\big|} & 0 & -1.0  \\
\hline
\end{tabular}
\caption{\label{tab:res2} Parameter estimation and $\chi^2$ values for $\Lambda$CDM and the RR model with $u_0=0$, using the CMB, BAO, SNe datasets discussed in Section~\ref{sect:datasets}, and the value $H_0 = (73.24 \pm 1.74) {\rm km}\, {\rm s}^{-1}\, {\rm Mpc}^{-1}$ from local measurements. The corresponding mean value of the derived parameter $\gamma= m^2/(9H_0^2)$ of the RR model [see below
\eq{syW}] is $\gamma=0.00917(13)$.
}
\end{table}

The prediction for the sum of the neutrino masses now becomes 
\be
\sum_{\nu}m_{\nu}=0.168_{-0.084}^{+0.078}\hspace{3mm}{\rm eV}\, .
\ee

\clearpage

\subsubsection{Structure formation}

\begin{figure}[t]
\centering
\includegraphics[width=0.6\columnwidth]{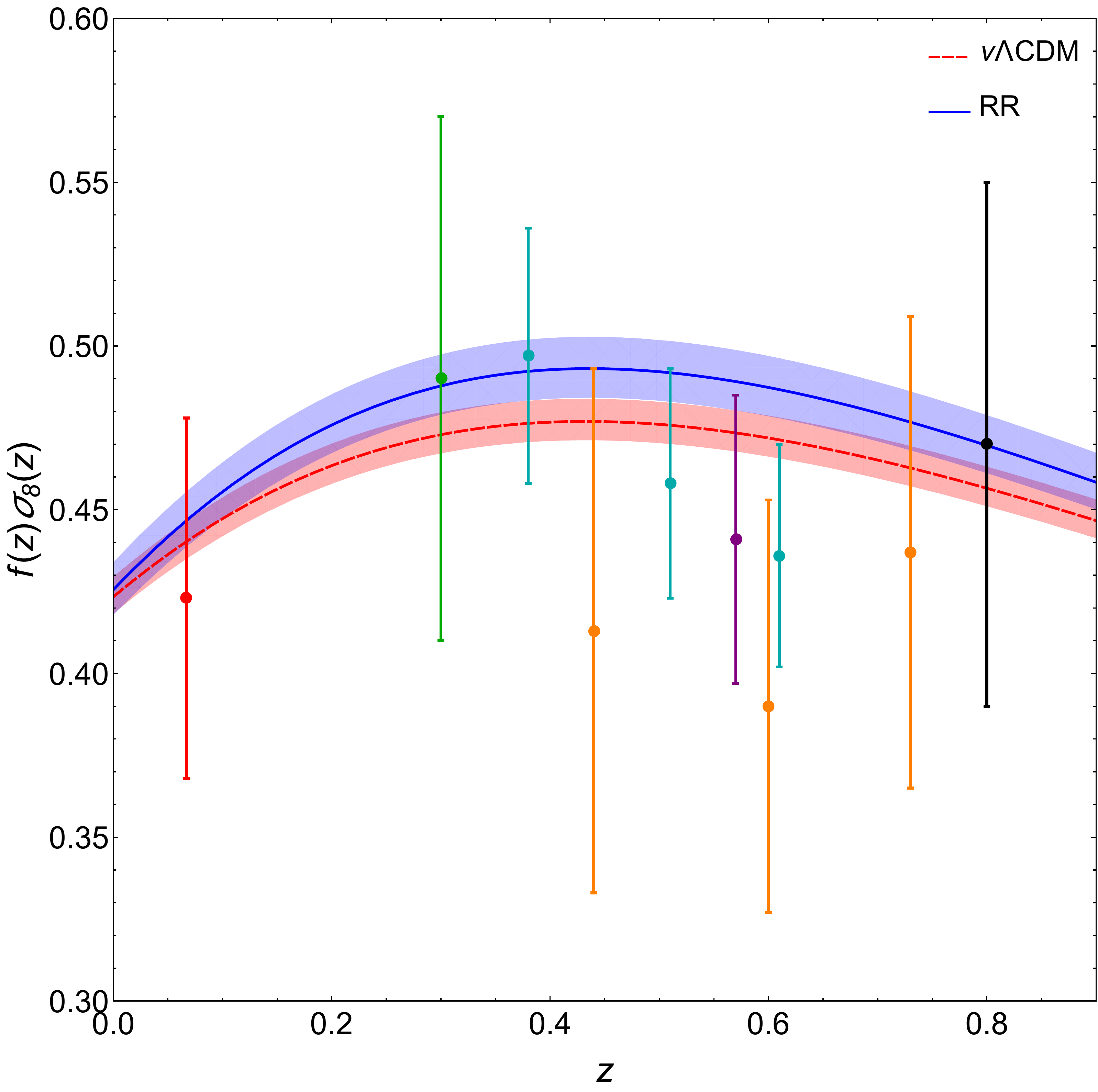}
\caption{\label{fsigma8} Measured values of $f\sigma_8$ at different redshifts, and comparison with the predictions of  the  $\nu\Lambda$CDM  and RR models. For each model we show the predictions obtained using the mean value of the parameters (red, dashed line for $\nu\Lambda$CDM, blue solid line for RR) and the $1\sigma$-level contours (red $\nu\Lambda$CDM, blue RR).
The prediction is obtained  using the mean values and errors  in Table~\ref{tab:res2}, which uses CMB+BAO+SNe+$H_0$. 
The data points are from
 6dF GRS \cite{Beutler:2012px} (red),  SDSS LRG \cite{Oka:2013cba} (green), BOSS CMASS \cite{Samushia:2013yga} (purple), WiggleZ \cite{Blake:2012pj} (orange), VIPERS \cite{delaTorre:2013rpa} (black) and  BOSS DR12 \cite{Alam:2016hwk} (cyan).}
\end{figure}

We next explore how well the minimal RR model and $\nu\Lambda$CDM fit structure formation data. In Fig.~\ref{fsigma8} we plot a compilation of measurements of $f\sigma_8$ at different redshifts and we compare with the predictions of  the minimal RR model and of $\nu\Lambda$CDM, using for the theoretical prediction the mean values  and errors
obtained from CMB+BAO+SNe+$H_0$ given in  Table~\ref{tab:res2}.
We find that  the $\chi^2$ is lower in 
$\nu\Lambda$CDM, compared to the minimal RR model, with a difference $\Delta\chi^2\simeq2.01$. Adding this to the value  $\Delta\chi^2=-1.0$ found from the comparison with 
{\em Planck}+BAO+JLA+$H_0$ data (see again Table~\ref{tab:res2}) we find that, overall, $\nu\Lambda$CDM is favored, with respect to the minimal RR model, with a $\Delta\chi_{\rm tot}^2\simeq 1.01 $, corresponding to statistical equivalence between the two models.
We have also repeated the analysis using the mean values and errors obtained from CMB+BAO+SNe, without $H_0$, given in Table~\ref{tab:res1}. In this case again the $\chi^2$ from structure formation is lower in 
$\nu\Lambda$CDM, with $\Delta\chi^2\simeq 1.33$, which, together with 
the value $\Delta\chi^2=3.62$ in Table~\ref{tab:res1}, gives $\Delta\chi_{\rm tot}^2\simeq 4.95 $  corresponding to weak evidence in favor of $\nu\Lambda$CDM.

\subsubsection{Speed and propagation of gravitational waves}\label{sect:cgw}

The recent observation by the LIGO and Virgo interferometers of the GWs from the neutron star binary coalescence 
GW170817~\cite{TheLIGOScientific:2017qsa} and of the associated $\gamma$-ray burst 
GRB~170817A by Fermi-GBM \cite{Goldstein:2017mmi} and INTEGRAL~\cite{Savchenko:2017ffs}
provides  a test of the speed $c_{\rm gw}$ of gravitational waves (GWs), at the level 
$|c_{\rm gw}-c|/c<O(10^{-15})$ \cite{Monitor:2017mdv} (where the exact lower limit on $c_{\rm gw}-c$ depends on the assumptions on the delay between the GW and GRB emission times). This puts a further significant constraint on modified gravity theories. For instance, this observation rules out a large class of Horndesky theories, leaving only those that are conformally coupled to gravity, i.e. of the form $f(\phi)R$~\cite{Creminelli:2017sry,Sakstein:2017xjx,Ezquiaga:2017ekz,Baker:2017hug}. Let us therefore see how nonlocal gravity performs from this point of view.

We recall, first of all, that in GR tensor perturbations over a FRW background satisfy 
\be\label{4eqtensorsect}
\pa_{\eta}^2\tilde{h}_A+2{\cal H}\pa_{\eta}\tilde{h}_A+k^2\tilde{h}_A=16\pi G a^2\tilde{\s}_A\, ,
\ee
where $\tilde{h}_A(\eta, \vk)$ are  the Fourier modes of the GW amplitude, $A=+,\times$ labels the two polarizations, $\eta$ denotes  conformal time,  
${\cal H}=\pa_{\eta}a/a$, and the source term $\tilde{\s}_A(\eta, \vk)$ is related to the anisotropic stress tensor. For the propagation in empty space we can set $\tilde{\s}_A=0$. 
It is convenient  to introduce a field $\tilde{\chi}_A(\eta, \vk)$ from
\be\label{4defhchiproofs}
\tilde{h}_A(\eta, \vk)=\frac{1}{a(\eta)}  \tilde{\chi}_A(\eta, \vk)\, .
\ee
Then \eq{4eqtensorsect} becomes
\be\label{4propchiproofs1}
\pa_{\eta}^2\tilde{\chi}_A+\(k^2-\frac{\pa_{\eta}^2a}{a}\) \tilde{\chi}_A=0\, .
\ee
On dimensional grounds, $\pa_{\eta}^2a/a\sim 1/\eta^2$. For sub-horizon modes $k\eta\gg 1$, and therefore $\pa_{\eta}^2a/a$ can be neglected compared to $k^2$. Observe that, for the GWs detected by LIGO/Virgo, the typical frequency around merger is $f\sim 10^2$~Hz, corresponding to a reduced wavelength
$\lbar=\lambda/(2\pi)\sim 500$~km. This is ridiculously small compared to the present Hubble scale $H_0^{-1}$, to the extent that $(k\eta)^{-2}\sim (\lbar/H_0^{-1})^2\sim 10^{-41}$. Therefore the term 
$\pa_{\eta}^2a/a$ in \eq{4propchiproofs1} is totally negligible with respect to $k^2$ even when we study deviations in the speed of GWs at the level   $|c_{\rm gw}-c|/c\sim 10^{-15}$, and we can write simply
\be\label{4propchiproofs2}
\pa_{\eta}^2\tilde{\chi}_A+k^2 \tilde{\chi}_A=0\, .
\ee
This shows that the dispersion relation of tensor perturbations is $\omega=k$, i.e. GWs propagate at the speed of light (that we have set to one). On the other hand, the factor $1/a$ in \eq{4defhchiproofs} combines with the factor $1/r=1/d_{\rm com}(z)$ (where $d_{\rm com}$ is the comoving distance) obtained when computing GW emission in the local wave zone (i.e., sufficiently far from the source that the $1/r$ behavior of the GW sets in, but still sufficiently close that the cosmological expansion can be neglected) to produce an overall behavior over cosmological distances
\be
\tilde{h}_A\propto \frac{1}{ d_{\rm com}(z) a }=\frac{1+z}{d_{\rm com}(z)}\, .
\ee
One then defines the luminosity distance from
\be\label{dLdcom}
d_L(z)=(1+z)d_{\rm com}(z)\, ,
\ee
where the factor $(1+z)$ is inserted to reabsorb the $(1+z)$ factors that appear when  passing from the radiated energy and time in the source frame to the corresponding quantities in the detector frame (see e.g. eq.~(4.158) of 
\cite{Maggiore:1900zz}). Then
\be
\tilde{h}_A\propto \frac{(1+z)^2}{d_L(z)}\, .
\ee 
In the waveform produced by the  inspiral of a compact binary the factor $(1+z)^2$ is then absorbed
into a redefinition of the chirp mass (together with analogous factors coming from the passage from the source-frame frequency and the detector-frame frequency of the signal; see  Section 4.1.4 of \cite{Maggiore:1900zz}), and the waveform is finally proportional to $1/d_L(z)$.

Let us now perform the same analysis for GWs in the RR model.
The equation governing the evolution  of tensor perturbations over FRW in the RR model has been derived in \cite{Dirian:2016puz}, and reads
\be
\( 1 - 3\gamma \bar{V} \) \(\pa_{\eta}^2\tilde{h}_A  +2{\cal H}\pa_{\eta}\tilde{h}_A+k^2\tilde{h}_A\)
- 3 \gamma \pa_\eta \bar{V} \pa_\eta \tilde{h}_A   =16\pi G a^2\tilde{\s}_A\, ,
\ee
where $\bar{V}$ is the background evolution of the auxiliary field defined in \eq{defVH0S}. Thus, for the free propagation  we have
\be
\pa_{\eta}^2\tilde{h}_A  +2 {\cal H}[1-\delta(\eta)] \pa_{\eta}\tilde{h}_A+k^2\tilde{h}_A=0\, ,
\ee
where
\be\label{defdeltacalH}
\delta(\eta)=\, \frac{3\gamma\bar{V}'}{2(1 - 3\gamma \bar{V})}\, ,
\ee
and, as usual, $\bar{V}'=d\bar{V}/dx$ (where $x=\ln a$, and we used $\pa_{\eta}\bar{V}={\cal H}\bar{V}')$.\footnote{Similar modified propagation equations have been found in other modified gravity models: in particular,  in the   DGP model \cite{Dvali:2000hr}, at cosmological scales gravity leaks into extra dimensions, affecting the propagation equation of tensor modes \cite{Deffayet:2007kf}. Modified GW propagation can be included  in the general framework of the effective field theory approach to dark energy \cite{Gleyzes:2013ooa,Gleyzes:2014rba,Nishizawa:2017nef}, and 
has  also been found  in some scalar-tensor  theories of the Horndeski class 
\cite{Saltas:2014dha,Lombriser:2015sxa,Arai:2017hxj}.}
We now introduce $\tilde{\chi}_A(\eta, \vk)$ from 
\be\label{4defhchiproofsRR}
\tilde{h}_A(\eta, \vk)=\frac{1}{\tilde{a}(\eta)}  \tilde{\chi}_A(\eta, \vk)\, ,
\ee
where 
\be\label{deftildea}
\frac{\pa_{\eta} \tilde{a}}{\tilde{a}}={\cal H}[1-\delta(\eta)]\, ,
\ee
and we get 
\be\label{4propchiproofsRR1}
\pa_{\eta}^2\tilde{\chi}_A+\(k^2-\frac{\pa_{\eta}^2\tilde{a}}{\tilde{a}}\) \tilde{\chi}_A=0\, .
\ee
Once again, inside the horizon we can neglect $\pa_{\eta}^2\tilde{a}/\tilde{a}$, and we see  that 
GWs propagate at the speed of light also in the RR model. The RR model therefore passes the test from
GW170817/GRB~170817A.

However, we see that there is an interesting  difference in the propagation of GWs in the RR model, with respect to GR, due to the fact that the GW amplitude  now scales as $1/\tilde{a}$ rather than $1/a$.
This means that, rather than being just proportional to $1/d_L(z)$, the GW amplitude observed today, after the propagation from the source to the observer, will have decreased by a factor 
$\tilde{a}_{\rm em}/\tilde{a}_{\rm obs}\equiv \tilde{a}(z)/\tilde{a}(0)$ instead of a factor $a_{\rm em}/a_{\rm obs}=a(z)/a(0)$, where the label refers to the emission time (at redshift $z$) and the observation time, at redshift zero. Therefore
\be
\tilde{h}_A\propto  \frac{\tilde{a}(z)}{\tilde{a}(0) }\, \frac{a(0)}{a(z) }\, 
\frac{1}{d_L(z)}=
 \frac{\tilde{a}(z)}{a(z)} \frac{1}{d_L(z)}
\, ,
\ee
where $d_L(z)\equiv d_L^{\,\rm em}(z)$ is the usual notion of luminosity distance appropriate for electromagnetic signals  and, since only the ratios $\tilde{a}(z)/\tilde{a}(0)$ and $a(z)/a(0)$ enter, without loss of generality we can choose the normalizations $\tilde{a}(0)=a(0)=1$.

Thus, we see that in the RR model there are two different notions of luminosity distance (or of comoving distance).  One appropriate for electromagnetic signals, which is given by the usual expression
\be\label{4dLem}
d_L^{\,\rm em}(z)=\frac{1+z}{H_0}\int_0^z\, 
\frac{d\tilde{z}}{\sqrt{\ora (1+\tilde{z})^4+\oma (1+\tilde{z})^3+\rde(\tilde{z})/\rho_0 }}\, ,
\ee
corresponding to the comoving distance given in \eq{4comovdistzRR}, and a GW luminosity distance

\begin{figure}[t]
\centering
\includegraphics[width=0.5\columnwidth]{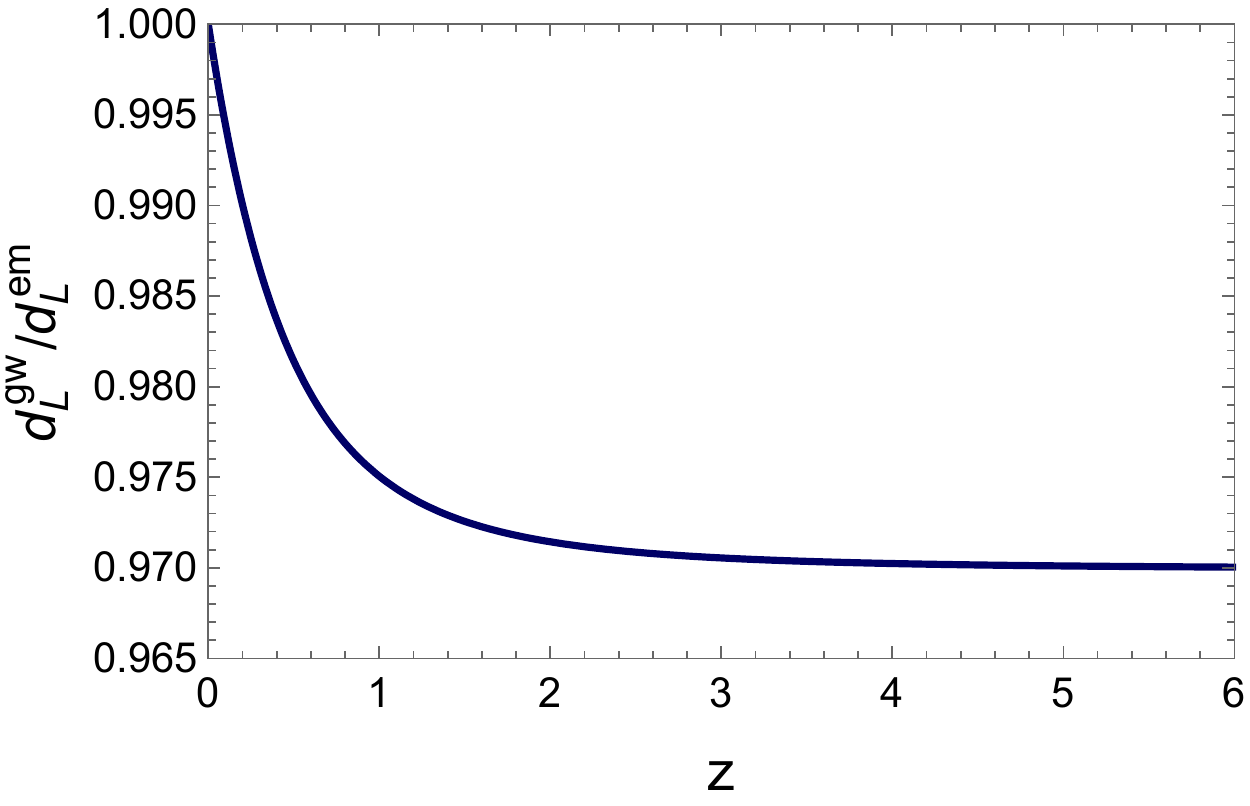}
\caption{The ratio $d_L^{\,\rm gw}(z)/d_L^{\,\rm em}(z)$ in the RR model.
\label{fig:dLgw_over_dLemz6}
}
\end{figure}

\be
d_L^{\,\rm gw}(z)=\frac{a(z)}{\tilde{a}(z)}\, d_L^{\,\rm em}(z)\, ,
\ee
appropriate for the propagation of GWs. The factor $a(z)/\tilde{a}(z) $ is obtained by integrating 
\eq{deftildea}, which gives
\be\label{dLgwdLemRR}
d_L^{\,\rm gw}(z)=d_L^{\,\rm em}(z)\, \sqrt{\frac{1-3\gamma\bar{V}(0) }{1-3\gamma\bar{V}(z) }}\, .
\ee
The background evolution $\bar{V}(z)$ is obtained from
the results presented  in Section~\ref{sect:backgRR} and the corresponding result for 
 the  ratio $d_L^{\,\rm gw}(z)/d_L^{\,\rm em}(z)$ in the RR model is shown in Fig.~\ref{fig:dLgw_over_dLemz6}.
We see  that,  at $z\, \gsim\, 1$,  the two notions of luminosity distance differ by about $3\%$.  This means that, in the RR model, 
 ``standard sirens" would not measure the same luminosity distance as standard candles, and this correction will have to be taken into account when using standard sirens to infer the cosmological parameters in the RR model. In a companion paper~\cite{Belgacem:2017ihm} we will elaborate on this point and discuss the possibility of using standard sirens to distinguish $\Lambda$CDM from the (minimal) RR model with third-generation GW interferometers such as the Einstein Telescope.
 
One can similarly show, using eq.~(27) of \cite{Dirian:2016puz}, that also in the RT model GWs propagate at the speed of light, but the GW luminosity distance is different from the electromagnetic luminosity distance. The same holds if, in the RR or RT models, we replace $\Box^{-1}$ with $(\Box-\xi R)^{-1}$. For the $\Delta_4$ non-local model (\ref{D4}), we will show in Appendix~\ref{sect:D4} that the situation is different, and GWs do not propagate at the speed of light, and in fact this rules out the $\Delta_4$ model.

\section{Conclusions}

The basic physical assumption of our approach is that, because of infrared quantum fluctuations,  gravity develops a mass scale, which modifies its IR behavior. We have explored different possible implementations of this idea, and we have found that the model that turns out to work well corresponds to giving a mass to the conformal mode;
see \eq{massconfmode}. Naively, one would think that the only way to give a mass to some component of the metric  is to break the invariance under diffeomorphisms. This is the route taken in massive gravity or in bigravity, where the spin-2  mode becomes massive because of the addition of terms that break the original diff invariance. However, this is  true only if we want to include a mass term at the level of the fundamental action. If a mass is generated dynamically by quantum effects, it will not appear in the fundamental action, but rather in the quantum effective action. The latter, as we have reviewed in Sections~\ref{sect:rem} and
\ref{sect:examples}, always include nonlocal terms whenever the fundamental theory has massless particles, such as the graviton. With nonlocal terms it is possible to construct diffeomorphism-invariant quantities that, expanded to second order in the field, correspond to a mass term. 
A completely similar situation takes place for gauge fields. Naively, one would think that the only way to give a mass to a gauge field is to break gauge invariance, and this is what is done, for instance, to give masses to the $W$ and $Z$ bosons. However, this is not the only possibility and,
indeed, in QCD the gluons are believe to get an effective mass  through the nonlocal term (\ref{Fmn2}) in the quantum effective action, generated non-perturbatively by the strong IR effects of QCD. Similarly, we have postulated that in GR the conformal mode gets an effective mass 
through the addition of a nonlocal term  to the gravity quantum effective action, of the form given in  \eq{RR}.\footnote{The fact that it is just the conformal mode that gets a dynamical mass matches well the result in \cite{Antoniadis:1986sb}, that show that, in de~Sitter space, the strongest IR divergences come from the long-distance behavior of the propagator of the conformal mode.} In this sense, strictly speaking, we believe that our proposal does not even belong to the class of ``modified gravity" theories. We are not modifying Einstein-Hilbert gravity as a fundamental theory, but we are rather trying to capture its leading quantum effects in the infrared.

Nonlocal terms might sound unfamiliar, particularly to most cosmologists, that are used to work with classical actions. However, it is a fact that they appear in the quantum effective action. Once properly understood, their use does not create any special conceptual nor technical difficulty, as we have discussed in detail in Section~\ref{sect:FAQ}. At most, one might have to face the problem of choosing the boundary condition for the nonlocal operator, or equivalently for the auxiliary fields introduced for writing the theory in a local form. In the RR model, we have seen that this just amounts to having to deal with one extra free parameter, $u_0$.
Indeed, once one overcomes the ``psychological barrier" of thinking in terms of quantum effective actions, and therefore allowing for nonlocal terms, one realizes that a model such as the RR model is much more elegant, simple and attractive than the typical modified gravity models that proliferated in the recent literature. It  has only one new parameter, the mass scale $\Lrr$, that has a physically clear and well-defined meaning as a mass term for the conformal mode, and that replaces the cosmological constant in $\Lambda$CDM  (plus a hidden parameter related to the initial conditions of the nonlocal operator). In contrast, modified gravity models that have been recently much studied in the literature typically involve several arbitrary functions of new hypothetical fields, with no physical clue on the form of these functions, and often a level of complication that eventually makes them look somewhat baroque, even more considering that they are meant to be an alternative to a simple cosmological constant.

At the phenomenological level, we have seen that the RR model works very well. If we  compare its performances to that of  $\Lambda$CDM by using CMB, BAO, SNe, local measurements of $H_0$  and structure formation data, the two models fits the data at a  statistically equivalent level. However, the RR model (in its minimal form, with $u_0=0$) has two particularly interesting predictions: a higher value of $H_0$, which reduces the tension with local measurement (bringing it down to the $2.0\sigma$ level if we estimate $H_0$ from CMB+BAO+SNe only), and a value for the neutrino masses which agrees well with the limit from oscillation experiments. In contrast, in $\nu\Lambda$CDM, where we let the neutrino masses as free fitting parameters,  
the one-dimensional likelihood for the sum of the neutrino masses is peaked in zero. These results are shown in Fig.~\ref{H0_mnu}, which displays the two most interesting predictions of the RR model, and  compares them  to the predictions of 
$\nu\Lambda$CDM. Observe, from   Fig.~\ref{H0_mnu}, that a direct measurement of the  neutrino masses from particle physics experiments could provide decisive evidence for the RR model, compared to 
$\Lambda$CDM. Near-future cosmological observations should also be able to discriminate between the RR model and $\Lambda$CDM, or at least put stringent bounds on the parameter $u_0$ which basically interpolates between the two. 
In separate publications will be discussed the forecasts for the RR model for future surveys such EUCLID, SKA and DESI~\cite{Casas:inprep}
and for a third-generation GW detector such as the Einstein Telescope~\cite{Belgacem:2017ihm}.

\vspace{5mm}\noindent
{\bf Acknowledgments.} We thank Giulia Cusin for very useful discussions.
The work  of the authors is supported by the Fonds National Suisse and  by the SwissMap National Center for Competence in Research.

\appendix

\section{The $\Delta_4$ model}\label{sect:D4}

The $\Delta_4$ non-local model (\ref{D4}) was introduced in \cite{Cusin:2016nzi}, where it was only studied at the background level. There it was found that its DE equation of state $\wde(z)$ is significantly different from $-1$, with a value $\wde(0)\simeq -1.34$. Comparison with the {\em Planck} limits on $\wde(z)$ \cite{Ade:2015rim} then already suggested that the model will not fit well the data, although, as was mentioned,  a full analysis of the perturbations is necessary to reach a definite conclusion. Here we perform such an analysis. We will see that, in fact, the model is already ruled out by the behavior of tensor perturbations, that do not propagate with the speed of light. However, the study of the scalar sector of this model is also methodologically interesting. In fact, in general, in modified gravity models, a DE equation of state  on the phantom side has the effect of rising the value of $H_0$ obtained from parameter estimation, and indeed the value of $H_0$ obtained  from local measurements could be obtained in a wCDM model with $w\simeq -1.3$ \cite{DiValentino:2016hlg}. So, the study of the scalar sector of the $\Delta_4$ model is interesting because its DE equation of state  is expected to be about the most phantom that one can have in order to fit reasonably the CMB+BAO+SNe data, and then its prediction for $H_0$ will give an idea of  the maximum value of $H_0$ that could be obtained from models of this type.

It is also conceptually interesting to observe that the $\Delta_4$ model (\ref{D4}) interpolates between the RR model and a model with a non-trivial form factor for Newton's constant. This can be seen more easily in de~Sitter space, where $R$ is constant and $\Rmn=(1/4)R\gmn$, so
\eq{defDP} simplifies to 
\be\label{D4BoxBoxR}
\Delta_4=\Box \(\Box-\frac{1}{6}R\)\, .
\ee
Then, after  integrations by parts, the effective action (\ref{D4}) becomes 
\be\label{D4deSitter}
\Gamma_{\Delta_4}=
\frac{\mplr^2}{2}\int d^{4}x \sqrt{-g}\, 
\[R-\frac{1}{6} m^2 \(\frac{1}{\Box}R\)\  \(\frac{1}{\Box-\frac{1}{6}R}R\)\]\, ,
\qquad {\rm (de Sitter)}
\ee
For Fourier modes such that $|\Box|\gg R/6$, this reduces to the RR model. In the opposite limit
$|\Box|\ll R/6$, in contrast
\be\label{D4runG}
\Gamma_{\Delta_4}\simeq
\frac{\mplr^2}{2}\int d^{4}x \sqrt{-g}\, 
\(1+\frac{m^2}{\Box}\)R
\qquad ({\rm de Sitter}, |\Box|\ll R/6)\, ,
\ee
corresponding to a running  Newton's constant. We next recall the main results on the background evolution of the model, and we work out its cosmological perturbations.

\vspace{2mm}

\noindent
{\bf Background evolution}.
The covariant equation of motion derived from (\ref{D4}) are\footnote{We have corrected a typo in eq.~(4.4) of \cite{Cusin:2016nzi}.
This typo did not affect any other equation or result in that paper.}
\bees
\label{eqmotD4}
&&G_{\alpha\beta}\left[1-\frac{m^2}{3} S + \frac{m^2}{9}(\n S)^2\right]
+\frac{m^2}{6}\bigg\{ \n_\alpha\n_\beta\left[2 S +\frac{1}{3}(\n S)^2\right]\nonumber\\
&&\hspace*{8mm}- g_{\alpha\beta}\left[2\Box S+\frac{1}{2}(\Box S)^2-R^{\rho\sigma}\n_\sigma S\n_\rho S-\frac{1}{6}\Box\left(\n S\right)^2\right]\nonumber\\
&&\hspace*{8mm}+S\left(\n_\rho R_{\alpha\beta}-\n_{(\alpha}R_{\beta)\rho}\right)\n^\rho S-4\left(\n^{\rho}S\right)R_{\rho(\alpha}\n_{\beta)} S+2\Box S \n_\alpha\n_\beta S\nn\\
&&\hspace*{8mm}-2\left(\n^\rho\n_{(\alpha} S\right)\n_{\beta)}\n_\rho S
+\frac{2}{3}R\n_{(\alpha}S \n_{\beta)} S+S\left(\n^\lambda R_{\lambda(\alpha\beta)\rho}\right)\n^\rho S\nonumber\\
&&\hspace*{8mm}
+2\left(\n^\lambda S\right)R_{\lambda(\alpha\beta)\rho}\n^\rho S-2\left(\n_{(\alpha}S\right)\left(\n_{\beta)}\Box S\right)\bigg\}=8 \pi G T_{\alpha\beta}\,.
\ees
where $S$ is defined by
\be\label{defSinD4}
S=\Delta_4^{-1}R\, . 
\ee
Specializing to FRW, the background evolution equations  are~\cite{Cusin:2016nzi}
\be\label{h2Y}
h^2(x)=\Omega_M e^{-3x}+\Omega_R e^{-4x}+\gamma Y(x)\, ,
\ee
where again $\gamma= m^2/(9H_0^2)$, while now

\be\label{defYD4}
Y=\frac{1}{2}W'(6-U'-2U)+W(3-6\zeta+\zeta U'+2\zeta U)+\frac{1}{4}U^2\, ,
\ee
and, following~\cite{Cusin:2016nzi}, we have introduced  two auxiliary fields
\bees
W&\equiv& H^2 S\, ,\label{defWinD4}\\
U&\equiv & a^{-2}\frac{\pa^2S}{\pa\eta^2}=H^2 \[ S''+(1+\zeta) S'\]\, ,\label{defUinD4}
\ees
where $\eta$ is conformal time (recall that, in contrast,  the prime denotes $d/dx$, where $x=\log a$).
The introduction of the two fields $U,W$ allows us to split the fourth-order equation $\Delta_4 S=R$ into a couple of second-order equations,
\bees
U''+(5+\zeta)U'+(6+2\zeta)U&=&6(2+\zeta)\, .\label{fun3}\\
W''+(1-3\zeta)W'+2(\zeta^2-\zeta-\zeta')W&=&U\, .\label{eqW}
\ees
Setting $\zeta(x)=\zeta_0$ constant  we find that the most general solution of  \eq{fun3} is
\be\label{solUD4}
U(x)=\frac{3(2+\zeta_0)}{3+\zeta_0}+u_1 e^{-(3+\zeta_0)x}+u_2 e^{-2x}\, .
\ee
Therefore both homogeneous solutions are decaying modes, in all cosmological epochs, and even the inhomogeneous solution is constant, rather than linearly growing in $x$ as in \eq{pertU}. The same holds for $W$, since the homogeneous equation $W''+(1-3\zeta_0)W'+2(\zeta_0^2-\zeta_0)W=0$ has the solutions
$W=e^{\beta_{\pm} x}$  with $\beta_{+}=2\zeta_0$ and $\beta_{-}=-1+\zeta_0$. Again, $\beta_{-}$ is negative in all three eras, while  $\beta_{+}$ is negative in RD and MD and vanishes, corresponding to a constant solution, in dS. Thus, there is no growing mode and the cosmological evolution is stable. 
Thus, even if we set the initial conditions of order one for $U$ and $W$ in a earlier inflationary epoch, 
$U$ and $W$ still enter the RD era with a value of order one. In RD $\zeta_0=-2$, so the inhomogeneous term in \eq{solUD4} vanishes, and for $U$ the de~Sitter solution is matched to the two decaying modes $e^{-x}$ and $e^{-2x}$. Thus, the solution is quickly attracted toward the one obtained setting $u_0=0$ deep in RD. Similarly, for $W$ the solution that emerges from de~Sitter is matched to its two decaying modes in RD. Thus,
the solution obtained setting the initial conditions $U=W=U'=W'=0$ at some initial time deep in RD is an attractor and, in the $\Delta_4$ model there is no free parameter associated to the boundary conditions (contrary to the RR model, where we have $u_0$). This makes the model  very predictive.\footnote{In fact, even if one set large  initial values for $U$ and $W$ at the beginning of RD, the exponential decay during RD brings them back, to high accuracy, to the solution obtained with vanishing initial conditions.}

\begin{figure}[t]
\centering
\includegraphics[width=0.42\columnwidth]{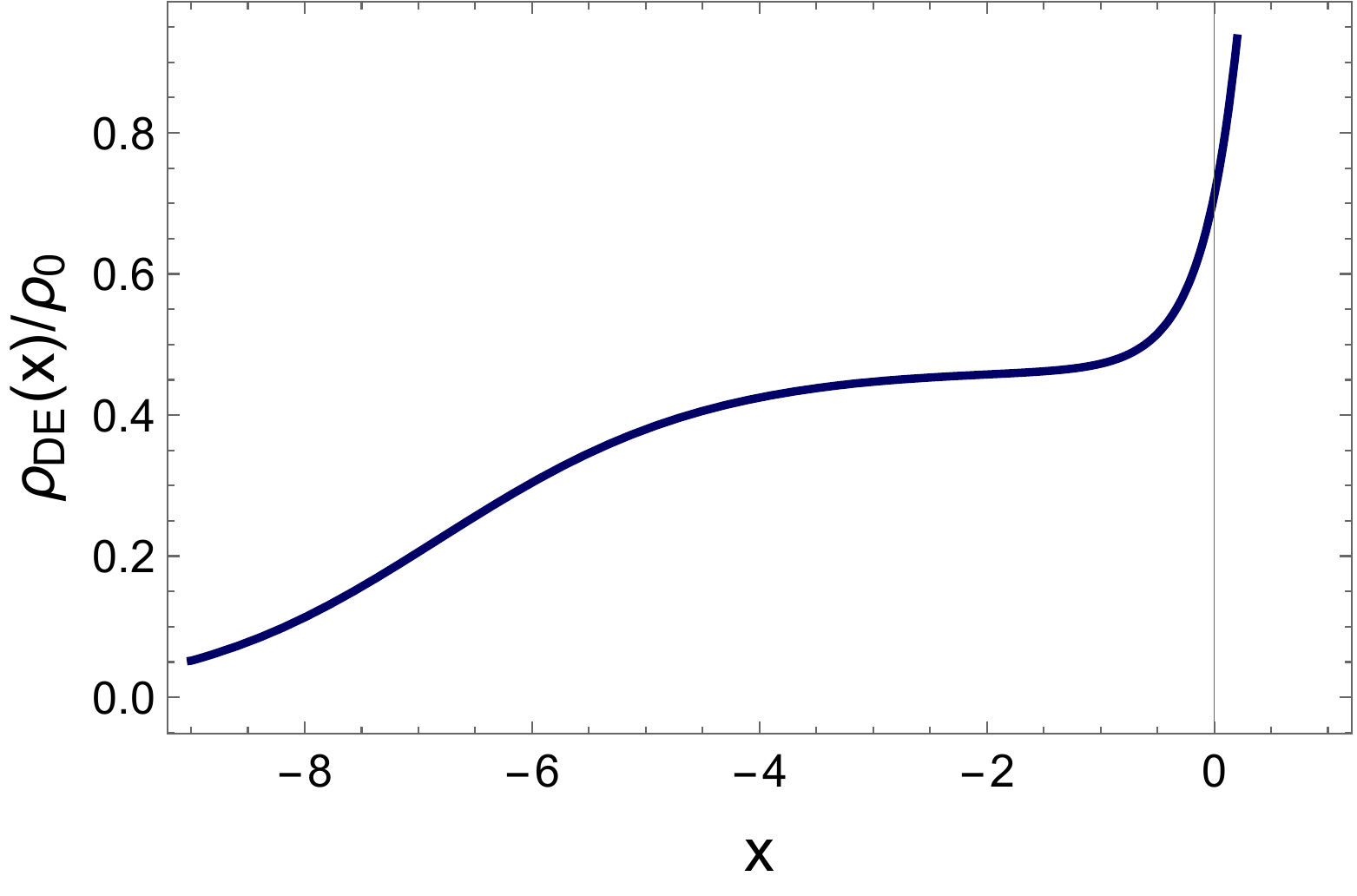}
\includegraphics[width=0.42\columnwidth]{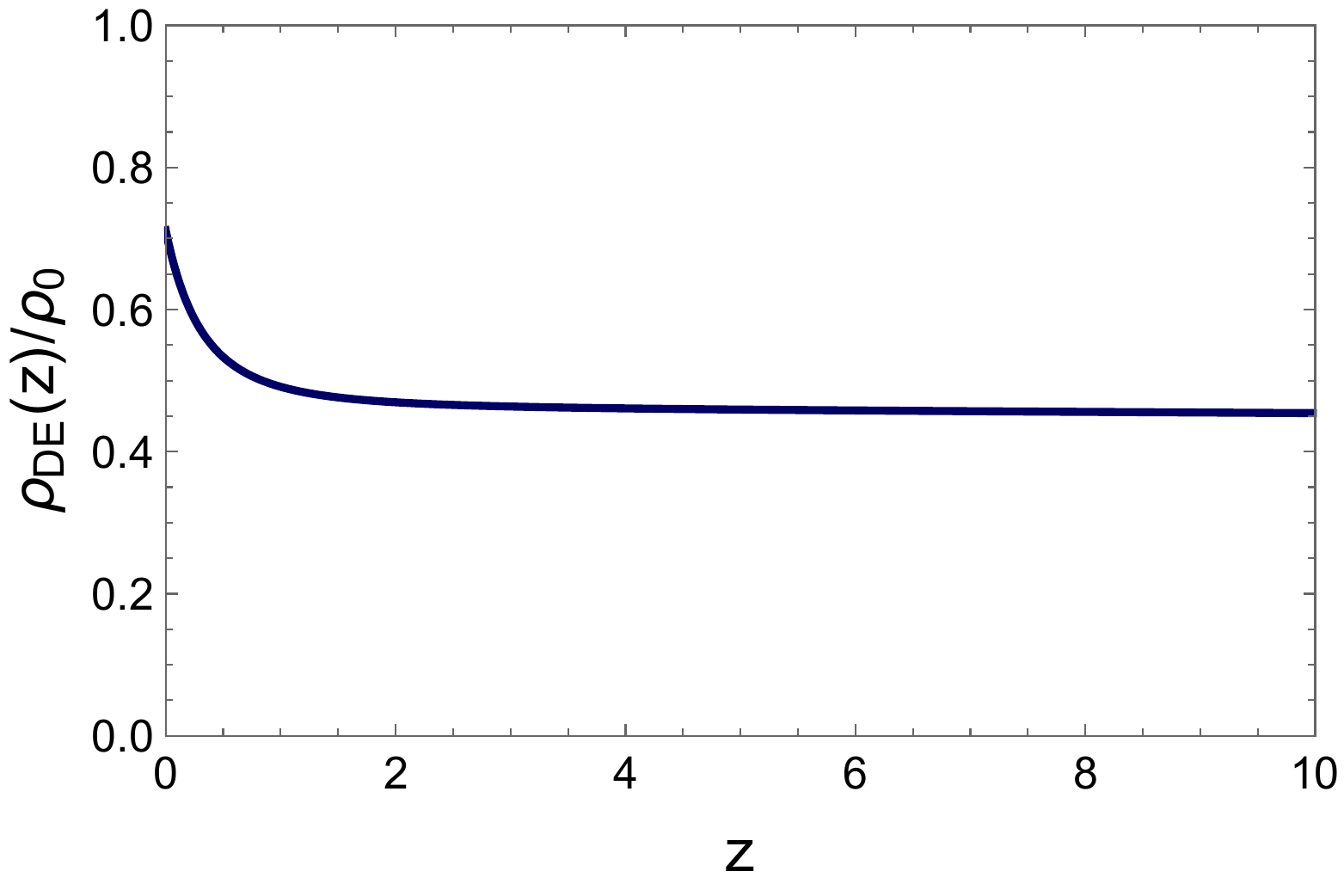}
\includegraphics[width=0.42\columnwidth]{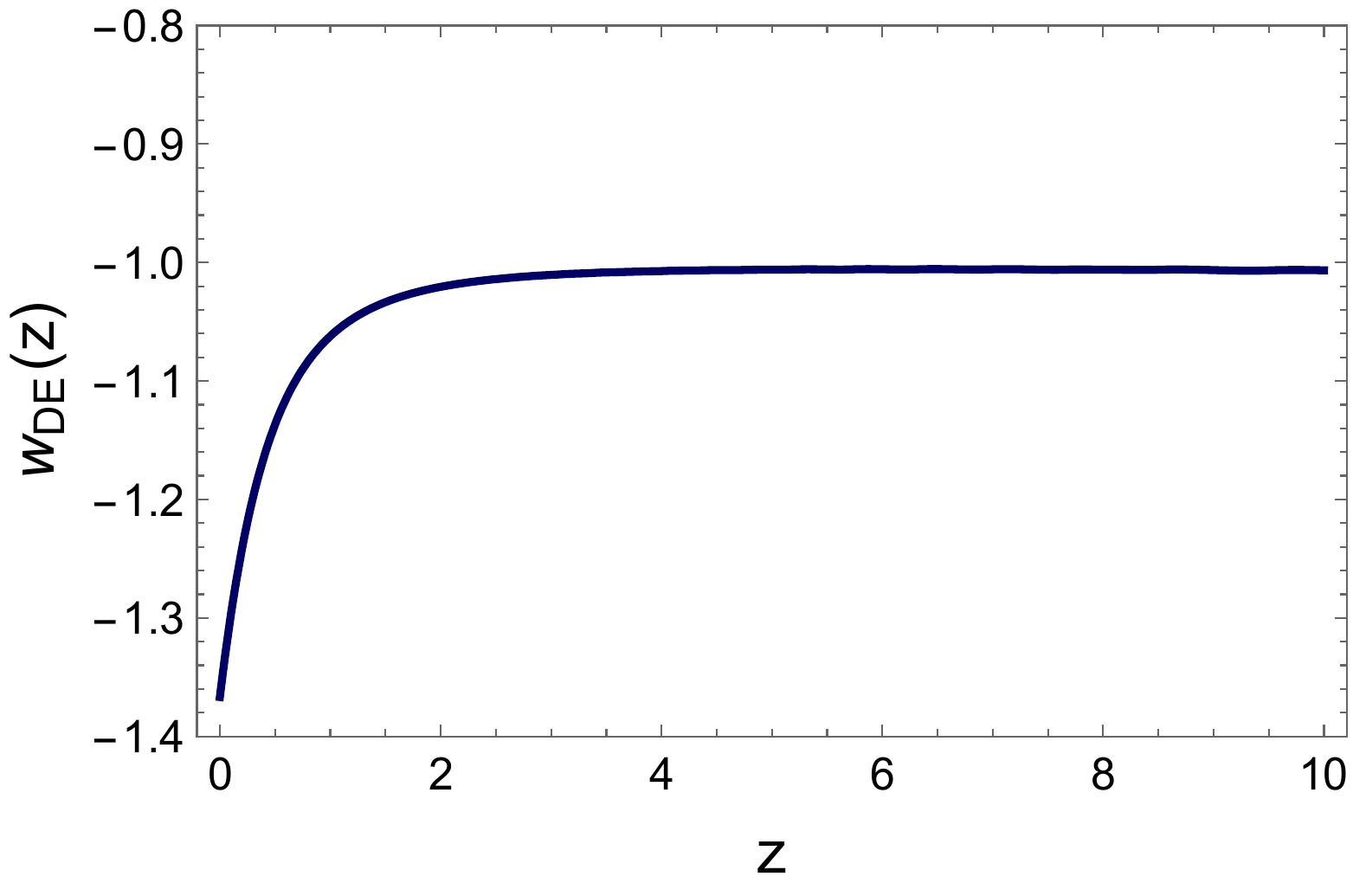}
\includegraphics[width=0.42\columnwidth]{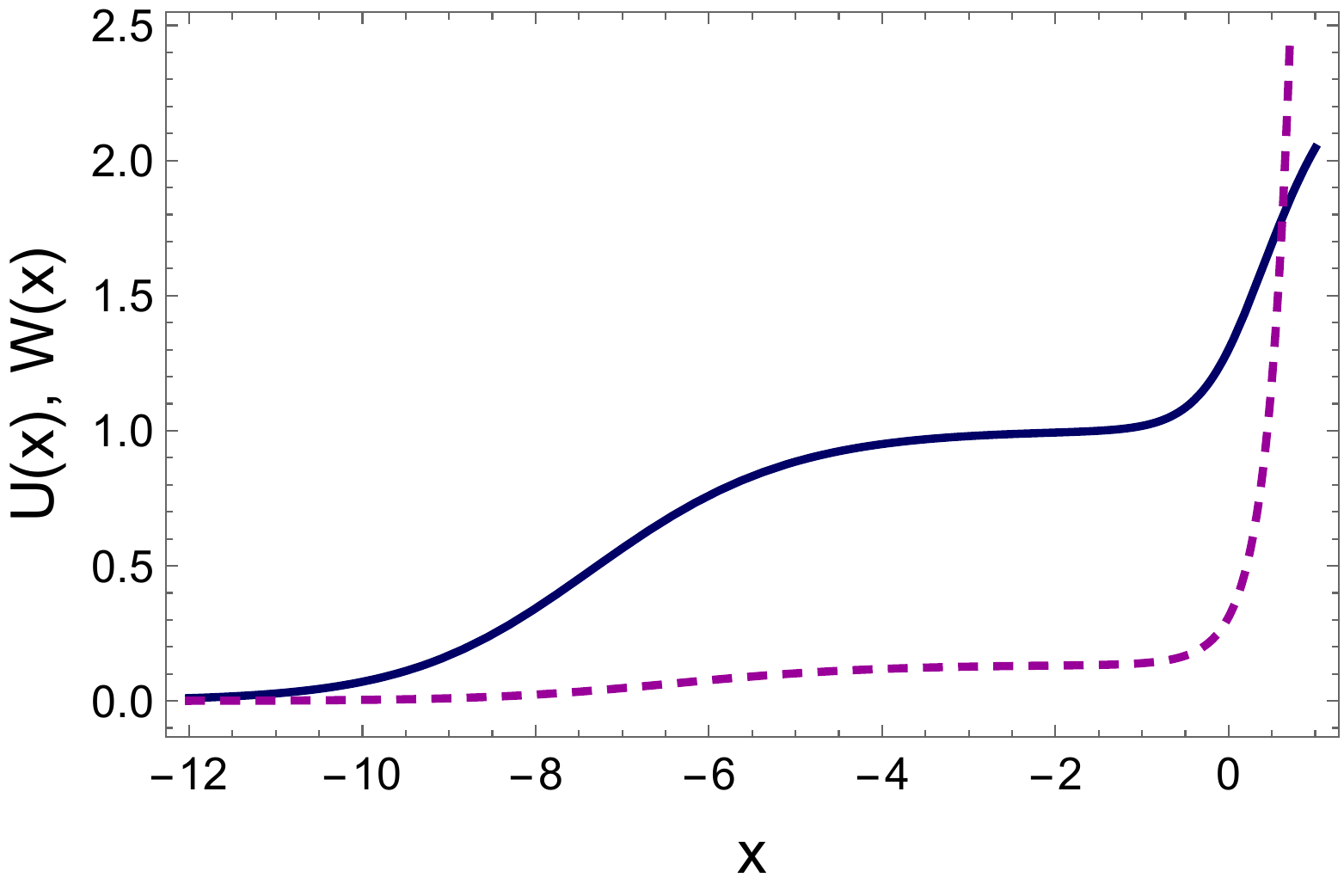}
\caption{As in Fig.~\ref{fig:backgRR}, for the $\Delta_4$ model, using 
$\oma\simeq 0.29$ and $h_0\simeq 0.71$, see Table~\ref{tab:resD42}.
\label{fig:backgD4}
}
\end{figure}

We can now integrate numerically the equations of motion, with  initial conditions $U=W=U'=W'=0$ at some initial time deep in RD. The result of the numerical integration  is shown in Fig.~\ref{fig:backgD4}~\cite{Cusin:2016nzi}.
As in the RR model, we have an effective  dark energy density $\rde=\rho_0\gamma Y$.  We see that, once again, the effective DE density vanishes deep in RD, and begins  to grow as we approach radiation-matter equality   (around $x_{\rm eq}\simeq -8.1$). In the $\Delta_4$ model, during MD $\rde(x)$ eventually stabilizes to a constant, leading to a long phase where the EoS parameter 
$w(z)\simeq -1$ as in $\Lambda$CDM. However, as the DE density starts to dominate over matter, $\rde$  finally grows again, leading near $z=0$ to an EoS which is substantially more phantom than in the RR model, with $\wde(z=0)\simeq -1.36$ (with the choice $\oma\simeq 0.29$ and $h_0\simeq 0.71$ obtained from parameter estimation in this model, see below). Using again the parametrization (\ref{ChevLind}) in the region $-1<x<0$,  for our best-fit values $\oma\simeq 0.29$ and $h_0\simeq 0.71$ we get
$w_0\simeq -1.33$ and $w_a\simeq 0.53$.

\vspace{2mm}

\noindent
{\bf Cosmological perturbations in the scalar sector}.
 At the perturbation level we write the metric  in the scalar sector as in \eq{defPhiPsi}. The auxiliary field $S$ defined in \eq{defSinD4} satisfies the fourth-order equation $\Delta_4S=R$. To keep the equations as close as possible to that of the RR model, we find convenient to split this fourth-order equation into a pair of second-order equation, introducing a second  auxiliary field. In a generic space-time, it is not possible to perform this split covariantly [although this would be possible in de~Sitter, see \eq{D4BoxBoxR}], so we simply keep the definitions (\ref{defWinD4}, \ref{defUinD4}) also when performing perturbations over an FRW background, i.e. we write $W(\eta,\vx) \equiv H^2 (\eta)S(\eta,\vx)$ and 
$U(\eta,\vx)\equiv a^{-2}\pa_{\eta}^2 S(\eta,\vx)$.
Similarly to the RR model, for studying the perturbations it is convenient to use  the 
variable $V=W/h^2=H_0^2S$ rather than $W$. We then expand  $V=\bar{V}+\delta V$. As a second perturbation variable for the auxiliary field, instead of $\delta U$, it is  convenient to chose
\be
\delta Z \equiv  h^2 \left[\delta V''+\left(1+\zeta\right)\delta V'+\bar{V}'\left(\Phi'-\Psi'\right)\right]+ 2 \left(\Phi-\Psi\right) \bar{U}\, ,
\ee
because then  higher-derivative terms drop out of the equations. The above equation can be taken as a dynamical equation for $\delta V$, while for $\delta Z$
one finds
\begin{align}
&\delta Z '' + \(5+\zeta\) \delta Z'+\left(6+2\zeta+2 \hat{k}^2 \right)\delta Z= -h^2 \hat{k}^4 \d V + 6 \left[\Phi''+(3+\zeta) \Phi'\right]  \\
& + \bigg(\bar{U}'+2\bar{U}-6 - \frac{\hat{k}^2} {3} h^2 \bar{V}' \bigg) \big( \Psi' - \Phi' \big) + 2 \hat{k}^2 \big(1 - \frac{2}{3} \bar{U} \big) \Psi +4\bigg[3 \big(\zeta +2\big)+\hat{k}^2 \bigg(1 + \frac{\bar{U}}{3}\bigg) \bigg]\Phi \, .\nn
\end{align}
The analogous of \eqst{ModPoisson}{MEE4x} are
\bees
&&\( 1 - 3 \gamma \bar{V} \) \( \hat{k}^2 \Phi + 3  \Phi' -3 \Psi  \) + \frac{3 \gamma}{2} \bigg\{ \left(\bar{V}'-\frac{\bar{U}}{2h^2}\right)\delta Z+\frac{1}{2}\bar{V}' \delta Z' \label{PoissonD4}\\
&&\hspace*{8mm}
-\left[3+\hat{k}^2\left(1+\frac{\bar{U}}{2}\right)\right]\delta V+6 \bar{V}' \Psi-3\bar{V}' \Phi'+\frac{\bar{U}^2}{h^2}\Phi
 +\left(\frac{\bar{U}'}{2}+\bar{U}-3+\frac{5}{6} \hat{k}^2 h^2 \bar{V}'\right)\delta V'\nn\\
 &&\hspace*{8mm}
 -\left(\bar{U}'+2\bar{U}\right) \bar{V}' \left(\Phi+\Psi\right) +\frac{2}{3}\hat{k}^2 h^2 \bar{V}'^2  \left(\Phi-\Psi\right)\bigg\}=  \frac{3}{2 h^2} \big(  \Omega_{\textsc{R}} e^{-4x} \delta_{\textsc{R}} + \Omega_{\textsc{M}} e^{-3x} \delta_{\textsc{M}} \big)\, ,\nn\\
&&\( 1 - 3 \gamma \bar{V} \)  \hat{k}^2 (  \Phi' - \Psi) + \frac{\gamma \hat{k}^2}{2}
\bigg[\left(\bar{U}-3\right) \left(\delta V'-\bar{V}' \Psi\right) +\bar{U} \bar{V}' \Phi
\label{MEE2xD4}\\
&&\hspace*{8mm}
+\frac{3}{2}\left(2-\bar{U}'-2\bar{U}-\hat{k}^2 h^2 \bar{V}'\right)\delta V
-\frac{1}{2}\bar{V}'\delta Z \bigg]=
 - \frac{3}{2 h^2} \bigg(  \frac{4}{3} \Omega_{\textsc{R}} e^{-4x} \hat{\theta}_{\textsc{R}} +  \Omega_{\textsc{M}} e^{-3x} \hat{\theta}_{\textsc{M}} \bigg)\,,\nn 
 \ees
 \bees
&&( 1 - 3 \gamma \bar{V} )\[   \Phi'' + (3+\zeta) \Phi' -  \Psi' - (3+2 \zeta) \Psi + \frac{\hat{k}^2}{3}(\Phi + \Psi) \] = - \frac{1}{2 h^2}  \Omega_{\textsc{R}} e^{-4x} \delta_{\textsc{R}}
 \nn \\ 
&&\hspace*{8mm} +\frac{\gamma}{2} \bigg\{
\left[3\Phi' - \left(6+\bar{U}'+2\bar{U}+\frac{2}{3}\hat{k}^2 h^2 \bar{V}'\right)\Psi +\left(\frac{2}{3}\hat{k}^2 h^2 \bar{V}'-\bar{U}'-2\bar{U}\right) \Phi\right]  \bar{V}'   \nn \\
&&\hspace*{8mm}
+\left(\bar{U}-6\right)\frac{\bar{U}}{h^2}\Phi + \left(\bar{V}'-\frac{\bar{U}}{2 h^2}+\frac{3}{h^2}\right)\delta Z +\frac{1}{2}\bar{V}' \delta Z'+ \left[3(3+2\zeta)+\left(2-\frac{\bar{U}}{2}\right)\hat{k}^2\right] \delta V\nn\\
&&\hspace*{8mm}
+ \left(\bar{U}'+2\bar{U}+6+\frac{5}{3}\hat{k}^2h^2\bar{V}'\right) \frac{\delta V'}{2}  \bigg\}\, ,
\label{MEE3xD4}\\
&&
( 1 - 3 \gamma \bar{V}+ \gamma \bar{V}'^2 h^2) \Psi + ( 1 - 3 \gamma \bar{V} - \gamma \bar{V}'^2 h^2)\Phi  - 3 \gamma \delta V (1- \bar{U})+\gamma h^2 \bar{V}' \delta V' =0\, .\label{MEE4xD4}
\ees
Finally, \eqst{dM1}{dtheta2} are of course unchanged, since they express the linearization of the energy-momentum tensor, and are therefore model-independent.
The results of the numerical integration shows  that the cosmological perturbations are again stable and relatively  close to those of $\Lambda$CDM.

\begin{table}[t]
\centering
\begin{tabular}{|l||c|c|} 
 \hline 
\multicolumn{1}{|l||}{ } & \multicolumn{2}{|c|}{CMB+BAO+SNe}  \\ \hline
Parameter &  $\nu\Lambda$CDM & $\Delta_4$    \\ \hline 
$H_0$ \phantom{\Big|}&   $67.60^{+0.66}_{-0.55}$ &  $70.27^{+0.95}_{-0.94}$ \\ 
\phantom{\Big|}$\sum_{\nu}m_{\nu}\ [{\rm eV}]$  & $< 0.10$  (at $1\sigma$) & $0.185^{+0.087}_{-0.096}$ \\
\phantom{\Big|}$\omega_c$ & $0.1189^{+0.0011}_{-0.0011}$  & $0.1202^{+0.0014}_{-0.0014}$  \\
\phantom{\Big|}100$\omega_b$  & $2.229^{+0.014}_{-0.015}$  & $2.217^{+0.017}_{-0.017}$ \\
\phantom{\Big|}$\ln (10^{10} A_s)$  & $3.071^{+0.026}_{-0.029}$  & $3.080^{+0.034}_{-0.036}$ \\
\phantom{\Big|}$n_s$ &   $0.9661^{+0.0043}_{-0.0043}$ & $0.9637^{+0.0050}_{-0.0050}$ \\
\phantom{\Big|}$\tau_{\rm re}$  & $0.06965^{+0.01393}_{-0.01549}$ & $0.07280^{+0.01769}_{-0.01927}$ \\
\hline
\phantom{\Big|}$\Omega_M$  & $0.3109_{-0.0084}^{+0.0069}$  & $0.2925^{+0.0096}_{-0.0101}$ \\
\phantom{\Big|}$z_{\rm re}$ & $9.150_{-1.355}^{+1.396}$ & $9.490^{+1.793}_{-1.656}$ \\
\phantom{\Big|}$\sigma_8$  & $0.8157^{+0.0135}_{-0.0104}$  & $0.8240^{+0.0199}_{-0.0177}$ \\
\hline
\phantom{\Big|}$\chi_{\rm min}^2$ & 13630.78  & 13649.98 \\
\phantom{\big|}$\Delta\chi_{\rm min}^2$  & 0  & 19.20 \\
\hline
\end{tabular}
\caption{\label{tab:resD41} Parameter estimation and $\chi^2$ values for $\nu\Lambda$CDM and the $\Delta_4$ model, using the CMB, BAO and SNe datasets.}
\end{table}

\begin{table}[t]
\centering
\begin{tabular}{|l||c|c|} 
 \hline 
\multicolumn{1}{|l||}{ } & \multicolumn{2}{|c|}{CMB+BAO+SNe+$H_0$}  \\ \hline
Parameter   & $\nu\Lambda$CDM & $\Delta_4$    \\ \hline 
\phantom{\Big|}$H_0$  &  $68.11_{-0.51}^{+0.57}$   & $70.81^{+0.87}_{-0.76}$ \\ 
\phantom{\Big|}$\sum_{\nu}m_{\nu}\ [{\rm eV}]$  & $<0.07$ (at $1\sigma$) & $0.149^{+0.071}_{-0.090}$ \\
\phantom{\Big|}$\omega_c$  &  $0.1183_{-0.0011}^{+0.0011}$  & $0.1201^{+0.0013}_{-0.0012}$  \\
\phantom{\Big|}100$\omega_b$  &  $2.235_{-0.015}^{+0.014}$  & $2.220^{+0.015}_{-0.015}$ \\
\phantom{\Big|}$\ln (10^{10} A_s)$  & $3.074_{-0.027}^{+0.026}$  & $3.077^{+0.032}_{-0.033}$ \\
\phantom{\Big|}$n_s$  &  $0.9676_{-0.0043}^{+0.0043}$ & $0.9642^{+0.0046}_{-0.0047}$ \\
\phantom{\Big|}$\tau_{\rm re}$  & $0.07159_{-0.01428}^{+0.01360}$  & $0.07123^{+0.01666}_{-0.01787}$ \\
\hline
\phantom{\Big|}$\Omega_M$  & $0.3045_{-0.0071}^{+0.0064}$ & $0.2870^{+0.0076}_{-0.0089}$ \\
\phantom{\Big|}$z_{\rm re}$  & $9.305_{-1.236}^{+1.334}$  & $9.333^{+1.689}_{-1.543}$ \\
\phantom{\Big|}$\sigma_8$  &  $0.8205_{-0.0099}^{+0.0119}$   & $0.8318^{+0.0180}_{-0.0150}$ \\
\hline
\phantom{\Big|}$\chi_{\rm min}^2$  & 13639.26  & 13651.86 \\
\phantom{\big|}$\Delta\chi_{\rm min}^2$  & 0  & 12.6 \\
\hline
\end{tabular}
\caption{\label{tab:resD42}Parameter estimation and $\chi^2$ values for $\nu\Lambda$CDM and the $\Delta_4$ model, using the CMB, BAO and SNe datasets, and the value $H_0 = (73.24 \pm 1.74) {\rm km}\, {\rm s}^{-1}\, {\rm Mpc}^{-1}$ from local measurements.}
\end{table}

\vspace{2mm}

\noindent
{\bf Parameter estimation for the $\Delta_4$ model}.
We have then implemented the perturbations in our Boltzmann code and performed Bayesian parameter estimation. The results are shown in Table~\ref{tab:resD41} (using  CMB+BAO+SNe) and  
Table~\ref{tab:resD42} (using  CMB+BAO+SNe +$H_0$). For easy of comparison, we write again the results for 
$\nu\Lambda$CDM already shown in Tables~\ref{tab:res1} and \ref{tab:res2}, and we give the difference in $\chi^2$ compared to $\nu\Lambda$CDM. We see that the $\Delta_4$ model indeed predicts a higher value of $H_0$, although not sensibly higher than that in the RR model. On the other hand, its $\chi^2$ is significantly worse than than that of $\nu\Lambda$CDM, even including $H_0$ in the dataset, so the model is already very strongly disfavored by the study of the scalar perturbations.

\vspace{2mm}

\begin{figure}[t]
\centering
\includegraphics[width=0.55\columnwidth]{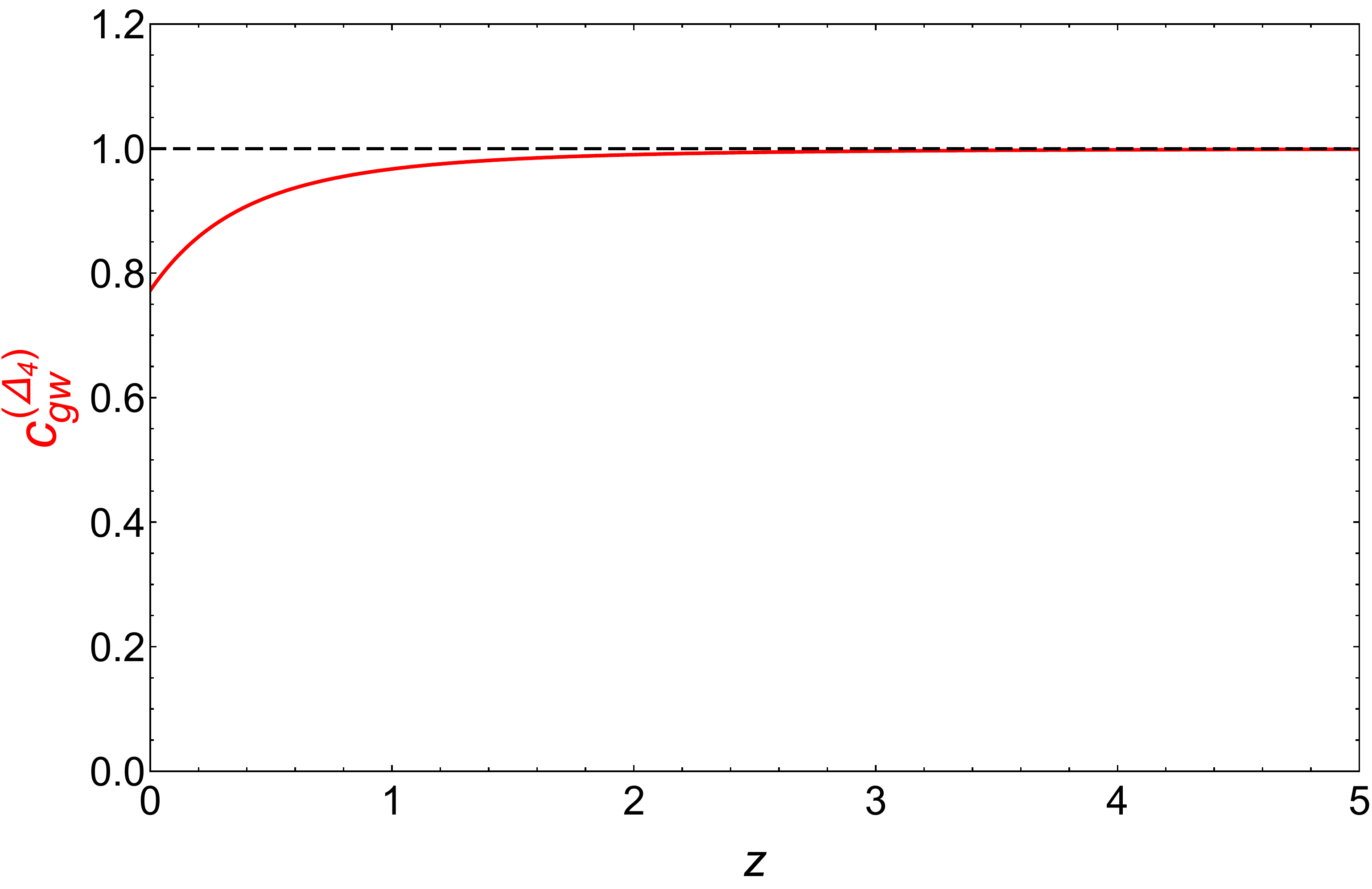}
\caption{\label{D4gw} The speed of gravitational waves in the $\Delta_4$ model as a function of the redshift.}
\end{figure}

\noindent
{\bf Tensor perturbations}. On top of this, we find that the model is ruled out by the fact that its tensor perturbations do not propagate at the speed of light. This is also interesting from the methodological point of view since it shows that, for these nonlocal models, it is not obvious a priori  to satisfy this constraint.
The  equation for tensor perturbations in the $\Delta_4$ model is
\bees
&& \( 1 - 3\gamma \bar{V} \) 
\(\pa_{\eta}^2\tilde{h}_A  +2{\cal H}\pa_{\eta}\tilde{h}_A+k^2\tilde{h}_A\)\nn\\
&&+\gamma \[ \frac{(\partial_\eta \bar{V})^2}{a^2 H_0^2}
\(2 \pa_\eta^2 \tilde{h}_A-k^2 \tilde{h}_A\)+\(4\bar{U}-3\) \pa_\eta \bar{V} \pa_\eta \tilde{h}_A \]   = 16\pi G a^2\tilde{\s}_A \, , \label{TensorD4}
\ees
and the corresponding speed of gravitational waves in a FRW background is

\be
c_{\rm gw}^{(\Delta_4)}=\sqrt{\frac{1-3\gamma \bar{V}- \frac{\gamma}{a^2 H_0^2}(\partial_\eta \bar{V})^2}{1-3\gamma \bar{V}+2 \frac{\gamma}{a^2 H_0^2}(\partial_\eta \bar{V})^2}}\,.  \label{GWspeedD4}
\ee
This quantity is always smaller than one, and in the recent epoch it differs from one significantly, see 
Fig.~\ref{D4gw}. Thus, the $\Delta_4$ model is ruled out by  the GW170817/GRB 170817A test.


\bibliographystyle{utphys}						
\bibliography{myrefs_massive}

\end{document}